# Light propagation and localization in modulated photonic lattices and waveguides

Ivan L. Garanovich[1], Stefano Longhi[2], Andrey A. Sukhorukov[1], and Yuri S. Kivshar[1]

[1]*Nonlinear Physics Centre and Centre for Ultra-high bandwidth Devices for Optical Systems (CUDOS), Australian National University, Canberra, ACT 0200, Australia*
[2]*Dipartimento di Fisica, Politecnico di Milano, I-20133 Milano, Italy*

**Abstract**

We review both theoretical and experimental advances in the recently emerged physics of *modulated photonic lattices*. Artificial periodic dielectric media, such as photonic crystals and photonic lattices, provide a powerful tool for the control of the fundamental properties of light propagation in photonic structures. Photonic lattices are arrays of coupled optical waveguides, where the light propagation becomes effectively discretized. Such photonic structures allow one to study many useful optical analogies with other fields, such as the physics of solid state and electron theory. In particular, the light propagation in periodic photonic structures resembles the motion of electrons in a crystalline lattice of semiconductor materials. The discretized nature of light propagation gives rise to many new phenomena which are not possible in homogeneous bulk media, such as discrete diffraction and diffraction management, discrete and gap solitons, and discrete surface waves. Recently, it was discovered that applying periodic *modulation* to a photonic lattice by varying its geometry or refractive index is very much similar to applying a bias to control the motion of electrons in a crystalline lattice. An interplay between periodicity and modulation in photonic lattices opens up unique opportunities for tailoring diffraction and dispersion properties of light as well as controlling nonlinear interactions. First, we review the linear effects in the modulated waveguides and waveguide arrays, including optical Bloch oscillations and optical dynamic localization, that are key to the understanding of the modulation-driven diffraction management of light. Then we analyze the effects of array boundaries and defects, and highlight a new type of modulation-induced light localization based on the defect-free surface waves. Finally, we discuss nonlinear properties of the modulated lattices with an emphasis on their great potential for all-optical shaping and switching.

## Contents

# 1. Introduction

## 1.1. Photonic lattices

Periodicity is one of the most fundamental physical and mathematical concepts, which also occurs frequently in nature. One of the primary examples is the regular arrangement of atoms in crystalline lattices in metals and semiconductors. The physics of such systems is very rich and has been extensively studied. For example, motion of electrons is strongly affected by periodic crystal potentials, allowing one to tailor the conducting properties of certain materials. This effect has lead to the invention of the transistor, which stands behind the development and ongoing advances of the microelectronics industry.

In optics, periodicity is associated with materials that exhibit a refractive index modulation in one or more spatial dimensions. Generic examples include optical gratings, slab waveguides, waveguide arrays and, introduced more recently, photonic crystals and photonic crystal fibers. Optical waves propagating in photonic crystals [99] behave in a way that is analogous to electrons traveling through a semiconductor crystal. Therefore, periodic photonic structures can be used to control the flow of light in a way quite similar to the case of electrons in semiconductor electronic devices, and thus hold an immense potential for the realization of optical integrated circuits, e.g., for optical computing, signal processing, and fiber communication systems. Much of today's research in optics is thus directed at exploring periodicity effects in photonic micro- and nanostructures.

Historically, the idea of discrete optical components emerged rather slowly in the field of optics. On the one hand, from a classical perspective, the electromagnetic field itself is a continuous function of both space and time, and it took scientists a while to realize that, on many occasions, light can exhibit behavior that is characteristic of that encountered in discrete systems [28]. Also, a more important barrier that prevented these ideas from becoming reality was the state of fabrication technologies. Although the first studies of optical coupling processes in waveguide arrays were performed in early 1960's [101, 266], it was only recently that high-contrast dielectric elements became available which made possible creation of photonic crystal structures [100].

Recently, discretized light propagation in *photonic lattices* has attracted a lot of interest [28]. Photonic lattices are periodic arrays of evanescently coupled optical waveguides, see examples in Fig. 1. In such structures, light can be readily confined at discrete sites in the weakly guiding waveguides, whereas at the same time light exchange among channels can occur via the evanescent coupling during propagation. Planar or one-dimensional waveguide arrays are periodic in one transverse direction, and translationally invariant with respect to the longitudinal direction of light propagation, while two-dimensional arrays are periodically modulated in both transverse directions. Thus, two-dimensional photonic lattices are similar to photonic crystal fibers [251], and some of the effects investigated in waveguide arrays can likewise be observed in photonic crystal fibers.

Waveguide arrays possess all the essential characteristics of a photonic crystal structure including Brillouin zones, allowed and forbidden bands and so on. As such, they support wave dynamics equivalent to the transport dynamics of electrons in semiconductors [319]. Bragg reflections and interference effects dominate the light propagation, especially for signals propagating across the array, and even linear dynamics in photonic lattices can be fundamentally different from that in homogeneous media.

There are several approaches which are used for the practical realization of one- and two-dimensional optical waveguide arrays. These include arrays of ridge waveguides etched in semiconductors [208] or created through titanium indiffusion in lithium niobate (LiNbO$_3$) crystals [202], lattices induced optically in photorefractive media [61], lattices created by periodic voltage biasing in liquid crystals [64], arrays of optical fibers [227, 247], optical fibers with multiple cores [17], and laser-written arrays in silica [222, 293]. The latter technique provides highly uniform arrays and holds much potential for the observation of effects in two-dimensional lattices [229].

Due to the periodic nature of the photonic lattices, many similarities with quantum-mechanics and solid state physics are found [168], what is often reflected in the terminology for their description. In particular, particles in periodic potentials, such as electrons in crystalline solids or custom-made semiconductor superlattices, Bose-Einstein condensates in optical lattices, and photons in photonic lattices, all share many common features which enable efficient knowledge exchange between different fields.

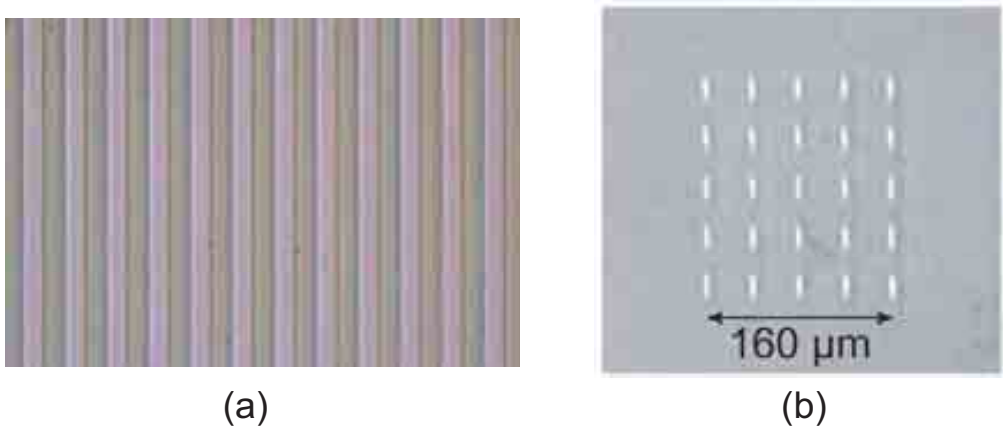

Figure 1: (a) Microscope image of a one-dimensional optical waveguide array (top view) with a waveguide separation of 14 $\mu$m fabricated by titanium indiffusion in LiNbO$_3$ crystals. (b) Microscope image of a two-dimensional cubic waveguide array (end facet view) with a waveguide separation of 40 $\mu$m fabricated by femtosecond laser direct-writing in glass. After Refs. [238, 283].

## 1.2. Discrete diffraction

### 1.2.1. Discretized light propagation

In an array of evanescently coupled waveguides, light hops from site to site through optical tunneling, and, in doing so, it profoundly alters its diffraction characteristics. There are two complementary approaches to the study of photonic lattices: coupled-mode description which considers lattice as a set of individual waveguides that are coupled together, and Floquet-Bloch analysis which treats lattice as a general type of periodic structure.

In the first approach, light propagation in waveguide array is primarily characterized by coupling due to the overlap between the fundamental modes of the nearest-neighboring waveguides. One can then effectively model the power exchange process in the array using coupled-mode theory, which is also known as the tight-binding approximation. This is possible if the waveguide sites are sufficiently separated, in which case the Floquet-Bloch functions belonging to the first band of the array can be described through the waveguide modes or bound states [28]. In this case, light propagation in a one-dimensional array can be described by a set of discrete equations for the mode amplitudes $\Psi_n(z)$ [28],

$$i\frac{d\Psi_n}{dz} + C(\lambda)\left[\Psi_{n+1} + \Psi_{n-1}\right] = 0. \qquad (1)$$

Here $z$ is the propagation distance along the waveguides, $n$ is the waveguide number, the coefficient $C(\lambda)$ defines the coupling strength resulting from the field overlap between the neighboring waveguides, which depends on the optical wavelength $\lambda$.

The infinite set of linear ordinary differential equations Eq. (1) is analytically integrable [275]. The general solution can be written using the Green's function, that defines the field evolution in the case when only one waveguide is excited at the input. Under such conditions, the solution for the electrical field in the $n$-th waveguide is given as

$$\Psi_n(z) = (i)^n \Psi_m(z=0) J_n(2Cz), \qquad (2)$$

where $i$ is the imaginary unit, $J_n$ is the Bessel function of the order $n$, and $m$ is the number of the excited guide. As the light propagates along the waveguides, the energy spreads into two main lobes according to Eq. (2), with several secondary peaks between them, see Fig. 2(b). Note that this diffraction is fundamentally different from that occurring in continuous systems as most of the optical energy is carried out along two major lobes far from the centre. This peculiar diffraction pattern became known as *discrete diffraction*.

For an arbitrary initial conditions, the diffraction pattern can be calculated as a linear superposition of the functions defined in Eq. (2), see examples of the resulting diffraction profiles in Figs. 2(c-e).

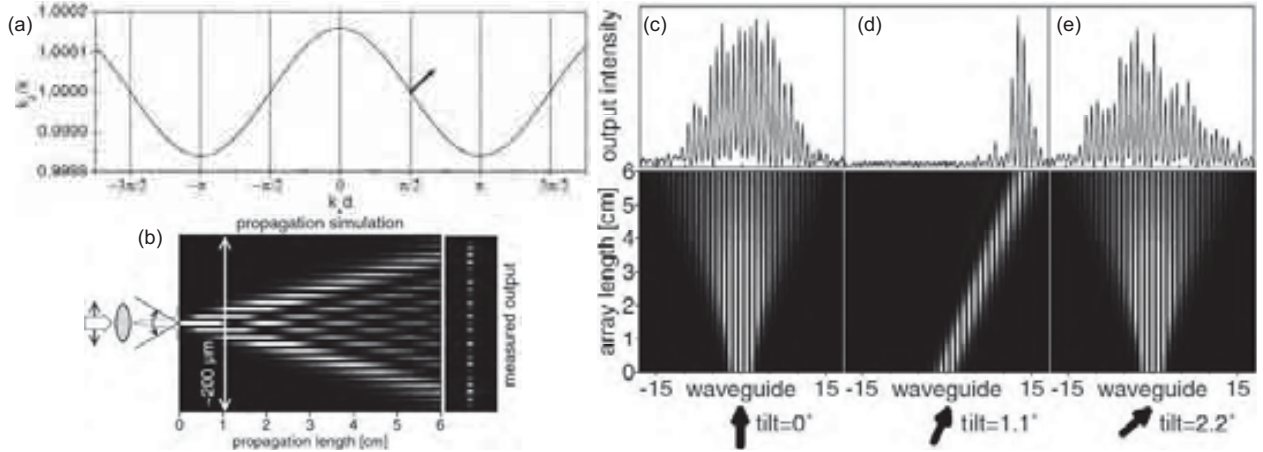

Figure 2: Discrete diffraction in photonic lattices. (a) Cosine diffraction curve in a one-dimensional waveguide array. The arrow shows the direction of beam propagation, which is normal to the diffraction curve. (b) A narrow input beam excites a single guide and produces a unique type of diffraction pattern, which is the Green's function of the array. (c-e) Measured output intensity profiles and simulated beam propagation shown for several input tilts for a broad Gaussian beam excitation. (c) Input tilt corresponds to the centre of the Brillouin zone, beam experiences normal diffraction. (d) Input tilt corresponds to the middle of the Brillouin zone, diffractionless propagation occurs. (e) Input tilt corresponds to the edge of the Brillouin zone, beam experiences anomalous diffraction. After Refs. [56, 230].

*1.2.2. Diffraction management*

Unlike a bulk system, where the magnitude of diffraction is fixed at a given wavelength, in photonic lattices, discrete diffraction can be deliberately engineered. The unusual dispersion properties of waveguide arrays make it possible to realize diffraction management of the light beams [56]. Each mode of the array is an extended Bloch wave with its own propagation constant and propagation direction given by the normal gradient to the transmission band [see Fig. 2(a)]. During linear propagation, each mode evolves independently of the others, acquiring its own individual phase. Any wave exciting the array is decomposed into the Bloch modes, and therefore the group dynamics will be determined by the band geometry. Since the modes remain the same but acquire different relative phases during propagation, the waveform may have a significantly different profile as it exits the array. Thus, photonic lattices can modify the spreading of light beams in the way much analogous to the dispersion management of optical pulses which is routinely done with gratings in optical fibers.

Let us consider the propagation of plane waves of the form

$$\Psi_n(z) = \Psi_0 \exp\left[ik_x nd + ik_z z\right], \qquad (3)$$

where $k_x$ and $k_z$ are components of the two-dimensional wavevector $k$, and $d$ is the distance between the centres of the adjacent waveguides. By substituting expression (3) into the coupled-mode equations Eq. (1), one can calculate the diffraction relation of the photonic lattice which reads

$$k_z = 2C \cos(k_x d) \qquad (4)$$

In photonic lattices, this diffraction relation, $k_z$ vs. $k_x$, plays the role of the dispersion relation, $\omega$ vs. $k$, in the temporal domain, where $\omega$ is the frequency. As with any periodic wave system, the linear modes in the waveguide array are extended Floquet-Bloch modes, with a transmission spectrum consisting of allowed bands separated by forbidden gaps, see Figs. 2(a) and 3(a). That is, for all the spatial modes, there is a range of propagation constants which are not allowed. Such a dispersion relation in photonic lattices differs dramatically form the dispersion relation in free space or in homogeneous media.

The second-derivative of the diffraction relation gives the relative spread or convergence of adjacent rays, so that band curvature determines the diffractive properties of a wavepacket. Diffraction strength for the light beams is determined by the diffraction coefficient [56],

$$D = \frac{\partial^2 k_z}{\partial k_x^2} = -2Cd^2 \cos(k_x d), \qquad (5)$$

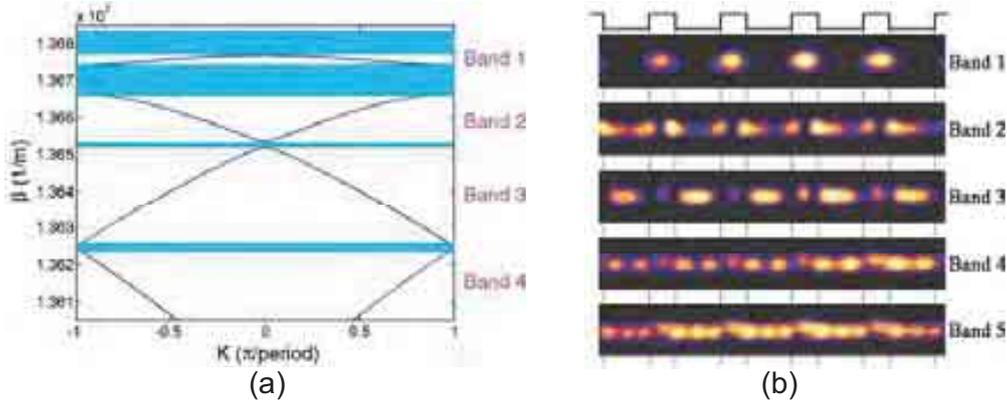

Figure 3: (a) Reduced band-gap diagram of a typical waveguide array folded into the first Brillouin zone, where the propagation constant is plotted as a function of the Bloch wave number. The shaded regions represent the gaps. The four lowest lying bands are shown. (b) Experimental images of the pure Floquet-Bloch modes excited in each band of the array. After Ref. [196].

which is derived for optical beams using the Eq. (4), in analogy with the definition of the temporal dispersion coefficient for light pulses.

Depending on the sign of the diffraction coefficient, beam diffraction can be normal or anomalous, or it even can completely disappear. Specifically, the propagation constants of the Bloch modes are governed by the transmission spectrum, so that linear dispersion is determined by the amount of relative phase acquired by the modal constituents of the beam during propagation. In regions of convex band curvature, the central mode propagates faster than its neighbors, and the beam acquires a convex wavefront during propagation [see Figs. 2(a) and 2(c)]. These are regions of normal diffraction, with wave behavior analogous to that in homogeneous media. By contrast, a group of modes in concave regions of band curvature will evolve anomalously, acquiring a concave wavefront during propagation [see Figs. 2(a) and 2(e)]. Note that there is an inflection point in each band, meaning that wave-packets propagating in this direction experience no diffraction [see Figs. 2(a) and 2(d)].

From an experimental perspective, simply changing the excitation angle, which determines the transverse momentum $k_x$ of a beam, controls its diffraction properties, as it is illustrated in Figs. 2(c-e).

We note, that lattice effects on wave propagation depend on the overall size of the input beam relative to the waveguide spacing. For example, a broad Gaussian beam launched on-axis into the array excites modes from different bands and stays mostly Gaussian as it propagates, as in the homogeneous case, with channels simply modulating the wave profile [see Figs. 2(c-e)]. By contrast, coupling a narrow beam into the fundamental guided mode of a single waveguide excites Bloch modes primarily from the first band. In this case, beam spreading follows the discrete diffraction pattern characterized by intense side lobes with little or no light in the central waveguide [see Fig. 2(b)].

More accurate description of waveguide arrays can be obtained using Floquet-Bloch theory, which provides a formal description not only of the on-site fundamental modes, but also of the high order waveguide modes, and radiation modes that propagate between the waveguides. Fig. 3(a) shows the calculated full band-gap diagram of a typical waveguide array. It relates the propagation constant to the Bloch wavenumber, which is the transverse component of the wavevector reduced to the first Brillouin zone. The propagation direction of each Bloch mode is given by the normal to the diffraction curve. The coupled-mode analysis describes only propagation within the first of these bands, where the energy is concentrated in the high-index waveguides [see Fig. 3(b)]. The diffraction curve of this first band is nearly sinusoidal, as predicted by the coupled-mode theory.

It is important to note that while many of the theoretical studies of modulated photonic lattices are based on the coupled-mode equations which are similar to Eq. (1), the full continuous simulations of the beam propagation are required to check the radiation losses at the waveguide bends.

*1.3. Discrete solitons*

*1.3.1. Optical nonlinearities*

In free space, the propagation of electromagnetic waves is governed by linear Maxwell's equations [96]. Therefore, the principle of superposition applies, and individual light beams cannot interact with each other. Also, once emitted from a source light cannot change its frequency.

However, when light propagates in some physical media, such as dielectric materials, the picture looks completely different. The interaction between the optical field and the atoms in the media can lead to a certain type of the nonlinear response. Such nonlinear interaction in general modifies the optical properties of the material which in turn affects the light propagation. That is, optical nonlinearity is a property of the material, rather than of the electromagnetic field itself. As a result, it becomes possible for an intense light beam to interact with itself or other beams in a nonlinear medium, the former phenomenon being known as the nonlinear self-action of light. In practice, it is often impossible to strictly categorize an optical system as being either linear or nonlinear. Typically, linearity represents an idealized first order approximation to a more complex nonlinear problem.

In the classical electromagnetic theory [96], light-matter interaction is characterized by the polarization density $P(E)$ of the medium, which is a function of the applied electric field $E$. The polarization density represents the medium response to the applied electric field.

Any departure from the linear relationship between $E$ and $P$ is a manifestation of a nonlinear interaction. In most situations, nonlinearity is relatively small and can be treated as a perturbation to the linear relation, which only becomes significant when $E$ is large. It is therefore possible to expand $P$ in a Taylor's series about $E = 0$, using only a few terms. In common notation, the expansion reads [18]

$$P = \epsilon_0\{\chi^{(1)}E + \chi^{(2)}E^2 + \chi^{(3)}E^3 + ...\}, \tag{6}$$

where $\epsilon_0$ is the permittivity of free space, $\chi^{(1)}$ is the linear susceptibility, and $\chi^{(n)}$ represents the $n$-th order nonlinear correction. Thus, in a nonlinear material, the susceptibility and hence the refractive index are not constant, but depend on the value of the electric field $E$. The nonlinear expansion coefficients, $\chi^{(n)}$, characterize the nature and strength of the nonlinear interaction between the optical field and a particular material.

In centro-symmetric media, the second order nonlinear susceptibility vanishes due to the inversion symmetry. In this case, the third order term $\chi^{(3)}$ becomes the leading nonlinear correction. The refractive index change in case of the third-order nonlinearity, known as the optical Kerr effect, is proportional to the intensity of the optical field,

$$I = \left|E^2\right|, \tag{7}$$

such that the refractive index can be represented as

$$n(I) = n_0 + n_2 I, \tag{8}$$

where $n_0$ is the linear refractive index, and the nonlinear material coefficient $n_2$ is proportional to $\chi^{(3)}$ [254].

The magnitude and sign of the parameter $n_2$ characterizes the strength and type of nonlinearity, respectively, where $n_2 > 0$ corresponds to self-focusing nonlinearity and $n_2 < 0$ corresponds to defocusing nonlinearity. An optical wave of intensity $I$ traveling through a nonlinear medium with an intensity dependent refractive index $n(I)$ experiences self-phase modulation according to

$$\Delta\phi = \frac{2\pi}{\lambda Ln(I)}, \tag{9}$$

where $\Delta\phi$ is the nonlinear phase shift acquired after a propagation distance $L$, and $\lambda$ is the wavelength. The higher the intensity, the larger the induced change in refractive index and rate of phase accumulation. A spatially finite laser beam with a non-uniform transverse intensity distribution $I(x, y)$ will experience uneven self-phase modulation over the beam cross section, and this strongly affects the evolution of the spatial beam profile upon propagation. For a typical optical beam the intensity and the nonlinear phase shift is the largest in the beam centre and decreases away from the propagation axis in the wings of the beam, leading to the formation of the effective optical lens. This lens acts as to either focus or defocus the beam upon propagation, depending on the sign of the nonlinearity. This effect is commonly known as beam self-focusing and self-defocusing.

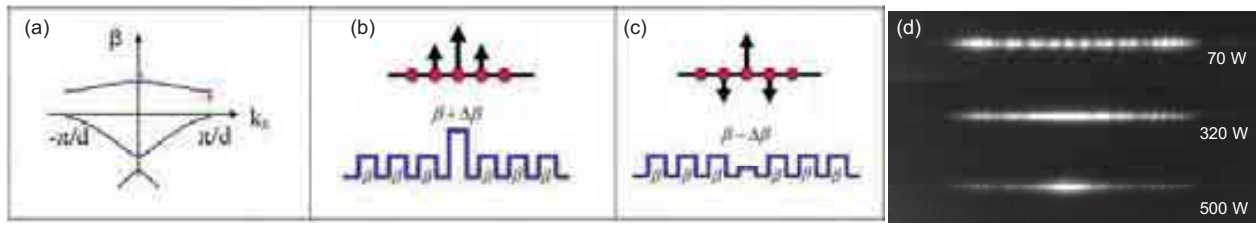

Figure 4: One-dimensional discrete solitons originating from the lattice first band. (a) Lattice transmission spectrum showing the nonlinear propagation constants for the fundamental lattice soliton (centre of the Brillouin zone) and spatial gap soliton (edge of Brillouin zone). (b) The fundamental lattice soliton has an in-phase structure that requires a self-focusing nonlinearity (positive defect). (c) The first-band gap soliton has a staggered phase structure that requires a defocusing nonlinearity (negative defect). (d) The first experimental observation of discrete optical solitons. Images of the array output facet are shown for different peak powers. At low power (70 W) linear discrete diffraction pattern is observed with two main lobes and a few secondary peaks in between. At intermediate power (320 W) the beam self-focusing takes place. At high power (500 W) a discrete soliton is formed. After Refs. [59, 57].

Different nonlinear optical effects have been of great fundamental and practical interest since the invention of the laser in the 1960's. Lasers can provide high light intensities required in order to observe experimentally nonlinear effects in most of the materials, and this made it possible to study numerous novel effects such as harmonic generation, frequency conversion, four-wave mixing, parametric generation and amplification, optical rectification, pulse generation, optical bistability, self- and cross-phase modulation, self-focusing, and soliton formation.

*1.3.2. Spatial solitons*

Diffraction is a fundamental phenomenon which leads to beam broadening upon propagation. In a nonlinear medium, the self-focusing effect reduces diffraction whereas the self-defocusing enhances the beam spreading. In a situation where the nonlinear self-focusing exactly balances diffraction, the beam can propagate as an optical spatial soliton, i.e. a self-trapped beam which preserves its shape upon propagation [122].

One can also understand the formation of spatial solitons through a waveguide analogy. In essence, an optical beam can create its own waveguide when it propagates in a nonlinear medium, and it can be trapped by this self-induced waveguide. The spatial soliton can be thought of as the fundamental mode of this waveguide. Moreover, such a nonlinear waveguide can even guide a weak probe beam of a different frequency or polarization.

In bulk media, self-defocusing nonlinearity does not support the formation of bright solitons which have an intensity maxima in their centre. Self-defocusing, however, can be used to generate dark solitons, which are formed by the nonlinear localization of a narrow dip in an otherwise uniform intensity background. Experimentally bright solitons can be excited by shining a narrow laser beam into a self-focusing material, and dark solitons can be observed in self-defocusing media by launching a broad beam with a zero intensity dip created, e.g., by imprinting a $\pi$ phase jump or a vortex phase singularity onto the beam.

The formation of spatial solitons by nonlinear compensation of the spatial diffraction broadening is analogous to the appearance of optical temporal solitons in nonlinear optical fibers, where self-phase modulation is used to balance the temporal dispersion broadening, such that the optical pulse preserves its shape upon propagation.

Self-focusing of optical beams in a bulk nonlinear medium was observed experimentally as early as in 1964 [23]. Later on, stable optical spatial solitons were observed using nonlinear media in which diffraction spreading was limited to only one transverse dimension [8]. Over the past several decades, the existence and unique properties of spatial optical solitons in homogeneous cubic, photorefractive, and quadratic nonlinear media have been studied extensively both theoretically and experimentally [122].

*1.3.3. Lattice solitons*

Nonlinear effects can considerably alter beam propagation in a waveguide array. In the case of a narrow beam, the light modifies the refractive index locally, thereby inducing a defect in the periodic structure of the lattice. In contrast to the extended Bloch modes of the perfect lattice, such a defect can support localized modes, which propagation constants lie off the linear transmission band, i.e. in a gap. When these nonlinear modes induce the defect and populate it self-consistently, the light beam becomes self-localized and its diffractive broadening is eliminated similar to the optical spatial solitons in homogeneous media. If the beam spatial profile remains constant and stable during propagation, then it is called a discrete soliton or a lattice soliton [136, 275].

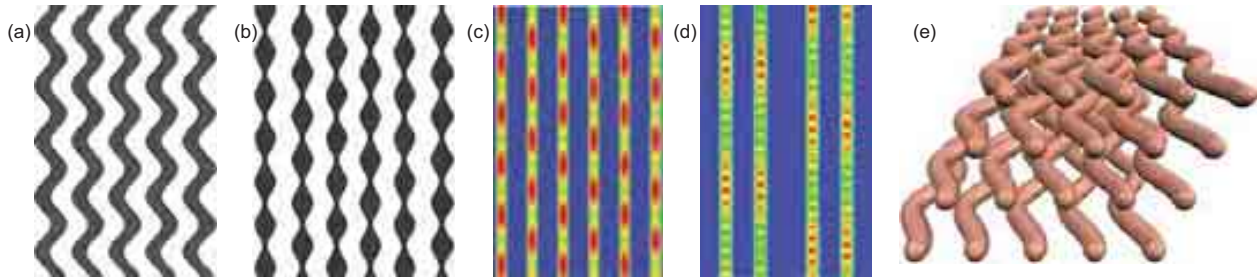

Figure 5: Different types of modulated photonic lattices. (a) Periodic axis bending. (b) Out-of-phase modulation of the waveguide width. (c) Out-of-phase modulation of the refractive index. (d) Combined in- and out-of-phase modulation. (e) Two-dimensional modulation. After Refs. [184, 112, 145, 291].

The influence of the nonlinearity, however, depends on where one is on the diffraction curve of the photonic lattice. In regions of normal diffraction, a focusing nonlinearity can compensate the convex curvature of the wavefront [26, 57] [see Figs. 4(a) and 4(b)], while regions of anomalous diffraction require a defocusing nonlinearity [121, 58, 214, 60] [see Figs. 4(a) and 4(c)]. Thus, depending on the type of local diffraction, discrete solitons can be formed in a lattice with both focusing and defocusing nonlinearity. This in a sharp contrast to the spatial solitons in homogeneous media, where only self-focusing nonlinearity can lead to the formation of the bright optical solitons. A spatial soliton is associated with a single propagation constant, and therefore it maintains its spatial profile and propagates as a single entity. In photonic lattices, this can happen only if the propagation constant of the beam deviates from the linear bands of the lattice transmission spectrum. That is, the propagation constant of a soliton is in a gap, and as such it represents a localized state, rather than the extended Bloch modes. For the fundamental on-axis discrete or lattice soliton [26], the propagation constant lies in the semi-infinite gap above the first band [see Fig. 4(a)]. On the other hand, the lattice soliton at the edge of the Brillouin zone [121], which is often called a spatial gap soliton [58], is formally equivalent to the spatiotemporal gap solitons observed in optical fibers [22, 27, 2, 262, 55]. Propagation constant of the gap soliton lies between the first and second lattice transmission bands [see Fig. 4(a)], while the Bragg condition $k_x = \pi/d$ gives this soliton a staggered phase profile [see Fig. 4(c)]. Gap solitons have no equivalent in homogeneous media. For modes originating from the first band, the curvature dictates that a defocusing nonlinearity is necessary for the formation of bright gap solitons, when the propagation constant goes down into the gap with increasing nonlinearity.

Fig. 4(d) shows the first experimental observation of a bright discrete soliton in a self-focusing AlGaAs waveguide array. At low optical powers, the optical field diffracts discretely in the array [cf. Fig. 4(d) and Fig. 2(b)]. At higher powers, however, the optical field self-localizes, which provides a clear indication of discrete soliton formation.

*1.4. Modulated photonic lattices*

In photonic lattices, the classical light tunneling between the neighboring waveguides [101, 266] closely resembles the quantum electron dynamics in crystalline potentials. Whereas the monitoring of the fast temporal evolution of the electron motion in crystals is a complicated problem, the spatial propagation of optical beams can be observed directly in real space, e.g. by means of fluorescence imaging [25, 39] or tunneling optical microscopy [33] techniques. This has stimulated the experimental studies of optical analogues for a variety of phenomena which were predicted originally for quantum systems including Bloch oscillations [226, 216, 25], Zenner tunneling [303, 301, 65] and dynamic localization [183, 95] (for a comprehensive review of various quantum-optical analogies in photonic structures see [168]).

The principle underlying the quantum-optical analogies in photonic lattices is based on the similarity between the wave equations describing spatial propagation of light beams in arrays of optical waveguides, and the temporal Schrödinger equation for quantum particles in a periodic potential driven by an external electromagnetic field [168]. Specifically, the effects associated with the presence of external driving field can be represented by a special modification of the photonic lattice by one of the following methods: (i) the transverse modulation of the photonic lattice, such as changing the waveguide strength and spacing [216] or the background refractive index [226, 301], or (ii) the bending of waveguides in the longitudinal direction [139]. Examples of different types of waveguide modulation, including periodic axis bending, out-of-phase modulation of the waveguide width, out-of-phase modulation of the refractive index, and combined in- and out-of-phase modulation are shown in Figs. 5(a-e).

Importantly, the characteristic longitudinal bending periods in photonic lattices are several orders of magnitude larger than the wavelength and the lattice transverse period, supporting adiabatic beam evolution along the structure in a broad spectral region. This is fundamentally different from the photonic crystals [99] which are featuring two- and three-dimensional modulation of the optical refractive index on the order of the wavelength. One important difference between the photonic crystals and photonic lattices is that the diffraction can only be optimized in photonic crystals in a narrow spectral window around certain resonant wavelength [126, 24, 243]. This is because the operation of photonic crystals is based on the scattering of light associated with the effect of Bragg reflection. This is a resonant process, which is highly sensitive to the optical wavelength. As a result, the typical bandwidth in photonic crystals is less than 10% of the central wavelength. In contrast, the resonant back-scattering of light is absent in photonic lattices, and therefore their wavelength dependence is rather smooth. This provides optimal conditions for diffraction management, as we will discuss in detail in Sec. 2 below.

The modulated photonic lattices not only provide a means to study the fundamental wave phenomena, but can also be directly used to control many aspects of light propagation in very efficient ways. Such photonic applications of modulated lattices constitute the focus of this review.

*1.5. Outline of the review*

After introducing the major concepts, we may outline the structure of this review. In Sections 2, 3, and 4 we review linear effects in modulated lattices and waveguides. First, in Sec. 2 we consider modulated lattices which are infinite in the transverse direction, when the boundary effects are negligible, and demonstrate that longitudinal lattice modulation can lead to new propagation phenomena, such as optical Bloch oscillations, Zener tunneling, Rabi oscillations, self-collimation, and dynamic localization effects. Then, in Sec. 3 we discuss new effects in finite sets of curved waveguides, such as adiabatic light transfer and broadband coupling. Finally, in Sec. 4 we analyze transmission and localization properties of modulated lattices with defects, including a new type of defect-free surface waves, light tunneling control, and optical Zeno effect. The second part of this review paper is devoted to the study of nonlinear effects. In particular, in Sec. 5 we analyze light propagation in nonlinear modulated lattices and discuss modulation instability, diffraction-managed solitons, interband nonlinear interaction, and nonlinear surface waves. Finally, Sec. 6 provides a summary of our work, and it also compares briefly modulated photonic lattices with photonic crystals. We also mention the applications of the basic concepts of this field to other photonic, plasmonic, and metamaterial structures. In Appendices, we derive generalized tight-binding model for one- and two-dimensional modulated photonic lattices.

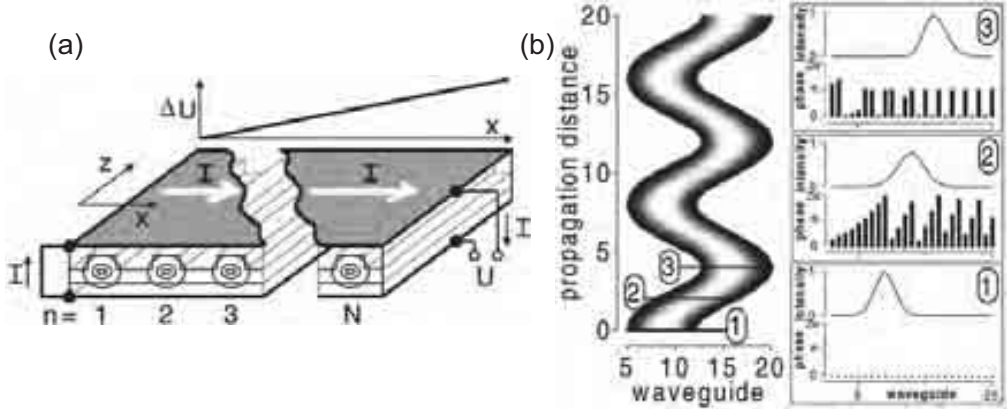

Figure 6: (a) Electro-optic waveguide array sandwiched between two current-leading electrodes. The changing voltage $\Delta U$ produces a linear variation of the propagation constants of the waveguides. (b) Numerical simulation of periodic motion (Bloch oscillation) of a discrete Gaussian beam in a waveguide array. The insets show the intensity and the respective phase distribution at three different distances ($\alpha = 0.5$). After Ref. [233].

## 2. Periodic modulated lattices

### 2.1. Optical Bloch oscillations

The motion of an electron in a crystal driven by an external dc electric field is a well known problem in solid-state physics which is at the basis of the early quantum theory of electrical conductivity. Around 1930s, Bloch and Zener predicted that a static electric field applied to a one-dimensional crystal should induce an oscillatory (rather than uniform) motion of the electrons, known as Bloch oscillations (BO) [16, 328]. The existence of BO is basically related to the circumstance that the energy spectrum of the electronic Hamiltonian changes from continuous (a band structure with delocalized Bloch eigenstates) in absence of the external field, to a discrete ladder energy spectrum and localized eigenfunctions (called Wannier-Stark spectrum) when the external field is applied. Optical analogues of electronic Bloch oscillations have been observed as temporal pulse oscillations in optical super-lattices [212, 32, 31, 255, 3, 78], or as spatial beam oscillations in optical waveguide arrays [226, 216, 28, 25, 300, 302, 63]. In waveguide arrays, the waveguides play the role of the attracting atomic potential in the crystal, whereas the external dc field is mimicked by a superimposed transverse refractive index ramp. As we describe below, optical waveguide arrays enable a direct visualization of Bloch oscillations in the spatial domain as an oscillatory beam path.

#### 2.1.1. Arrays with index gradient

Peschel and collaborators [233] showed that an array in which the difference of the propagation constants between any two adjacent waveguides is constant exhibits peculiar dynamics: light undergoes periodic motion and linear localization instead of the conventional discrete diffraction. These effects can be understood in terms of an interplay between total internal and Bragg reflection and can be identified as optical Bloch oscillations.

Following Ref. [233], we apply the coupled-mode theory to describe the evolution of the modal amplitudes $\Psi_n$ in the $n$-th waveguide by the scaled equation

$$i\frac{d\Psi_n}{dz} + C[\Psi_{n+1}(z) + \Psi_{n-1}(z)] + \alpha n \Psi_n(z) = 0, \qquad (10)$$

where $C$ is the linear coupling constant, $\alpha = \Delta\beta$ defines the linear index gradient, and $\Delta\beta$ is the wave-number spacing between two waveguides.

Field evolution in the array can be related to beating of localized eigenmodes with equidistant wave numbers, called a Wannier-Stark ladder. A discrete Fourier transform of Eq. (10) can be used to find the Wannier-Stark states of the array of the form $\Psi_n^m(z) = u_n^m \exp(i\beta_m z)$. This notation means that the supermode $m$ has the wave number $\beta_m$ and its profile is characterized by a superposition of the constant amplitudes $u_n^m$ of the modes of the individual waveguides. It was shown in Ref. [233] that the wave number $\beta_m$ can attain only multiple integers of $\alpha$, $\beta_m = m\alpha$. This eigenmode spectrum constitutes the Wannier-Stark ladder [31]. It is discrete and unbounded, in contrast with the continuous and bounded spectra in uniform arrays. The constant mode spacing $\alpha$ results in a complete recovery of the initial field after a propagation distance of $z = L_B$, where the Bloch period is

$$L_B = \frac{2\pi}{\alpha}. \tag{11}$$

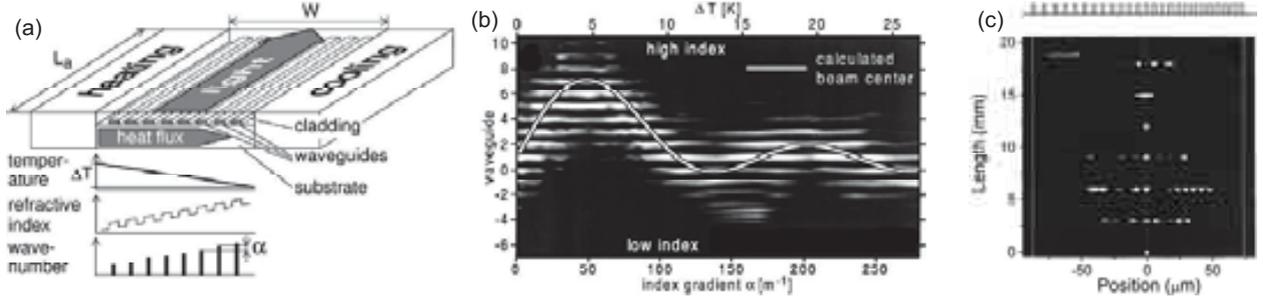

Figure 7: Observation of optical Bloch oscillations. (a) Schematic of a thermo-optic polymer optical waveguide array sandwiched between two copper blocks at different temperature and (b) output intensity for a broad Gaussian beam excitation($C = 105$ m$^{-1}$). (c) Bloch oscillations for a single waveguide excitation: the power distribution is shown as a function of the propagation length. (Solid line: array boundaries. Dashed line: input guide. Arrow: direction of growing index. Insert on the top shows transverse profile of the AlGaAs waveguide array.) After Refs. [226, 216].

Indeed, numerical solution of Eq. (10) shows that at an initial stage of evolution the excitation experiences discrete diffraction and spreads over several waveguides, but in contrast with the behavior in both a homogeneous array the beam refocuses and returns to the initial position [see Fig. 6(b)], and a periodic motion occurs that is reminiscent of Bloch oscillations. The excitation remains localized, and no unconfined radiation is observed.

There is a simple physical explanation for these optical Bloch oscillations. They can be understood in terms of an interplay between the total internal and Bragg reflection. The linear index gradient $\alpha$ causes two principal effects: an increasing phase difference between adjacent waveguides [phase profiles 1 and 2 in Fig. 6(b)] and a lateral shift of the beam position. The discrete nature of the system comes into play if this phase difference approaches $\pi$ [phase profile 3 in Fig. 6(b)]. When the tilt of the beam corresponds to the angle for the Bragg reflection at the array, the field evolution reverses toward decreasing wave numbers, until it experiences the total internal reflection and is deflected toward increasing wave numbers again.

To observe optical Bloch oscillations, several approaches were used to create waveguide arrays with linear variation of the propagation constant. Pertsch and collaborators [226] obtained the required linear variation of the propagation constant across the thermo-optic polymer array by applying a temperature gradient [see Fig. 7(a)]. By heating and cooling the opposite sides of the array a transverse linear temperature gradient was established, leading to a linear variation of the refractive index as well as the propagation constants of the waveguides across the array. Simultaneously a homogenous evanescent coupling constant $C$ was preserved. The period and amplitude of the oscillations can be controlled by varying the temperature gradient. While increasing the temperature gradient, the oscillating transverse motion of the output beam center was measured, which is displayed in Fig. 7(b) for a broad Gaussian beam excitation. The decrease of the beam broadening in the region from $\alpha = 70$ m$^{-1}$ to $\alpha = 210$ m$^{-1}$ is due to the stronger localization of the Wannier-Stark states for higher $\alpha$. Furthermore, for increasing $\alpha$ the spreading of the excitation is reduced down to a minimum at $\alpha = 140$ m$^{-1}$ and $\alpha = 140$ m$^{-1}$, where the integer number (one and two) of Bloch oscillations periods occurs at the structure length and the initial Gaussian distribution of the excitation is therefore recovered.

Morandotti and collaborators [216] fabricated ridge waveguides [for a schematic drawing see top of Fig. 7(c)] etched on top of an AlGaAs slab waveguide. To obtain a linear increase of the effective index the rib width was varied from 2 to 3.4 mm, corresponding to an index difference of $\delta n = 1.3 \times 10^{-4}$ between adjacent guides. To ensure constant coupling also the spacing between the guides was varied from 6.6 to 3.3 mm [see top of Fig. 7(c)]. The sample was cleaved into pieces of different length varying from 3 to 18 mm to allow for an insight into the field evolution. Figure 7(c) shows the experimentally observed field evolution for a narrow input beam (3 mm) launched into the central waveguide only. Similarly to the case of a homogeneous array, the field first spreads into both directions. In contrast to the broad beam excitation [see Fig. 7(b)], the action of the linear potential does not result here in shifting into the direction of growing indices but in a subsequent refocusing of the beam.

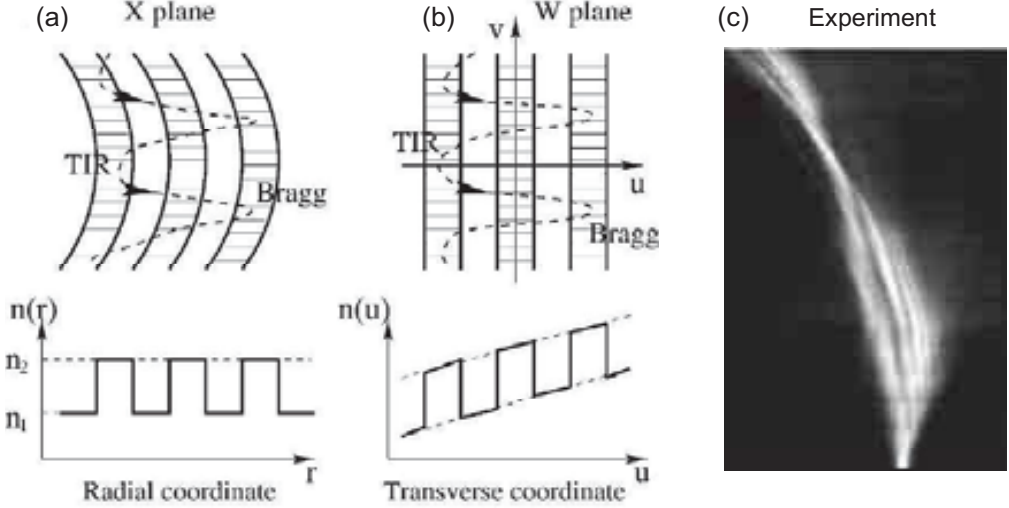

Figure 8: (a,b) Top: Schematic of (a) a curved waveguide array and (b) its conformal transformation. The dashed arrows illustrate the light confinement mechanisms - Bragg reflection on one side of the structure and total internal reflection on another side. Bottom: (a) Refractive index in the curved array, and (b) the transformed refractive index. The refractive indices of the guides and the surrounding material are $n_2$ and $n_1$, respectively. (c) Measured fluorescence pattern corresponding to the central waveguide excitation of a curved array with radius of curvature $R = 7.7$ cm and period $d = 8$ $\mu$m. After Refs. [139, 25].

The striking feature is the apparent symmetry of the power distribution in Fig. 7(c), which has a simple physical explanation [233]. Evidently, if $\Psi_n(z)$ is a solution of Eq. (10), $\tilde{\Psi}_n(z) = (-1)^n \Psi^*_{-n}(z)$, where the asterisk stays for the complex conjugation, is likewise a solution that has the same intensity profile but exhibits a phase jump of $\pi$ between the adjacent waveguides. As a consequence one solution tends to move toward increasing and the other one toward decreasing wave numbers. For the single waveguide excitation $\Psi_n(0) = \tilde{\Psi}_n(0)$ and the field is evenly decomposed into both solutions, so that the intensity distribution stays symmetric on propagation, $|\Psi_n(z)|^2 = |\Psi_{-n}(z)|^2$ [Fig. 7(c)].

### 2.1.2. Circular arrays

Imposing a uniform linear variation in the index profile is technologically challenging [226, 216]. Lenz and collaborators [139] proposed a simple and practical alternative optical structure that can demonstrate optical Bloch oscillations. The structure consists of an array of periodically spaced, curved optical waveguides. The curvature plays a central role here, leading to an effective linear ramp in the refractive index distribution. This ramp is superimposed on the periodic index distribution due to the equal spacing of the guides, as required by Eq. (10). The advantage to observing BO in a curved waveguide array is that fabrication of a curved waveguide array is not more difficult than any standard photonic circuit. The linear potential is controlled by changing the bending radius which can be very accurately defined.

To analyze light propagation in curved waveguides, we follow the approach of Lenz and collaborators [139], based on conformal transformations, and consider a coordinate system $u\,v$ which is defined with respect to the original coordinates $y$ and $z$ by $W = u + iv = f(X) = f(y + iz)$. As shown in Ref. [139], a logarithmic conformal transformation,

$$W = R \ln\left(\frac{X}{R}\right), \tag{12}$$

where $R$ is a radius of the curvature, converts curved boundaries in the $X$ plane [see Fig. 8(a)] into straight ones in the $W$ plane [see Fig. 8(b)]. The transformed refractive index in this equivalent structure is a product of $\exp(u/R)$ and the refractive index in the appropriate region of the curved guide. For a typical waveguide array, the coordinate $u$ varies on a micrometer scale, whereas $R$ varies on a millimeter scale. Thus, the difference between $\exp(u/R)$ and its Taylor expansion $[1 + (u/R)]$ is virtually indistinguishable. Therefore, the ramp in the refractive index distribution can be approximated as linear with the ramp parameter $\alpha = n_{\text{eff}}/R$ where $n_{\text{eff}}$ is the effective index [139], making this geometry suitable for observation of optical Bloch oscillations. The spatial period of the optical Bloch oscillations, $L_B$, depends linearly on $R$ [139]:

$$L_B = \frac{\lambda R}{d n_{\text{eff}}}, \quad (13)$$

where $\lambda$ is the wavelength, and $d$ is the array period. It was demonstrated numerically in Ref. [139] that the light propagation in this geometry exhibits spatial Bloch oscillations with the longitudinal period depending on the radius of the curvature and the wavelength. The wavelength dependence is a unique feature of optical Bloch oscillations which has no analogy in the solid state case.

Direct visualization of optical Bloch oscillations in curved waveguide arrays by exploiting the green upconversion fluorescence of an Er:Yb-doped glass substrate was reported by Chiodo and collaborators [25]. For the experiment, a set of circularly curved waveguide arrays was designed and manufactured by use of a femtosecond laser writing technique in an active Er:Yb-doped phosphate substrate. Each array consisted of 11 curved channel waveguides with different combinations for the radius of curvature $R$ and waveguide separation $d$. Owing to the relatively modest number of waveguides in each array, to avoid boundary effects in the experiment BO dynamics was visualized by single-waveguide excitation. Bright green upconversion luminescence was monitored and recorded from the top of the sample by a CCD camera connected to a microscope. Fig. 8(c) shows a typical fluorescence pattern observed when curved array was excited. Note that a clear breathing pattern is observed which corresponds to Bloch oscillations.

*2.1.3. Summary*

Prediction and successful experimental demonstration of optical Bloch oscillations in photonic lattices stimulated further studies of Bloch dynamics in waveguide systems. Bloch oscillations in photonic lattices with *next-nearest-neighbor interaction* where recently considered by Wang and collaborators [315] using an example of a quasi-one-dimensional zig-zag waveguide array (for dynamic localization and diffraction management in a similar zig-zag structure see Sec. 2.4.2). It was shown in Ref. [315] that BOs in the lattices with the second-order coupling exhibit complex spatial oscillation patterns featuring a double turning-back which occurs when the beam approaches the band edge, mimicking the corresponding dispersion relation.

A *quantum theory* of optical Bloch oscillations in arrays of coupled waveguides was theoretically presented by in Ref. [166], where the classical particle-like behavior of photons was highlighted. Bloch oscillations in complex photonic lattices with gain or loss regions were investigated in Ref [170]. It was found that BO in complex lattices with $\mathcal{PT}$ symmetry behave differently than in ordinary lattices owing to nonreciprocity of Bragg scattering, and the conventional wisdom of BO and wave transport needs to be modified when dealing with complex crystals [170].

For polychromatic multi-color beams, periodic beam reconstruction is exactly attained for each spectral component in the conventional settings considered in Sec. 2.1.1 and Sec. 2.1.2. However, the self-imaging Bloch period $L_B$ [see Eqs. (13) and (11)] turns out to be dependent on the wavelength, and therefore Bloch oscillations are smeared out for polychromatic beams. Usually, for a beam spectrally broadened by less than $\sim 10\%$ around its carrier field reconstruction is fully lost just after few BO cycles. It was shown in Ref. [174] that *polychromatic optical BOs* can be approximately achieved over a broad spectral range by suitable insertion of tailored lumped phase slips into the array, which combat the dispersion of BO period with wavelength. In practice, a lumped phase gradient can be introduced by a sudden tilt of waveguide axis by a small angle, by local waveguide segmentation, channel narrowing or index change modulation. As a general rule, the increase of the number of phase gradients makes the wavelength-flattening procedure more flexible, resulting in the polychromatic imaging of higher quality or applicable to a broader spectral range.

Spatial Bloch oscillations of light waves of purely *topological origin* were shown to exist in weakly deformed slab waveguides in Ref. [164]. The wave diffraction was shown to be strongly affected by the topology of the deformed surface, which can be tailored to simulate the effect of a tilted periodic refractive index.

*2.2. Dynamic localization and tunneling control*

*2.2.1. Dynamic localization effect*

The quantum motion of an electron in a periodic potential subjected to an external field provides a paradigmatic model to study fascinating and rather universal coherent dynamical phenomena. These include Bloch oscillations for

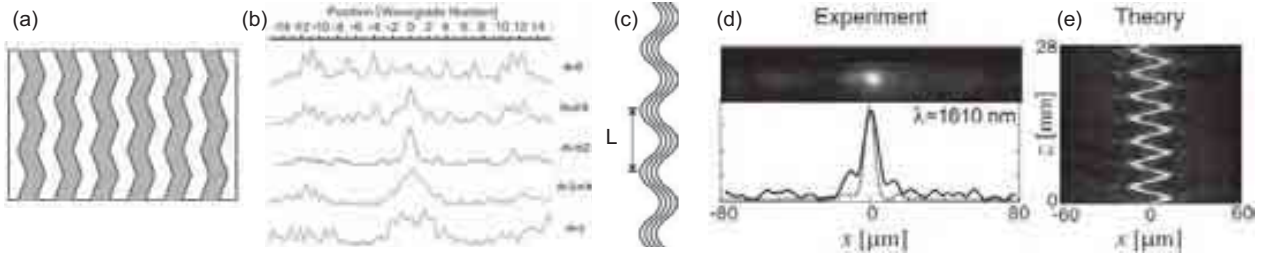

Figure 9: Diffraction management in zig-zag and sinusoidal arrays. (a) Sketch of a zig-zag array. (b) Experimentally measured output intensity profiles after propagation of a 5 $\mu$m wide input beam through the zig-zag array for various tilt angles. (c) Schematic of a sinusoidally-curved array with bending period $L$. (d) Output intensity pattern recorded on the IR camera for a single waveguide excitation of the sinusoidal array (top), and corresponding cross section intensity profile (bottom). The dashed curve is the numerically-predicted intensity profile. (e) Beam evolution predicted by numerical simulations. After Refs. [56, 183].

*dc fields* discussed in Sec. 2.1] above, and the more recently-predicted dynamic localization (DL) for *ac fields* [49], in which a localized particle periodically returns to its initial state following the periodic change of the driving field. The condition for DL in the nearest-neighbor tight-binding (NNTB) approximation and for a sinusoidal driving field was originally predicted by Dunlap and Kenkre [49], and DL effect has been shown to be related to the collapse of the quasienergy minibands [89]. Subsequently, the general conditions for DL beyond the NNTB approximation for generalized ac fields have been identified [37], and DL under the action of both ac and dc fields has been studied [334] [see Sec. 2.2.2]. DL effect was observed experimentally in a number of physical systems, including semiconductor superlattices [118, 188] and Bose-Einstein condensates [187, 50]. Recently, it has been suggested [138, 152] and demonstrated [183] that optical waveguide arrays with a periodically-bent axis may provide an ideal laboratory system for an experimental realization of DL in optics, where the local curvature of the waveguide provides the optical equivalent of an applied electric field [139, 138, 152].

We follow Ref. [152] and consider an array of single-mode waveguides, in the NNTB approximation and assuming that the lowest Bloch band of the array is excited. The following coupled-mode equations can be derived for the amplitudes $\Psi_n$ of the field in the individual waveguides number $n$,

$$i\dot{\Psi}_n + C(\Psi_{n+1} + \Psi_{n-1}) = \omega \ddot{x}_0(z) n, \qquad (14)$$

where $z$ is the paraxial propagation distance (the dot indicates the derivative with respect to $z$), $C > 0$ is the coupling constant,

$$\omega = 2\pi n_0 d\lambda \qquad (15)$$

is the normalized optical frequency, $n_0$ is the substrate refractive index, $d$ is the array transverse period, and $x_0(z)$ is the waveguide bending profile. It can be shown that light propagation in the periodically curved array with the bending period $L$ can be described using the effective coupling coefficient (see, e.g., Ref. [152]),

$$C_{\text{eff}} = \frac{C}{L} \int_0^L \cos[\omega \dot{x}_0(\xi)] d\xi, \qquad (16)$$

which defines light diffraction after each complete bending period ($z = NL$, $N = 0, 1, 2, \ldots$). The condition for DL is $C_{\text{eff}} = 0$, which corresponds to periodic self-imaging at planes $z = NL$.

Pioneering demonstration of diffraction management was made by Eisenberg and collaborators [56] using zig-zag arrays [see Fig. 9(a)]. It was suggested to use the cascading of short segments of tilted at angles $\pm\alpha$ waveguide arrays in order to achieve a desired average diffraction. The effective coupling coefficient calculated for such a zig-zag array is

$$C_{\text{eff}} = C \cos(\theta), \qquad (17)$$

where $\theta = \alpha\omega$ is the normalized tilt angle, such that $\theta = \pi$ corresponds to the Bragg angle. Therefore, for a zig-zag sequence of segments with alternating tilt angles $\theta = \pm\pi/2$ the effective coupling vanishes, $C_{\text{eff}} = 0$, which implies a complete cancellation of diffraction [56]. Experiential results for zig-zag arrays with different tilt angles $\theta$ are presented in Fig. 9(b). The absence of diffraction at $\theta = \pi/2$ is clearly observed.

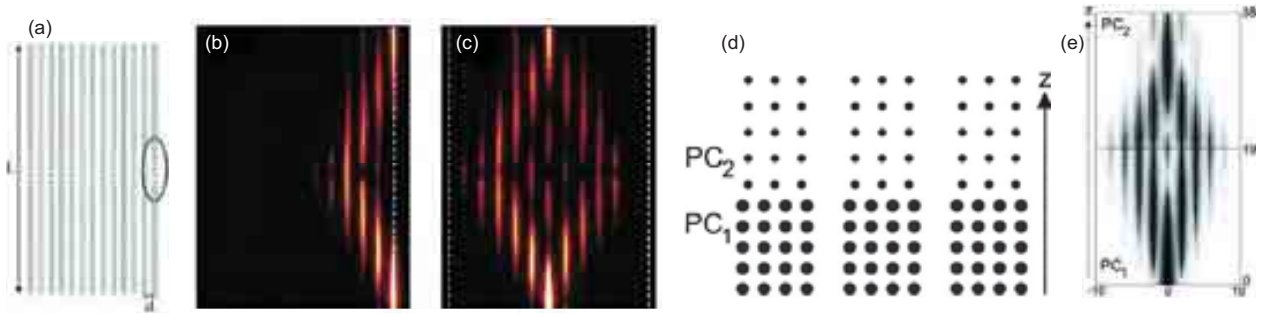

Figure 10: Self-imaging in segmented arrays and photonic crystal waveguides. (a) Sketch of the segmented array configuration with the segmentation region highlighted. The diffraction is reversed so that the input plane is imaged on the output facet regardless of (b) surface or (c) centre waveguide excitation. Vertical dashed lines mark sample edges. (d) Schematic view of the junction between four-rod and three-rod two photonic crystal waveguides arrays. Note the change of the lattice constant that permits straight waveguides to be obtained. (e) Simulation of the time-averaged intensity of the field in the diffraction-managed device. Central horizontal line marks the interface between the four-rod and the three-rod arrays. The units of the axes are micrometers. After Refs. [286, 147].

For a sinusoidally-bent array [see Fig. 9(c)] with period $L$ and amplitude $A$, $x_0(z) = A\sin(2\pi z/L)$, which corresponds to the sinusoidal driving force considered originally by Dunlap and Kenkre [49], the effective coupling coefficient is

$$C_{\text{eff}} = C J_0\left(\frac{2\pi\omega A}{L}\right), \tag{18}$$

where $J_0$ is the Bessel function of the first kind. When the bending parameters are such that $2\pi\omega A/L = \xi_1$, where $\xi_1 \simeq 2.405$ is the first the Bessel function root, the condition for DL is satisfied, $C_{\text{eff}} = 0$, and an effective suppression of waveguide coupling is attained.

Direct experimental observation of DL of light in sinusoidally-curved Lithium-Niobate waveguide arrays electric fields was reported by Longhi and collaborators [183]. The measured output beam profile for a single waveguide excitation of the curved array is shown in Fig. 9(d), together with the beam evolution along the array as predicted by a numerical analysis [Fig. 9(e)]. Note that a single spot is observed at the output plane in the DL regime, in good agreement with the theoretical predictions.

A similar DL effect was later observed by Dreisow and collaborators [39] in curved femtosecond laser written sinusoidal waveguide arrays. The light propagation inside the array was directly observed by monitoring fluorescence of color centers induced during the laser writing process. In addition to monochromatic excitation the spectral response of the arrays was investigated by launching white light supercontinuum into the arrays [39].

When the modulation amplitude varies with the propagation distance, $A = A(z)$, one can realize flexible control of the beam trajectory and spreading law. A general method for flexible control of the path of discretized light beams in homogeneous waveguide lattices, based on longitudinal modulation of the coupling constant, was proposed in Ref. [177]. It was shown [177], for instance, that to realize a parabolic (cubic) path for a discretized beam, a linear (parabolic) change of $A(z)$ is required. In the case of the linear or parabolic change of the amplitude $A(z)$, regimes of superdiffraction can is realized, when beam broadens faster than the conventional linear ballistic spreading which takes place in lattices without modulation. As compared to beam steering and refraction control achievable in graded-index waveguide arrays, this method [177] allows to synthesize rather arbitrary target paths and could of potential interest for beam-steering applications in discrete photonics.

Behavior of a Bloch particle in complex lattices with loss and gain subjected to a sinusoidal ac force was recently theoretically investigated in Ref. [171]. For unbroken parity-time ($\mathcal{PT}$) symmetry and in the single-band approximation, it was shown that time reversal symmetry of the ac force preserves the reality of the quasienergy spectrum. Like in ordinary lattices, the dynamic localization is attained for a sinusoidal band shape (i.e., in the NNTB approximation). For certain threshold values of the loss/gain, the wave packet dynamics turns out to be deeply modified at the $\mathcal{PT}$ symmetry breaking point, where band merging occurs and Bragg scattering in the lattice becomes highly non-reciprocal [171].

It should be mentioned that the DL phenomena rely on boundary-free light propagation, which means that a truncation in the array leads to a considerable distortion of the self-imaging effect. As suggested by Gordon [79], this problem can be overcome by precisely tailoring coupling constants of each waveguide to obtain propagation constants of the array equally spaced by $\Delta\beta$, such that the input distribution reappears periodically after propagating the beat length of $2\pi/\Delta\beta$. However, this approach requires an extraordinarily high precision of the waveguide fabrication technique, which is not readily available.

S. Longhi suggested [165] that a segmentation of appropriate lattice sites yielding particular phase shifts [see Fig. 10(a)], an approach conceptually similar to the diffraction management in the zig-zag arrays, can result in a perfect image reconstruction irrespective of the excitation position in finite and infinite arrays. Following Ref. [165], we consider a spatially finite one-dimensional array made of a number $N$ of straight waveguides, which can be described by Eq. (14) with $x_0(z) \equiv 0$, and $n = 1, 2, 3, \ldots N$. Such system admits a set of $N$ propagation-invariant orthonormal supermodes, for which it can be shown [165] that a sudden phase shift of $\pi$ in every second waveguide in the middle of the array (at the position $z = L/2$) results in a pair-wise conversion of the supermodes into their complex-conjugate counterpart, which yields $|\Psi_n(0)|^2 = |\Psi_n(L)|^2$ for an arbitrary input field distribution, so that an exact image reconstruction at $z = L$ is achieved for the field intensity. The required phase shift of the propagating modes can be achieved by insertion of short segmentation regions in the middle of every second waveguide [see Fig. 10(a)], such that the accumulated additional phase shift in the segmented region is $\pi$. The length of the segmentation region should be small compared to the coupling length, so that the condition of an abrupt phase shift is satisfied.

Image reconstruction in segmented waveguide arrays created using femtosecond writing technique was observed by Szameit and collaborators [286]. For the direct monitoring of the propagation pattern, the fluorescent light of nonbridging oxygen-hole color centers which were formed during the writing process was used. The exact image of the input light distribution was reproduced at the output facet of the sample for both centre and edge waveguide excitation [see Figs. 10(b) and 10(c)]. Image reconstruction by segmentation was also observed [286] in a two-dimensional waveguide array with a square topology. This is possible as the light propagation in a two-dimensional square array can be considered as a superposition of horizontal and vertical one-dimensional lattices. Consequently, the same mechanism as discussed above will hold in square geometries. It should be noted that the inclusion of diagonal coupling terms for the two-dimensional array[see Sec. 2.4 below], as well as of higher-order coupling terms for the one-dimensional array [see Sec. 2.2.2], would make the self-imaging imperfect. However, for image plane distances $L$ comparable or shorter than the characteristic coupling length associated with such higher-order coupling terms, image reconstruction should be produced with good accuracy as long as the influence of the diagonal coupling is negligible.

Tailored phase shifts were also used for diffraction engineering in photonic crystal (PC) waveguide arrays by Locatelli and collaborators [147]. In contrast to the conventional waveguide arrays where the coupling coefficient $C$ is always positive, in PC waveguides it is possible to control the sign of coupling coefficient $C$ by varying the number of rods $N$ between the waveguides [Fig. 10(d)]. Following Ref. [147], we consider PC waveguides composed of different number of rods [see Fig. 10(d)]. The fundamental mode, which has the largest propagation constant, can be even (symmetric) or odd (antisymmetric), depending on the number of rods $N$. In the lowest gap of the band structure, if $N$ is even the fundamental mode is even, whereas if $N$ is odd the fundamental mode is odd. The coupling coefficient $C$ can be related to the propagation constants $\beta_{\text{even, odd}}$ of the modes of the coupled waveguides as [147]

$$C = \frac{\beta_{\text{even}} - \beta_{\text{odd}}}{2}. \tag{19}$$

When the fundamental mode is even (i.e., $N$ is even) the coupling coefficient is positive. When the fundamental mode is odd (i.e., $N$ is odd) $C$ becomes negative. Since the sign of $C$ can be engineered in PC waveguide arrays, the diffraction regime (normal or anomalous) can be designed as desired even for a *normal beam incidence*, providing additional flexibility compared to the conventional waveguide arrays where a waveguide or beam tilt is required for the diffraction management [56]. Thus, diffraction management can be realized in a PC device which consists of cascaded sections of PC waveguides with the opposite sign of diffraction. For example, beam self-imaging can be achieved in a PC device composed of a cascade of four-rod (PC$_1$) and three-rod arrays (PC$_2$), see Fig. 10(d). The time-domain vector finite-element simulation of the full Maxwell equations [147] shows that the input excitation spreads during propagation in the first section, whereas it exhibits an opposite behavior in the second section, and the initial field distribution is recovered [see Fig. 10(e)].

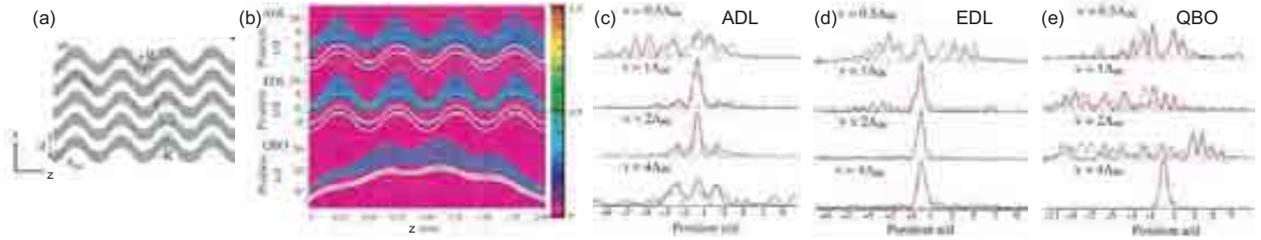

Figure 11: (a) Schematic of the curved waveguide array structure. The waveguide coordinate system, $(u, v)$, is highlighted. (b) Simulations of light propagation in structures with the approximate dynamic localization (ADL), exact dynamic localization (EDL), and quasi-Bloch oscillations (QBO). Some neighboring waveguides are marked white for visibility. The plots are in the $x-z$ coordinate system. (c-e) Measured (solid lines) and calculated (dashed lines) transverse beam profiles shown at different propagation distances $v$ for (c) ADL, (d) EDL, and (e) QBO. After Ref. [102].

*2.2.2. Lattices with long-range interactions*

The DL localization effect was introduced in Sec. 2.2.1 for arrays of weakly coupled waveguides, when each lattice site is only coupled to its *two nearest neighbors,* the one on the left and the one on the right [see Eq. (14)]. For waveguide arrays with a stronger coupling, when each waveguide is coupled to its $N$ nearest neighbors on both sides, the $N$-th order NNTB approximation should be used instead,

$$i\dot{\Psi}_n + \sum_{-N \leq m \leq N\, m \neq 0} C_{|m|}\Psi_{n-m} = \omega \ddot{x}_0(z) n, \tag{20}$$

where $C_m$ are the high-order long-range coupling coefficients ($C_1 \equiv C$).

It has been shown by Dignam and collaborators [37] that dynamic localization with any *continuous* periodic ac driving (including the sinusoidal waveguide bending discussed in Sec. 2.2.1) can only occur in the first order NNTB approximation with $N = 1$. We will further follow Ref. [37] and refer to this type of DL as the approximate dynamic localization (ADL), since for such driving the beam does not relocalize in general structures with the long-range interactions [see Figs. 11(b), top, and 11(c)]. In contrast, it was predicted [37] that exact dynamic localization (EDL) and rigorous beam relocalizations can occur in these general structures only if the effective driving force $F_{\text{eff}} = \omega \ddot{x}_0(z)$ is *discontinuous,* and the radius of the waveguide curvature profile, $R(v) \approx 1/\ddot{x}_0(v)$, where $v$ is the curvilinear propagation coordinate [see Fig. 11(a)], satisfies the two necessary and sufficient conditions,

$$\beta_p(\Lambda) = 0 \qquad \text{for all } p \neq 0, \tag{21}$$

and

$$\gamma(\Lambda) = 2\pi q, \tag{22}$$

where $\Lambda$ is the bending period in the waveguide system, $p$ and $q$ are integers,

$$\beta_p(v) \equiv \frac{1}{c}\int_0^v e^{-ip\gamma(v')}dv', \tag{23}$$

$c$ is the speed of light, and

$$\gamma(v) = \omega \int_0^v \frac{1}{R(v')}dv'. \tag{24}$$

A simple physical interpretation of the EDL condition (21) was given by Wan and collaborators [313]. It was shown that the EDL regime for electrons in crystals corresponds to the requirement that, during one period of the applied ac electric field, the electron spends an equal amount of time at each position in the first Brillouin zone. One can draw two conclusions from the requirement that electrons spend equal time everywhere in the Brillouin zone. The first is that the initial condition does not matter. The second is that, since the electron spends equal time at each position, the contribution to the electron dynamics associated with a position $k$ in the Brillouin zone is exactly canceled by that contribution at $-k$. Therefore, irrespective of the initial condition, after an integer number of field periods each electron ends up at its initial position, and the electron is thus dynamically localized [313].

Of the many different curvature profiles that can produce EDL [37], the simplest is the square-wave field, with curvature period $\Lambda$, where $R(v) = +R_0$ for $(1/4)\Lambda < v \leq (3/4)\Lambda$, and $R(v) = -R_0$ for $(3/4)\Lambda < v \leq (5/4)\Lambda$ [see Fig. 11(a)]. The required radius of curvature amplitude, $R_0$, using Eq. (22), is found to be

$$R_0 = \frac{\Lambda\omega}{4\pi q}. \tag{25}$$

The EDL in the square-wave case is easily understood by considering the beam evolution over one full curvature period: the beam undergoes one BO in the first half of the curvature period (where $R(v) = +R_0$), and undergoes another BO in the second half of the curvature period (where $R(v) = -R_0$) [see Figs. 11(b), middle, and 11(d)]. EDL for the square-wave driving was observed experimentally in ALGaAs waveguide arrays by Iyer and collaborators [95] who have chosen $q = 1$, in order to maximize $R_0$ and thereby minimize bending losses due to the coupling to radiation modes.

Due to the requirement of discontinuities for the EDL regime [37], it is not possible to obtain such electric fields exactly in experiments on an electronic system. It is thus important to quantify just how much smoothing of the discontinuities can be tolerated if good dynamic localization is to be achieved. A detailed examination of the properties of continuous ac electric fields that result in the dynamic localization of electrons in periodic potentials was presented by Domachuk and collaborators [38]. Localization arising from square-wave ac electric fields was found to be surprisingly tolerant to smoothing [38]. Other functions leading to EDL, which have larger discontinuities, were found to be less tolerant. It also found that even a square wave with very significant smoothing results in much better localization than a pure sine wave.

Domachuk and collaborators [38] also showed that all symmetric ac electric fields are guaranteed to exhibit ADL (i.e., DL in the first order NNTB limit) for some value of the field amplitude as long as the vector potential $\gamma$ [Eq. (24)] and the driving force $F_{\text{eff}} = \omega\ddot{x}_0(z)$ are never zero simultaneously. This demonstrates that the earlier results showing that ADL can occur for sinusoidal [49] and triangular [142] driving fields are not isolated cases, but can be considered examples of a more general result.

The conditions for DL for general combined ac-dc fields were analyzed by Wan and collaborators [312], who examined localization dynamics in a one-dimensional periodic potential in the presence of general periodic driving,

$$\ddot{x}_0(z) = x_{ac}(z) + x_{dc}, \tag{26}$$

where $x_{ac}(z)$ is a $z$-periodic ac component, and $x_{dc}$ is a constant $z$-independent dc part. It was found that, similar to the EDL, the necessary and sufficient condition for DL to occur in the combined ac-dc field (26) is given by Eq. (21), combined with the additional requirement that the ratio of the Bloch oscillations period $L_B$ due to the dc component [see Eqs. (11) and (13)] to the period of the ac driving field $L_{ac}$ must be a rational number,

$$\frac{L_B}{L_{ac}} = \frac{N}{Q}, \tag{27}$$

where $Q$ and $N$ are positive integers with no common factor [312]. Two distinct types of DL in the presence of combined ac and dc fields were further identified. The first type of DL is similar to the conventional EDL or ADL in a pure ac field without a dc component, and occurs if the ratio $L_B/L_{ac}$ is a rational number and the total field meets the conditions for DL; this, therefore, requires that the ac part of the field has a particular shape and amplitude to meet the EDL or ADL conditions [see Eqs. (16) and (21)]. The second type of DL can occur with the period $L = NL_{ac} = QL_B$ (where $N > 1$) if the ratio $L_B/L_{ac}$ is a noninteger rational number [312]. This type of DL can occur for *any shape and amplitude of the ac part* of the field. Just as for conventional Bloch oscillations [see Sec. 2.1], the occurrence of this type of DL only depends on the dc component of the field. Therefore, this form of DL is refered to as *quasi-Bloch oscillations* (QBOs). We note that QBOs only occur within the tight-binding approximation. However, QSOs are more robust than ADL in that QBOs can occur beyond the nearest-neighbor approximation in most cases (if $N \geq 2$) and in fact are valid up to the $(N-1)$-th nearest-neighbor approximation [312].

To demonstrate QBOs and show their robustness beyond the NNTB approximation, Joushaghani and collaborators [102] fabricated three curved waveguide arrays that differ only in their curvature profiles: one to exhibit QBOs and one each to examine ADL and EDL [see Fig. 11]. The optical intensity profiles in the three non-NNTB curved waveguide arrays calculated using beam propagation method are shown in Fig. 11(b). The solid lines in Figs. 11(c-e) show measured power profiles at different propagation distances $v$. Consistent with numerical simulations, the light in the EDL structure is relocalized after each oscillation period, while the light in the ADL structure only partially relocalized. In the QBO structure, light relocalizes only after four ac periods at $v = \Lambda = 4\Lambda_{ac} = \Lambda_B$, demonstrating that QBOs occur even for non-NNTB structures. By resolving the spectral response of the QBOs and comparing it to other DL structures, it was shown that QBOs can have a larger relocalization wavelength bandwidth than other DL schemes [102].

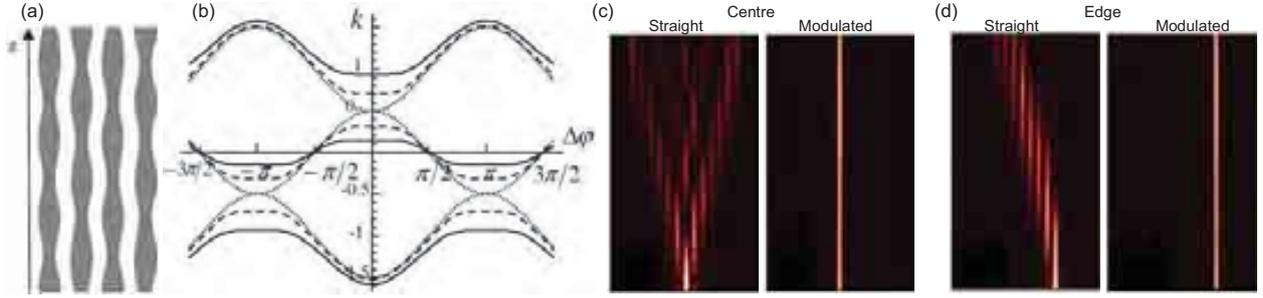

Figure 12: (a) Qualitative representation of the modulated waveguide array. The grey regions indicate the thickness of the waveguides. Notice the alternating periodic variation of the thickness along the $z$-direction. (b) Spatial dispersion curve: the wavenumber $k$ is plotted vs. the tilt of the Bloch modes $\Delta\phi$. The dotted curves correspond to the limit of small modulation, $\Delta\beta \to 0$. The parameters are $\Delta\beta = 0.2$ (dashed) and $\Delta\beta = 0.4$ (solid), and correspond to the case when the uncoupled curves just touch one another. In this situation, an increase of the modulation depth $\Delta\beta$ broadens the subdiffractive plateau. (c,d) Fluorescence images of sub-diffractive light propagation recorded in straight (left panels) and modulated (right panels) laser-written arrays. Excited are (c) centre and (d) edge waveguides. After Refs. [270, 292].

It should be mentioned that EDL has been generalized to potentials without inversion symmetry by Wan and collaborators [313]. It was shown that the symmetry of the periodic potential has no effect on the existence of dynamic localization or on the electric fields required to achieve it, and the condition for EDL is identical for symmetric and asymmetric potentials. However, the dynamics of the electrons can depend strongly on the inclusion of new terms that enter the evolution equations when the potential is asymmetric. Although these terms vanish if maximally localized Wannier functions are used, they are not negligible in general [313].

As the effective ac driving field depends explicitly on the optical frequency $\omega$ [see Eq. (20)] and DL only occurs for particular field amplitudes, DL in curved waveguide arrays only occurs at resonant wavelengths. This idea has been exploited to propose that such structures could be used as optical filters. A thorough study of the wavelength dependence of the transmission properties of curved optical waveguides in the DL regime was performed by Wan and collaborators [314]. It was found that the transmission spectra of these novel optical filters depend on the curvature profiles employed [314]. For example, for waveguides with square-wave curvature profiles it was shown that for bandpass filters with lengths of 2 cm, rejection ratios better than 30 dB, and FWHM less than 10 nm at $\lambda = 1.5$ $\mu$m are achievable, and smaller bandwidth can be possible with more general curvature profiles [314]. It was also demonstrated that the transmission can be further tailored via the periodic insertion of apertures [314]. These filters can have physical dimensions and rejection ratios that are comparable to Fabry-Perot and Bragg grating filters [314], although their bandpass is not as narrow.

### 2.2.3. Sub-diffractive propagation and tunneling control

Diffraction cancellation mediated by the DL effect (Sec. 2.2.1) bears a close connection with the self-collimation effect in photonic crystals (PC), which was first predicted and observed by Kosaka and collaborators [126] in a three-dimensional PC, followed by other experiments including the recent demonstration of super-collimation over a very large number of structure periods in a two-dimensional PC with a square lattice of holes in air [244]. In such PC structures, self-collimation phenomena arise owing to the flattening of the isofrequency PC band surfaces at frequencies usually close to a band edge (see, for instance, Refs. [126, 324, 185, 148, 267, 271]), which strongly weakens the spatial dispersion for light waves. In the self-collimating photonic crystals [267, 272], circular sectors in the dispersion curves (the iso-frequency lines in the wavevector space) indicate the normal dispersion, and convex segments in these curves indicate negative diffraction, the flat segments that arise for specific PCs architectures, indicate zero diffraction.

A scheme for the sub-diffractive propagation of light in a one-dimensional array of coupled waveguides, which imitates the phenomena occurring in sub-diffractive photonic crystals, was proposed by Staliunas and Masoller [270]. The key ingredient of this scheme is that the propagation constant in the waveguides is modulated along the longitudinal (propagation) direction $z$,

$$\beta_n(z) = \beta_0 + 2\Delta\beta(-1)^n \cos(z), \tag{28}$$

where the factor $(-1)^n$ is responsible for the alternating character of the longitudinal modulation, $n$ is the waveguide number, and $\Delta\beta$ is the modulation depth. One can implement the modulation of the propagation constant by periodically changing the thickness of the individual fibers in an alternating order, as depicted in Fig. 12(a). The periodic variation of the propagation constant along the $z$-direction can also be obtained, e.g., by fine temperature control, or it can be realized by modifying directly the refractive index of the waveguides (incorporating in the waveguides periodic profiles of impurities). The advantage of the proposed configuration is that it allows eliminating not only the first-order diffraction (second order spatial derivatives), but due to symmetry, also all the odd-order derivatives [270]. The remaining terms can be approximated, at the lowest order, by the fourth-order spatial derivative (recently, a similar effect was shown to happen in symmetric PCs [268]). The dispersion diagram for the arrays with the modulation of the propagation constant according to Eq. (28) is shown in Fig. 12(b). The appearance of the horizontal plateaus in the dispersion curves shown in Fig. 12(b) indicates the existence of regimes of sub-diffractive propagation. Indeed, these plateaus mean that the longitudinal wavevectors $k$ of the corresponding Bloch modes do not depend on the tilt angle $\Delta\phi$, i.e., that the components that are tilted at different angles do not dephase during the propagation. This means that the beam (or any arbitrary pattern), being a Fourier composition of differently tilted components of Bloch modes, does not broaden or blur during the propagation. Obviously, there are high-order terms of diffraction that do not vanish, therefore, we refer to *sub-diffractive* rather than to non-diffractive propagation. A variation of the modulation amplitude $\Delta\beta$ results in a change in the width of the plateau, as can be seen from Fig. 12(b). Another interesting point is that there are two plateaus [see Fig. 12(b)]. In the case of subdiffractive PCs, where only one plateau has been found [267], the initial beam projects essentially on two modes - diffractive and sub-diffractive. The first one diffracts quickly and vanishes, while the second one continues to propagate collimated. Here, the initial beam projects into two sub-diffractive modes, which propagate together without broadening. As these two modes have different propagation constants a beating, i.e., periodic pulsations with respect to the longitudinal coordinate, can be expected.

The sub-diffractive propagation in modulated arrays was used by Szameit and collaborators [292] to experimentally demonstrate control of light tunneling in waveguide arrays. It was demonstrated that a longitudinal out-of-phase modulation of the linear refractive index along the propagation direction in the waveguides,

$$\Delta n_n(z) = (-1)^n \mu \sin(\Omega z), \tag{29}$$

where $\mu$ is the index modulation amplitude, and $\Omega$ is the modulation frequency, within the frame of the tight-binding approximation is equivalent to a reduction in the coupling constant $C$ by the factor of $J_0(2\mu/\Omega)$, where $J_0$ is the zero-order Bessel function [292]. Coupling thus vanishes for $2\mu/\Omega = \xi_i$, with $\xi_i \approx 2.4, 5.5, \ldots$ being roots of the Bessel function. When the frequency and amplitude of the modulation are properly chosen, the band of quasienergies is considerably narrowed (such tunneling inhibition cannot be exact in contrast to the DL effect), forcing the light to remain in the excited channel.

The experimental samples [292] were fabricated using a femtosecond writing method. Since the index modulation of the single lattice sites crucially depends on the writing velocity, one can particularly introduce an out-of-phase longitudinal harmonic modulation of the trapping channels by varying slightly the writing speed for each waveguide. The direct observation of the light propagation inside the arrays was done using a special fluorescence technique [292]. Figures 12(c) and 12(d) compare the light propagation in non-modulated and optimally modulated waveguide arrays for the central [Fig. 12(c)] and edge channel [Fig. 12(c)] excitations. As a result of the modulation of the refractive index [Eq. (29)] the Floquet-Bloch modes exhibit almost identical quasienergies, irrespective of the number of waveguides in the system or the position of excitation [292], and tunneling is inhibited almost completely at any propagation distance $z$ irrespective of the beam input position [see Figs. 12(c) and 12(d)]. This is in a contrast to the dynamic localization in periodically curved waveguides, where *periodic* self-imaging occurs in *infinite* arrays (see Sec. 2.2.1). The possibility of the sub-diffractive light localization in the bulk or at the surface of arrays expands the opportunities for diffraction control.

It was shown by Kartashov and Vysloukh [113] that out-of-phase longitudinal refractive index modulation can also lead to the tunneling inhibition in waveguide arrays in which defect channels with a reduced refractive index are spaced periodically and the light guiding is achieved because of the Bragg reflection. This is a non-trivial generalization since the guiding due to the total internal reflection is essentially different from that due to the Bragg reflection.

The specific features of tunneling inhibition in linear and weakly nonlinear regimes for higher-order complex modes incorporating multiple bright spots were studied by Lobanov and collaborators [146], who considered the evolution of multichannel excitations in longitudinally modulated waveguide arrays where the refractive index either oscillates out-of-phase in all neighboring waveguides or when it is modulated in phase in several central waveguides surrounded by out-of-phase oscillating neighbors. Both types of modulations were found to allow resonant inhibition of light tunneling, but only the modulation of the latter type was shown to conserve the internal structure of multichannel excitations. It was shown that parameter regions where light tunneling inhibition is possible depend on the symmetry and structure of multichannel excitations. Antisymmetric multichannel excitations were found to be more robust than their symmetric counterparts.

Light tunneling inhibition in an array of couplers with longitudinal refractive index modulation was studied in detail by Kartashov and collaborators [112]. The periodic array of optical couplers serves as an example of a photonic structure where a rapid local energy exchange between guides in each coupler coexists with a slow global energy tunneling into adjacent couplers. The presence of two characteristic energy tunneling scales in this system makes the problem of light localization in a single channel especially challenging, since one may expect that simple longitudinal waveguide bending usually adopted in waveguide arrays to achieve the dynamic localization regime (see Sec. 2.2.1) will not yield inhibition of tunneling. Kartashov and collaborators [112] suggested that light tunneling inhibition in a periodic array of optical couplers can be achieved by a special design of longitudinal and transverse modulations of the refractive index. It was shown that by properly selecting the law of longitudinal modulation a relatively slow energy exchange between adjacent couplers in the array and a rapid energy exchange between channels of individual couplers can be inhibited simultaneously that results in the diffractionless propagation of single-channel excitations at certain resonant values of the modulation frequency.

It should be mentioned that a fundamental difference between the sub-diffraction propagation (self-collimation) and the dynamic localization discussed in Sec. 2.2.1 was revealed by Longhi and Staliunas [184]. While the DL effect in periodically-curved waveguide arrays may bear a close connection with self-collimation phenomena and these two phenomena have been sometime referred using the same terminology, there are some deep differences, both conceptually and practically. In particular: (i) self-collimation is generally related to local flattening of a portion of the isofrequency curve in the reciprocal $k$-space and thus diffraction cancellation occurs solely at low orders; conversely, dynamic localization is a more stringent requirement as it implies diffraction suppression at any order and thus needs a full collapse of the quasi-energy band of the modulated array, (ii) dynamic localization does not necessarily mean beam spreading suppression at any distance, rather a periodic refocusing of the beam at the periodicity of the modulated array along the propagation direction, which may be larger than the diffraction length of the beam, and (iii) self-collimation is typically a non-resonant effect with respect to the amplitude of the modulation, whereas dynamic localization it is [184].

Recently, it was shown in Ref. [177] that an effective $z$-modulation of the coupling constant, $C = C(z)$, can be realized by either waveguide axis bending or by out-of-phase modulation of the propagation constants of adjacent waveguides. The longitudinal modulation of the coupling constant was used to suggest a new general method for flexible control of the path of discretized light beams, which enables the synthesis of rather arbitrary beam paths, and also results in an effective engineering of the discrete diffraction. For example, regimes of super-diffraction were demonstrated using linear or parabolic change of the waveguide bending amplitude $A(z)$ [177].

*2.2.4. Broadband diffraction management*

In this Section, we discuss propagation of polychromatic light in modulated photonic lattices. Advances in the generation of light with broadband or supercontinuum spectrum [see an example in Fig. 13(a)] in photonic-crystal fibers [245, 251, 48] opened many new possibilities for a range of applications including optical frequency metrology [305], spectroscopy [335], tomography [236], and optical characterization of photonic crystals [237]. While periodic photonic structures offer many new unique opportunities to engineer and control the fundamental properties of light propagation, the commonly studied structures are primarily optimized for beam shaping and deflection in a narrow-frequency range. Indeed, the physics of periodic structures is governed by scattering of waves from modulations of the refractive index and their subsequent interference. This is a resonant process, which is sensitive to both the frequency and propagation angle. Strong dependence of the beam refraction on the optical wavelength known as *superprism effect* was observed in photonic crystals [127]. In optical waveguide arrays, the mode overlap and the coupling strength increase substantially with the wavelength, and the beam diffraction broadening is substantially larger at longer wavelengths [see an example in Fig. 13(d)]. For example, it was found in the recent experiments [243, 183] that the effect of beam dynamic localization was restricted to a spectral range of less than 10% of the central frequency. Thus, it is important to explore the potential of periodic photonic structures for tunable spatial shaping of the polychromatic light beams.

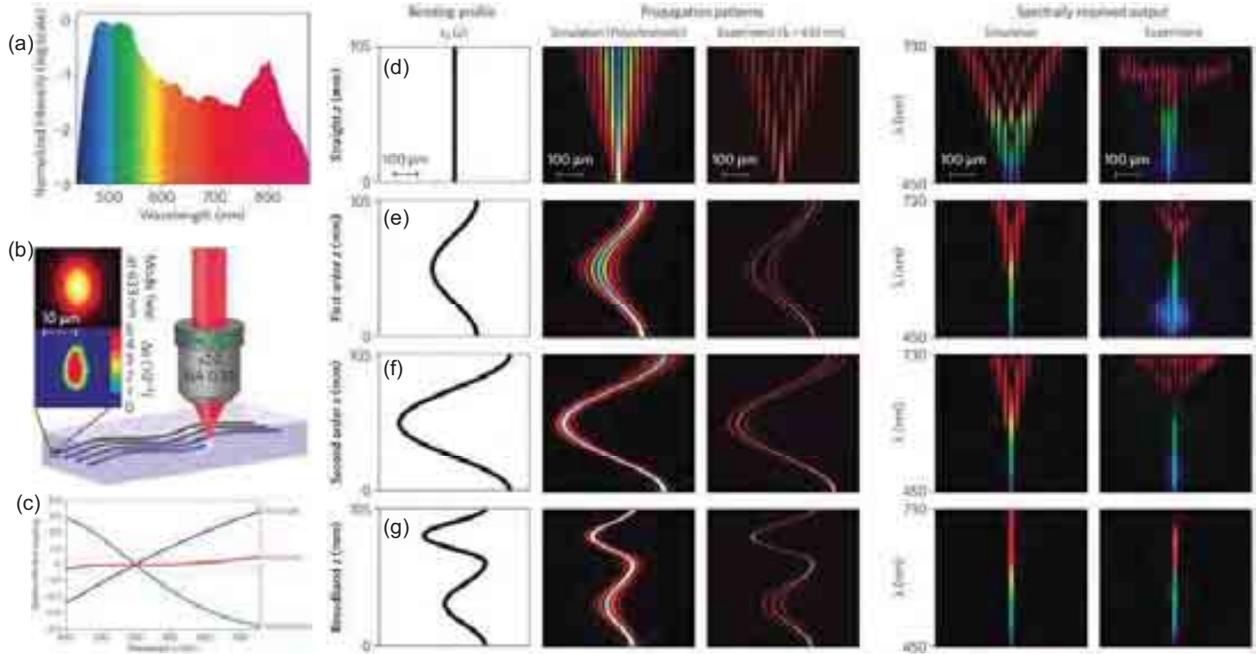

Figure 13: Polychromatic dynamic localization. (a) The broadband supercontinuum spectrum spanning over the entire visible range. (b) Schematic of the femtosecond writing procedure. Insets show the resulting waveguide refractive index modulation $\Delta n$ (bottom) and mode profiles (top). (c) Effective couplings in the curved arrays shown in (d-g) as a function of the wavelength. (d) Diffraction in straight waveguide array. (e) First-order DL at the wavelength $\lambda = 550$ nm. (f) Second-order DL at the wavelength $\lambda = 550$ nm. (g) Broadband DL in the spectral region $450 - 730$ nm in a two-segment curved array. First column: waveguide axis bending profiles. Second and third columns: numerical simulations of the polychromatic beam propagation, and corresponding fluorescent images measured at the wavelength $\lambda = 633$ nm. Fourth and fifth columns: calculated numerically and measured experimentally spectrally resolved output beam profiles. After Ref. [289].

It has been suggested in Ref. [72] that the intrinsic wavelength-dependence of diffraction strength in waveguide arrays can be compensated by geometrically-induced dispersion resulting from the periodic waveguide bending, making possible *broadnand diffraction management* in a very broad frequency range covering a spectral range up to 50% of the central frequency. This opens up novel opportunities for efficient control of white-light beams and patterns, offering enhanced flexibility compared to the spatial-spectral reshaping recently demonstrated experimentally in arrays of straight optical waveguides [278, 219, 279].

We follow Ref. [72] and study propagation of beams emitted by a continuous white-light source in an array of coupled optical waveguides, where the waveguide axes are periodically curved in the propagation direction $z$. When the tilt of beams and waveguides at the input facet is less than the Bragg angle at each wavelength, the beam propagation is primarily characterized by coupling between the fundamental modes of the individual waveguides. We consider the case of the weak coupling, when each waveguide is coupled to its nearest neighbors only. Then, the beam evolution can be described by a set of tight-binding equations for different frequency components, $\Psi_n(z; \omega)$, which in the linear regime take the form [72],

$$i\frac{d\Psi_n}{dz} + C(\omega)\exp\left[-i\omega\dot{x}_0(z)\right]\Psi_{n+1} + C(\omega)\exp\left[i\omega\dot{x}_0(z)\right]\Psi_{n-1} = 0, \quad (30)$$

where $n$ is the waveguide number, $x_0(z) \equiv x_0(z + L)$ is the periodic waveguide bending profile, $L$ is the modulation period, coefficient $C(\omega)$ defines a coupling strength between the neighboring waveguides (it characterizes diffraction strength in a straight waveguide array with $x_0 \equiv 0$ [101, 266]), the dots stand for the derivatives, and $\omega$ the dimensionless frequency [see Eq. (15)]. Similar to the monochromatic DL discussed in Sec. 2.2.1, it can be shown by analyzing the plane-wave solutions of Eq. (30) that after the full bending period ($z \to z + L$) the polychromatic beam diffraction in the periodically curved waveguide array is the same as in a straight array with the effective coupling coefficient

$$C_{\rm eff}(\omega) = \frac{C(\omega)}{L}\int_0^L \cos\left[\omega\dot{x}_0(\zeta)\right]d\zeta. \quad (31)$$

Therefore, diffraction of multi-color beams is defined by an interplay of the additional *bending-induced dispersion* introduced through the frequency dependence of the integrand in Eq. (31), and the intrinsic frequency dependence of the coupling coefficient in a straight waveguide array $C(\omega)$. It was suggested in Ref. [72] that spatial evolution of all frequency components can be synchronized allowing for shaping and steering of multi-color beams, when effective coupling remains constant around the central frequency $\omega_0$,

$$\left.\frac{dC_{\rm eff}(\omega)}{d\omega}\right|_{\omega=\omega_0} = 0, \quad (32)$$

and we demonstrate below that this condition can be satisfied by introducing special waveguide bending profiles.

As an example, we demonstrate *dynamic localization of white-light beams*, where all the wavelength components experience periodic self-imaging. DL regime is realized when the effective coupling coefficient vanishes at the central frequency $\omega_0$, $C_{\rm eff}(\omega_0) = 0$, and simultaneously the wavelength-independence condition (32) is satisfied. We follow Ref. [72] and demonstrate that this becomes possible in hybrid structures with a periodic bending profile that consists of alternating segments [see example in Fig. 13(g)],

$$\begin{aligned}x_0(z) &= A_1\left\{\cos\left[2\pi\frac{z}{z_0}\right]-1\right\} \quad \text{for } 0 \leq z \leq z_0,\\ x_0(z) &= A_2\left\{\cos\left[2\pi\frac{z-z_0}{L-z_0}\right]-1\right\} \quad \text{for } z_0 \leq z \leq L,\end{aligned} \quad (33)$$

where $A_1$ and $A_2$ are bending amplitudes, and $L$ is bending period. Effective coupling in such hybrid structure can be calculated analytically [72],

$$C_{\rm eff}(\omega) = \frac{C(\omega)}{L}\left[z_0 J_0(\xi_1) + (L-z_0)J_0(\xi_2)\right], \quad (34)$$

where $J_m$ is the Bessel function of the first kind of the order $m$, and

$$\xi_1 = \frac{2\pi A_1\omega}{z_0}, \quad \text{and} \quad \xi_2 = \frac{2\pi A_2\omega}{L/2 - z_0}. \quad (35)$$

In order to realize broadband DL, we choose the structure parameters such that $\xi_1(\omega_0) = \tilde{\xi}_1$, and $\xi_2(\omega_0) = \tilde{\xi}_2$, where $\tilde{\xi}_i \simeq 2.40, 5.52, \ldots$ are roots of the Bessel function $J_0$, which correspond to different orders of dynamic localization resonances [72, 289]. Then, the dynamic localization condition is exactly fulfilled at the central frequency $\omega_0$, $C_{\rm eff}(\omega_0) = 0$, and simultaneously the condition of wavelength-independent coupling in Eq. (32) is satisfied for the following modulation parameters,

$$A_1 = \frac{\tilde{\xi}_1\tilde{\xi}_2 J_1(\tilde{\xi}_2)}{2\pi\omega_0\left[\tilde{\xi}_2 J_1(\tilde{\xi}_2) - \tilde{\xi}_1 J_1(\tilde{\xi}_1)\right]}L, \quad A_2 = -\frac{J_1(\tilde{\xi}_1)}{J_1(\tilde{\xi}_2)}A_1, \quad z_0 = \frac{2\pi\omega_0 A_1}{\tilde{\xi}_1}. \quad (36)$$

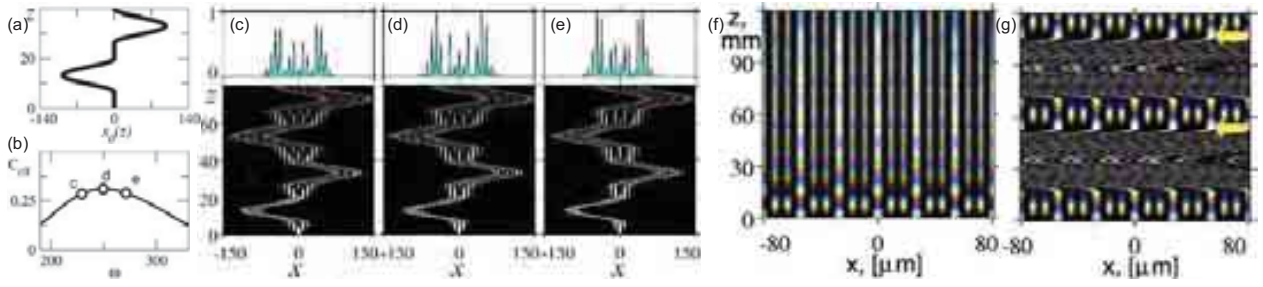

Figure 14: Wavelength-independent diffraction and multi-color Talbot effect in an optimized periodically curved waveguide array. (a) Waveguide bending profile with the period $L = 40$ *mm* and (b) corresponding effective coupling normalized to the coupling in the straight array at the central wavelength $C_0 = C(\lambda_0)$. (c-e) Evolution of beam intensity and output intensity profiles after propagation of two full bending periods for the wavelengths (c) $\lambda_r = 580$ *nm*, (d) $\lambda_0 = 532$ *nm*, and (e) $\lambda_b = 490$ *nm*, which correspond to points 'c', 'd', and 'e' in (b). Propagation of polychromatic pattern $\{\ldots 1, 0, 0, 1, 0, 0, \ldots\}$ in (f) straight and (g) optimized curved arrays. Arrows mark points at which input pattern is completely restored in the modulated array. Polychromatic light has flat spectrum covering $490 - 580$ nm. After Refs. [72, 69].

As a result, one obtains an extremely flat coupling curve shown in Fig. 13(c), red curve. The effective coupling remains close to zero in a very broad spectral region of up to 50% of the central frequency. Importantly, the modulation period $L$ is a free parameter, and it can always be chosen sufficiently large to avoid scattering losses due to waveguide bending since the maximum waveguide curvature is inversely proportional to the period, $\max|\ddot{x}_0(z)| \sim L^{-1}$.

Broadband DL in curved photonic lattices fabricated by laser direct-writing in silica glass [see schematic of the writing setup in Fig. 13(b)] was studied by Szameit and collaborators [289]. The samples were 105 mm long with the waveguide spacing $d = 26\,\mu$m. Three curved waveguide arrays consisting of 19 waveguides each were fabricated. The arrays for the first- and second-order DL contained one full bending period equal to the sample length ($L = 105$ mm), and had bending amplitudes $A = 93\,\mu$m and $A = 214\,\mu$m, respectively. The curved array for the broadband DL consisted of two successive segments of the length $L_1 = 63$ mm and $L_2 = 42$ mm with bending amplitudes $A_1 = 56\,\mu$m and $A_2 = 86\,\mu$m [see Eq. (36)], respectively. In order to characterize comprehensively light propagation in the fabricated arrays, Szameit and collaborators [289] employed two complementary methods: (i) direct visualization of the laser beam propagation at $\lambda = 633$ nm via fluorescence imaging [284, 39], and (ii) spectrally resolved imaging of the output optical field distribution for the broadband supercontinuum excitation. A white light continuum light source which spans over the entire visible ($\lambda = 450 - 800$ nm) [see Fig. 13(a)] was used to characterize the samples.

Experimental results for curved arrays which satisfy the first- and second-order DL resonance condition at the wavelength of $\lambda_0 = 550$ nm are presented in Figs. 13(e) and 13(f), respectively. One can see that, in both structures, the spectral components in a narrow spectral region around $\lambda_0$ return to a single waveguide at the output, while significant diffractive broadening occurs for other wavelengths. In order to understand the difference between the DL of different orders, wavelength-dependencies of the corresponding effective coupling coefficients are presented in Fig. 13(c). For the first-order DL, $C_{\rm eff}$ changes sign from negative to positive at the resonance when we move from shorter to longer wavelengths, corresponding to the transition from normal to anomalous discrete diffraction. In contrast, the sign changes from positive to negative around the second-order DL resonance.

The difference between DL of different orders suggests that approximate DL in an extremely broad spectral region by combining two successive segments of the curved waveguide array which bending amplitudes are tuned to the first and second order DL at the same wavelength [see Eq. (36)]. The corresponding waveguide bending profile is shown in Fig. 13(g). Then, although different colors detuned from the exact DL wavelength exhibit non-zero diffraction after the propagation in the first segment, the input beam profile is restored through the reversed diffraction in the second segment. The overall effective coupling in the two-segment structure almost completely vanishes in a very broad spectral region around the DL resonance wavelength [Fig. 13(c), red line]. In Fig. 13(g), light propagation and output beam profiles are shown for the two-segment curved waveguide array. One can see that all spectral components, from $\lambda = 450$ nm to $\lambda = 730$ nm, are localized in a single waveguide at the array output [Fig. 13(g), fourth and fifth columns], despite the highly nontrivial evolution of different spectral components inside the structure [Fig. 13(g), second and third columns].

We now follow Ref. [72] and analyze the conditions for *wavelength-independent diffraction*, which may find applications for reshaping of multi-color beams. In order to reduce the device dimensions, it is desirable to increase the absolute value of the effective coupling and simultaneously satisfy Eq. (32) to achieve broadband diffraction management. It was found in Ref. [72] that Eq. (32) can be satisfied in the two-segment hybrid structure defined by the Eq. with

$$z_0 = \frac{L}{2} \text{ and } A_1 = \frac{\xi}{4\pi\omega_0}L. \tag{37}$$

The set of possible parameter values $\xi$ is determined from the relation

$$\frac{J_0(\xi)}{J_1(\xi)} = \frac{\xi}{\omega_0}\frac{C_0}{C_1}, \tag{38}$$

where coefficients $C_0 = C(\omega_0)$ and $C_1 = (\dot{C}(\omega)/d\omega)_{\omega=\omega_0}$ characterize coupling dispersion. It was shown in Ref. [72] that it is possible to obtain both normal and anomalous diffraction regimes for normally incident beams, corresponding to positive and negative effective couplings $C_{\text{eff}}(\omega_0) = C_0 J_0(\xi)$ depending on the chosen value of $\xi$. For example, for the waveguide array considered in Ref. [72], at the normalized central frequency $\omega_0 = 250$ (corresponding physical wavelength is $\lambda_0 = 532$ nm) the coupling dispersion parameters were $C_0 \simeq 0.13$ $mm^{-1}$ and $C_1 \simeq -0.0021$ $mm^{-1}$. Then, constant positive coupling around the central frequency $C_{\text{eff}}(\omega_0) \simeq 0.25 C_0$ can be realized for $\xi \simeq 6.47$ and constant negative coupling $C_{\text{eff}}(\omega_0) \simeq -0.25 C_0$ for $\xi \simeq 2.97$.

Comprehensive analytical and numerical analysis performed in Ref. [72] revealed that a hybrid structure with the bending profile consisting of one straight (i.e $A_1 \equiv 0$) and one sinusoidal segment can provide considerably improved performance compared to a pure sinusoidal bending if

$$\omega_0 \frac{C_1}{C_0} > \xi_{cr} \frac{J_1(\xi_{cr})}{J_0(\xi_{cr})}, \tag{39}$$

where value $\xi_{cr} \simeq 5.84$. Under such conditions, larger values of positive effective coupling can be obtained in a hybrid structure with

$$A_1 \equiv 0, \quad A_2 = \frac{C_1 C_{\text{eff}}(\omega_0)}{4\pi C_0^2 J_1(\tilde{\xi}_2)}L, \quad z_0 = \frac{C_{\text{eff}}(\omega_0)}{2 C_0}L, \tag{40}$$

where the effective coupling around the central frequency is

$$C_{\text{eff}}(\omega_0) = \frac{\tilde{\xi}_2 C_0^2 J_1(\tilde{\xi}_2)}{\tilde{\xi}_2 C_0 J_1(\tilde{\xi}_2) + \omega_0 C_1}. \tag{41}$$

Example of such a hybrid structure which provides strong wavelength-independent diffraction is shown in Fig. 14(a), and the corresponding effective coupling is plotted in Fig. 14(b). The diffraction rate in this optimized structure is almost the same in a broad spectral region, see examples for three different wavelengths in Figs. 14(c-e).

We note that the polychromatic dynamic localization in modulated tight-binding lattices with long-range interaction is theoretically proposed and experimentally demonstrated in Ref. [180], where an efficient suppression of discrete diffraction over the whole white-light spectral region (450-750 nm) was observed in femtosecond-laser-written curved triangular-waveguide lattices with first- and second-order coupling.

It was suggested in Refs. [72, 69] that the modulated lattices with broadband diffraction management can be used to achieve periodic self-imaging of polychromatic light patterns for integrated diffractive optics devices, optical interconnections, optical array illumination, and optical measurement. Usually, light patterns do not maintain their shape constant over long propagation distances as they are quickly distorted because of diffraction. Periodic reconstruction of input light patterns can be achieved using Talbot effect, when any periodic monochromatic light pattern reappears upon propagation at certain equally spaced distances

$$L_T = \frac{D^2}{\lambda}, \tag{42}$$

where $D$ is the pattern period [297]. Being a direct consequence of Fresnel diffraction, Talbot effect is one of the most fundamental phenomena in wave optics, and similar wave phenomena have been observed in other physical systems.

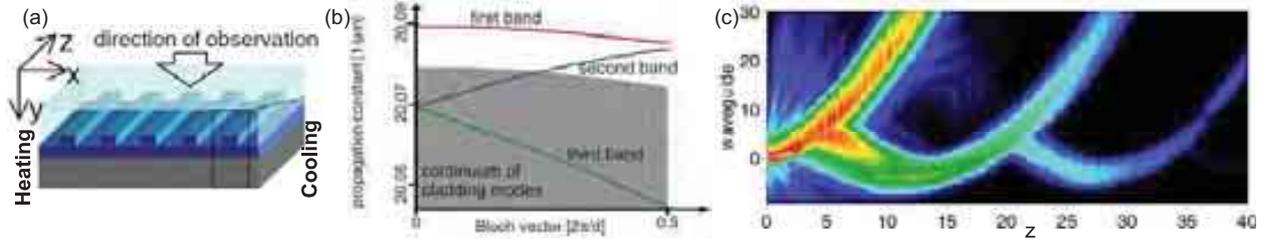

Figure 15: (a) Schematic drawing of a thermo-optical polymer waveguide array used in the experiment. (b) Corresponding band structure of the waveguide array where the second band already intersects with the spectral continuum of the cladding modes. The gap between the first and second band is minimal at the edge of the Brillouin zone. (c) Bloch oscillations and Zener tunneling in the waveguide array when several waveguides are excited by a 25 $\mu$m wide elliptical beam. After Ref. [302].

Recently, it was was shown by Iwanow and collaborators [94] that the Talbot effect is also possible in waveguide arrays when the period $N$ of the input pattern belongs to the set $N \in \{2, 3, 4, 6\}$. For example, for $N = 3$ Talbot revivals take place at the distance

$$L_T^{(3)} = \frac{2\pi}{3C(\omega)}. \tag{43}$$

However, classical Talbot effect is only possible for light sources with a narrow spectral width [299], because the Talbot distance $L_T$ depends on the wavelength. In straight waveguide arrays, coupling strength $C(\omega)$ depends strongly on the wavelength [see Fig. 13(d)]. Consequently, Talbot revivals are not possible for polychromatic light patterns, as the Talbot distance $L_T^{(N)}$ is inversely proportional to the coupling strength [94], and it is therefore different for different spectral components, see Fig. 14(f).

It was demonstrated in Refs. [72, 69] that multi-color Talbot effect can be realized in modulated arrays if one chooses the bending period $L$ to be equal to the Talbot distances $L_T^{(N)}$. An example of Talbot revivals for polychromatic patterns with periods $N = 3$ is shown in Fig. 14(g), where the Talbot distance is given by

$$L_T^{(3)} = \frac{2\pi}{3C_{\text{eff}}(\omega_0)}. \tag{44}$$

Self-imaging of light patterns with the discrete Talbot effect has some intrinsic limitations [72]. Discrete Talbot effect is only possible for just a few pattern periodicities which belong to the set $N \in \{2, 3, 4, 6\}$, while for any other pattern period its propagation becomes aperiodic [94]. Also, the pattern must be necessarily periodic and infinite (or very wide in practical conditions). For polychromatic light patterns in modulated lattices discussed above, there is a further limitation on the period of the multi-color Talbot recurrences. Once the waveguide array parameters and its bending profile are fixed, the effective coupling becomes fixed as well, and there is only one single value of the bending period $L$ which coincides with the Talbot distance. This imposes restrictions on the device size and also on the bending losses [72], which can make some bending profiles impractical. It was demonstrated in Ref. [69] that imaging of arbitrary narrow or non-periodic polychromatic patterns can be achieved in modulated waveguide arrays with periodically canceled diffraction, similar to those used to demonstrate white-light dynamic localization [see Fig. 13(g)].

It should be noted that Talbot effect functionality may be extended by introduction of optical nonlinearities. Nonlinear Talbot effect was recently demonstrated by Zhang and collaborators [333] using periodically poled LiTaO$_3$ crystals. No real grating was used in the experiment. Instead, periodic intensity patterns appeared on the output surface of the crystal due to the periodic domain structures, i.e., the modulated second-order nonlinear susceptibility. The formation of second-harmonic self-imaging was observed with spatial resolution improvement by a factor of 2 as a result of the frequency doubling [333].

### 2.3. Multi-band dynamics
### 2.3.1. Zener tunneling

In the previous Sections, waveguide arrays were usually analyzed in the framework of the coupled-mode theory [see, e.g., Eq. (10)], where the array mode was described as a collective excitation (a "supermode") of the modes of the individual waveguides evanescently coupled to each other. A more general approach, in which waveguide arrays are regarded as an example of a general one-dimensional periodic optical structure, is the Floquet-Bloch (FB) analysis (see, e.g, Ref [197]). It predicts that the propagation-constant spectrum of the array's eigenmodes (the FB waves) is divided into bands, separated by gaps in which propagating modes do not exist. Figure 15(a) shows the calculated diffraction relation (band-gap diagram) of a typical waveguide array. It relates the propagation constant to the Bloch wave number which is the transverse component of the wave vector reduced to the first Brillouin zone. In contrast, only the first band is correctly described by the coupled-mode equations (10).

Presence of the high-order bands can strongly affect propagation of optical waves. In particular, Bloch oscillations become damped owing to the field-induced tunneling into other bands at high field strengths, the phenomenon known as *Zener tunneling* (ZT)] [16, 328]. The breakdown of BOs is expected to happen when the energy difference imposed on a period of the lattice by the linear potential reaches the order of the gap to the next band. This has been proven by ZT observation in spectral and time-resolved transmission measurements in photonic superlattices composed of a Bragg mirror with chains of embedded defects of linearly varying resonance frequency, where both enhanced transmission peaks and damped BOs due to ZT have been observed by Ghulinyan and collaborators [78].

On the other hand, direct visual observation of BOs and ZT is possible in photonic lattices [see Sec. 2.1]. The major difference to experiments in photonic superlattices [78] is that instead of the temporal photon dynamics a spatial pattern of light can be observed on top of the waveguide array, avoiding resolution of fast temporal oscillations or transmission spectra. The eigenmodes of a waveguide array are Bloch waves with a Bloch vector $k$ pointing in the $x$-direction. Respective propagation constants $\beta$ in the $z$-direction are arranged in bands with respect to $k$ [see Fig. 15(b)]. The band structure shown in Fig. 15(b) was calculated by fully accounting for the periodic potential and not using the tight-binding model (10) which is restricted to nearest-neighbor interactions. For an increasing transverse force higher order bands come into the play. The index gradient can cause the propagation constants of different bands to overlap in adjacent parts of the array, thus inducing phase matching between the bands. As a result, an excitation consisting of first band modes will couple to the second and higher bands. Ultimately, ZT will lead to the decay of lowest order band BOs. Obviously, this tunneling increases with decreasing band gap. Tunneling between the first and the second band therefore happens predominantly at the Brillouin zone edge [see Fig. 15(b)]. For small transverse gradients and narrow band gaps the tunneling rate upon each Bloch oscillation $|R|^2$ can be approximated by [302]

$$|R|^2 = \exp\left(-[\text{Im}(k)]_{max}\frac{\Delta\beta\pi}{2\alpha}\right) \qquad (45)$$

where $\Delta\beta$ is the band gap width, $\alpha$ is the transverse refractive index ramp, and $(\text{Im}k)_{max}$ is the maximum of the imaginary part of the wave number $k$ in the gap center.

Trompeter and collaborators [302] reported the first direct visual observation of Zener tunneling and the associated BO decay by monitoring the evolution of light in a photonic lattice. Light was fed into a thermo-optical waveguide array [see Fig. 15(a)], where a temperature drop between the simultaneously heated and cooled opposite array sides was used to create the index gradient [Eq. (11)] required for Bloch oscillations. The light propagation monitored in the experiments shows that light performs BOs, the shape of which is notable [see Fig. 15(c)]. It can be clearly recognized from Fig. 15(c) that light escapes from BOs at the high-index turning points corresponding to the Brillouin zone edge. This is the very manifestation of Zener tunneling from the first to the second band. At these edges the band gap attains its minimum [see Fig. 15(b)] and the tunneling rate has its maximum with about 8%.

Optical Zener tunneling effect was also directly observed in a one-dimensional lattice of undoped liquid crystalline waveguides by Fratalocchi and collaborators [65, 63]. A wide, intense Gaussian pump beam was used to impresses a non-adiabatic acceleration onto the lattice in transition regions around the pump beam centre. A second probe beam, initially a superposition of Bloch modes that belong to an upper band, was transferred to a lower band in the transition region, and ZT was observed as a change in the propagation angle of the probe beam.

*2.3.2. Binary superlattices*

A paradigmatic system showing two-band dynamics, for which generalized coupled-mode equations [see Eq. (46)] can be used to capture the main dynamics, is a binary array consisting of periodically alternating wide and narrow waveguides [see Fig. 16(a)], which represents an optical analog of a quantum two-level system. In general, two types of lattice modulation can be applied that mimic either ac or dc external driving force, which were used to demonstrate optical Bloch oscillations in Sec. 2.1 and dynamic localization in Sec. 2.2.1, respectively.

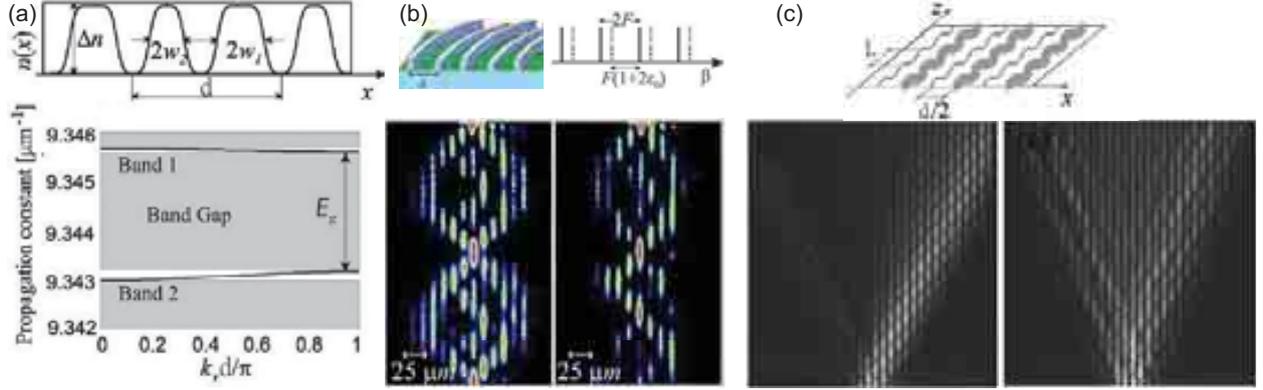

Figure 16: Multi-band effects in modulated binary arrays. (a) Top: transverse profile of a binary array composed of wide and narrow waveguides. Bottom: numerically computed band-gap diagram. (b) Bloch-Zener oscillations in binary arrays with dc modulation. Top: array sketch (left) and corresponding Wannier-Stark ladder spectrum (right). Bottom: experimentally measured fluorescent images of beam propagation for synchronous (left) and asynchronous (right) Zener tunneling. (b) Multi-band diffraction in binary arrays with ac modulation. Top: array sketch. Bottom: simulated propagation of a tilted broad Gaussian beam for even (left) and odd (right) resonances. Double refraction occurs for the odd resonance. After Refs. [154, 45].

First, we describe light propagation in binary arrays with *dc-type modulation* [see Fig. 16(b)]. Usually, cascading of Zener tunneling to higher-order bands results in irreversible dumping of optical Bloch oscillations in multi-band waveguide arrays [Sec. 2.3.1]. However, it was suggested by S. Longhi [157] and demonstrated experimentally by Dreisow and collaborators [45], that a coherent dynamics instead of an irreversible decay process is expected when only two bands, coupled by Zener tunneling, are involved. In this case the dynamics of occupation probabilities in the two bands shows a complex coherent behavior, and Zener tunneling manifests itself as an aperiodic or periodic splitting and recombination of propagating light beam superimposed to the BO motion [157]. Such *Bloch-Zener oscillations* may offer potential applications for the realization of tunable beam splitters and interferometers for light waves.

Following Refs. [157, 45], we consider light propagation in a modulated binary array [see sketch in Fig. 16(a)], which in the tight binding limit can be described by the following set of coupled equations for the modal amplitudes $\Psi_n$ of light waves trapped in the various waveguides,

$$i\frac{d\Psi_n}{dz} + (-1)^n \Delta\beta + C(\Psi_{n+1} + \Psi_{n-1}) = -nF\Psi_n, \qquad (46)$$

where $2\Delta\beta$ and $C$ are the propagation constant mismatch and the hopping rate between two adjacent waveguides of the array, respectively, and $F$ is the modulation function related to the waveguide bending. For $F = 0$, i.e., for straight unmodulated waveguides, inserting a plane wave ansatz $\Psi_n \propto \exp[i(\beta z - nkd)]$ into Eq. (46) yields the following dispersion relation for the two minibands:

$$\beta = \pm\sqrt{\Delta\beta^2 + 4C^2 \cos^2(kd)}, \qquad (47)$$

where where $\beta$ and $k$ are the longitudinal and transverse propagation constants, respectively. Note that the two minibands are separated by a gap $E_g = 2\Delta\beta$ [see Fig. 16(a), bottom].

When the waveguides are bent $F = \omega/R$, where $\omega$ and $R$ are the normalized optical frequency and waveguide radius of curvature, respectively, and the two minibands are replaced by two interleaved Wannier-Stark ladders, $\beta_n^{(1)} = (2n + \epsilon_0)F$ and $\beta_n^{(2)} = (2n + 1 - \epsilon_0)F$, where the parameter $\epsilon_0$ determines the relative distance between the two WS ladders [see Fig. 16(b), top]. The offset $\epsilon_0$ turns out to be a function of $C/F$ and $\Delta\beta/F$ solely, and it can be numerically computed from a spectral analysis of Eq. (46). For a single site illumination, of all Bloch modes of the two minibands are excited with a flat Fourier spectrum. According to the acceleration theorem, the Bloch vectors drift through the reciprocal space, and Zener tunneling occurs between the minibands at the band edge $kd = \pm\pi/2$. As a gap between the two minibands exists in binary arrays with $\Delta\beta \neq 0$ [see Eq. (47)], ZT at the band edges is not complete, and the breathing motion becomes generally aperiodic and can be characterized by two spatial periods [45]. The first one, $L_1$, is determined by mode spacing of each WS ladder and equals half of the Bloch period, i.e., $L_1 = L_B/2 = \pi/F$. The second one, which will be referred to as the Zener period, is determined by the shift of the two interleaved WS ladders and is given by $L_2 = 2\pi/|F(1 - 2\epsilon_0)|$ [see Fig. 16(b), top]. If $L_1$ and $L_2$ are commensurate, ZT is with the BO motion, and perfect wave packet reconstruction is achieved [45]. Figure 16(b) (bottom, left) shows an example of synchronous Bloch-Zener oscillations with $L_2 = 2L_1 = L_B$, as directly observed using fluorescence imaging in femtosecond laser-written arrays in the experiments by Dreisow and collaborators [45]. Conversely, in Fig. 16(b) (bottom, right), different array parameters yield *asynchronous* ZT, and wave reconstruction is not observed after one Bloch period.

A similar synchronous and asynchronous Bloch-Zener oscillations were also observed in Ref. [45] for a broad Gaussian excitation, when at normal incidence a narrow spectrum of Bloch modes, centered around $\beta = 0$ and belonging to the first miniband, is excited at the input plane. Then the linear potential accelerates the spectrum from $\beta = 0$ at $z = 0$, to $\beta = Fz/d$ at the distance $z$. Since ZT occurs when $\beta$ is close to the band edges $\beta = \pm\pi/(2d)$, tunneling zones can be defined in propagation coordinate $z$ around $L_T 1 = L_b/4$, $L_T 2 = 3L_b/4$, $L_T 3 = 5L_b/4$, ... [157, 45].

We now turn our attention to binary arrays (46) with *ac-type modulation* [see Fig. 16(c), top] which can be used for *multi-band diffraction and refraction* control as suggested by S. Longhi [154]. To study the role of periodic waveguide axis bending on the diffraction and refraction properties of the binary array, we follow Ref. [154] and consider as an example a sinusoidal modulation function,

$$F = \frac{4\pi^2 \omega A}{L^2} \sin\left(\frac{2\pi z}{L}\right), \tag{48}$$

where $A$ and $L$ are the bending amplitude and period, respectively (note that a similar modulation was used to realize DL in Sec. 2.2.1). It was highlighted in Ref. [154] that the quantum analog of Eq. (46) with the modulation (48) corresponds to the motion of a charged particle in a two-site crystalline potential under the action of an external ac field of frequency $\Omega = 2\pi/L$, and the condition $E_g = n\Omega$ ($E_g = 2\Delta\beta$ is the gap width) corresponds to field-induced $n$-photon resonances between the two minibands. At the $n$-photon resonance condition light dynamics can be captured by the effective equations [154],

$$i\frac{d\tilde{\Psi}_n}{dz} + C_n(\tilde{\Psi}_{n+1} + \tilde{\Psi}_{n-1}) = 0, \tag{49}$$

where the effective coupling coefficient is $C_n = CJ_n(\pi\omega A/L)$ for even resonances ($n$ even), and $C_n = (-1)^n CJ_n(\pi\omega A/L)$ or odd resonances ($n$ odd) Here $J_n$ is the bessel function of the order $n$. Thus, by varying the value of the modulation depth $A$ discrete diffraction can be consistently tuned; in particular, diffraction is suppressed close to the zeros of the Bessel function $J_n$. Note that for even resonances, coupled-mode equation (49) is similar to the one describing discrete diffraction within a single-band model, whereas for odd resonances it has a different form. In fact, for the even resonances one has a single band for plane waves, whereas for odd resonances one has two bands. This circumstance is responsible for a very different propagation behavior in the odd and even resonance cases when the array is excited by a broad beam. For the odd resonances, double refraction and beam splitting occurs due to different refraction angles of the two bands [see Fig. 16(c), bottom, right], in a similar way to what happens for straight binary arrays [276]. In contrast, for even resonances multi-band effects, such as beam splitting due to double refraction, are quenched [see Fig. 16(c), bottom, left].

### 2.3.3. Rabi oscillations

Since the famous work of Rabi in 1936 [242], it has been well known that in quantum systems a periodic modulation can stimulate a resonant transition from one energy level to another and back. A typical example is a two-level atomic system, where an electromagnetic wave which frequency is tuned to the energy gap between the two states causes periodic population exchanges accompanied by emission and reabsorption of a photon. In one scenario, the oscillations involve *direct transitions* only; that is, either the states possess the same momentum or, in a periodic potential, the momentum difference is an integer quanta of lattice momentum. Also, the momentum difference between the two states can be supplied by a phonon, or any other kind of wave carrying momentum, facilitating *indirect Rabi*

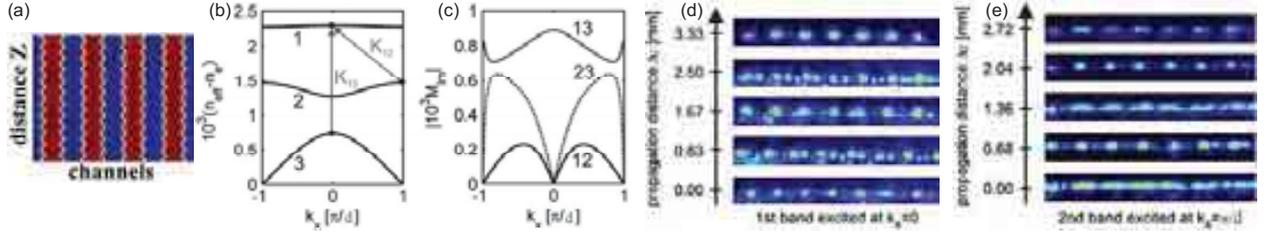

Figure 17: (a) Schematic of modulated waveguide array where the waveguide width varies periodically (b) Band structure of the array where the first three bands arise from guided modes. The arrows indicate both direct transitions ($K_{13}$, coupling bands 1 and 3 at $k_x = 0$) and indirect transitions ($K_{12}$, coupling bands 1 and 2 via a tilted grating with $\Delta k_x = \pi/d$). (c) Spatial overlap integral $M_{lm}$ of possible direct Floquet-Bloch mode transitions as a function of Bloch momentum. (d) Direct Rabi transition of Floquet-Bloch modes between bands 1 and 3. Shown are photographs of the output intensity of the Floquet-Bloch modes, after different propagation distances $\Delta z$, when a pure mode either in band 1 is excited with the prism coupler. (e) Indirect Rabi transitions of Floquet-Bloch modes between bands 1 and 2. After Refs. [190, 257].

*transitions.* It should be noted that Rabi oscillations occur as a result of parametric mixing and thus are different from Zener tunneling described in Sec. 2.3.1.

Makris and collaborators [190] predicted that similar Rabi interband oscillations can take place in optical lattices. When the unit-cell is periodically modulated along the propagation direction [see Fig. 17(a)], direct power transfer between two Floquet-Bloch modes can occur under phase-matching conditions. We follow Refs. [190, 257] and consider light propagating in a modulated one-dimensional waveguide array described by the paraxial wave equation:

$$i\frac{\partial E}{\partial z} + \frac{1}{2k}\frac{\partial^2 E}{\partial x^2} + \frac{k}{n_0}[\Delta n(x) + \delta n(x,z)]E = 0, \quad (50)$$

where $k$ is the wave number, $n_0$ is the substrate index, $\Delta n(x)$ is the transverse periodic modulation defining the lattice, and $\delta n(x,z) = \delta n(x)\delta n(z)$ describes the additional periodic modulation stimulating the Rabi oscillation. In this sense, $\delta n(x,z)$ is equivalent to the modulation terms in quantum mechanics, for example, the time-varying dipole moment induced by a photon. In the experiments by Shandarova and collaborators [257], $\Delta n(x)$ was realized by Ti indiffusion in photorefractive LiNbO$_3$, defining the waveguide array. The additional modulation $\delta n(x,z)$ was realized by two-beam holographic recording of an elementary refractive-index grating. The grating vector $K$ was aligned at an angle $\phi$ relative to the $z$ direction, i.e., $\tan(\phi) = K_x/K_z$, with $K_x$ and $K_z$ being the transverse and longitudinal components of $K$, respectively [see Fig. 17(b)]. Solving the related linear eigenvalue problem leads to the solutions of the form $E_l(x,z) = A_l(z)U_l(x)\exp(i\beta_l z)$, where $A_l$ is the amplitude, $U_l$ is the normalized transverse mode profile, and $\beta_l$ is the propagation constant of mode $l$. Here the index $l$ covers both band index and transverse Bloch momentum of the modes. By using the related orthogonality relations of these modes, adiabatic transitions among different Floquet-Bloch modes $l$ and $m$ can be described by a system of coupled equations [257],

$$i\frac{\partial A_l}{\partial z} = -\frac{1}{2}M_{lm}A_m\exp(-i[\beta_l - \beta_m - K_z]z)\delta(k_x^{(l)} - k_x^{(m)} - K_x), \quad (51)$$

which are similar to equations for probability amplitudes for a quantum system driven by a classical field. In Eq. (51), $k_x^{(i)}$ is the wave number of mode $i$, and the delta function accounts for the conservation of transverse momentum. The exponential function in Eq. (51) describes the longitudinal phase matching condition, so that only for $K_z = \beta_l - \beta_m$ full power exchange between modes $l$ and $m$ can be expected. The integral

$$M_{lm} = \int_{-d/2}^{d/2} U_l(x)\delta n(x)U_m(x)dx, \quad (52)$$

where $d$ is the lattice constant of the waveguide array, accounts for the transversal overlap of modes $l$ and $m$ with the modulation $\delta n(x)$; it defines the coupling strength between modes $l$ and $m$. When $\delta n(x)$ is symmetric, mode coupling is limited to those with equal parity. However, when $\delta n(x)$ is asymmetric, as is the case for a tilted grating ($\phi \neq 0$), modes of different parity are coupled, too. Figure 17(b) presents the calculated band structure (effective refractive indices $n_{eff}$ of Floquet-Bloch modes versus Bloch momentum $k_x$). Possible band-to-band transitions among different

Floquet-Bloch modes include both direct (for example, $K_{13}$) and indirect (for example, $K_{12}$) transitions [Fig. 17(b)]. For the case of direct transitions (non-tilted grating, $K_x = k_x^{(l)} - k_x^{(m)} = 0$), Fig. 17(c) shows the calculated overlap integral (52) as a function of Bloch momentum $k_x$. Interband transitions are most effective between the first and third band in the whole Brillouin zone. It should be mentioned, that similar periodic oscillations between two different lattice solitons in the nonlinear domain are also possible [190].

Direct Rabi oscillations ($K_{13}$) observed by Shandarova and collaborators [257] are presented in Fig. 17(d), where the intensity structure at the lattice output is shown for five different propagation distances $\Delta z$, for excitation at band 1 with a prism coupler. An almost complete conversion of the power carried by modes is visible at a (full) coupling length of about 1.67 mm. In another experiment [see Fig. 17(e)], which corresponds to the indirect Rabi oscillations, tilted grating $K_{12}$ was recorded [257], where the transverse part $K_x = \pi/d$ acts as a "coherent phonon". Periodic coupling between band 2 excited at $k_z = \pi/d$ and band 1 at $k_z = 0$ was observed [Fig. 17(e)].

It should be mentioned that resonant Rabi-like oscillations and adiabatic transitions between confined light modes in properly modulated multimode waveguides were predicted by Kartashov and collaborators [117]. It was shown that the application of a suitable shallow longitudinal refractive index modulation with the frequency equal to the difference between the propagation constants of two eigenmodes with the same parity can stimulate the mode conversion, in a manner similar to that in Rabi oscillation between quantum atomic states in multilevel quantum systems. This phenomenon was shown to take place in complex confining multimode structures in both the linear and the nonlinear regimes.

Recently, Rabi oscillations between Bloch modes of an optical waveguide array with subwavelength periodicity were analyzed by Alfassi and collaborators [4], who showed that both the oscillations frequency and the field amplitude diverge when the optical wavelength approaches a mathematical exceptional point at which the Bloch mode becomes self-orthogonal and field varies rapidly at subwavelength scale. This effect is unique to Rabi oscillations in optical systems, as described by Maxwell's equations, and it has no equivalent in quantum systems where the physical potentials are real. It was found that small changes in the optical wavelength can dramatically affect the dynamics, offering an effective tool for light manipulation in nano-arrays.

## 2.4. Two-dimensional lattices

In Secs. 2.1 - 2.3 our review of light propagation in modulated photonic lattices has been limited to one-dimensional (1D) geometries. On the other hand, one can expect that new unique features may arise in higher dimensions due to the additional degrees of freedom. Moreover, the recent advances in fabrication of high-precision two-dimensional (2D) waveguide arrays of arbitrary topology for coherent light propagation (see, e.g., Refs. [247, 222, 229, 282, 281, 86] and references therein) made it possible to observe a number of multi-dimensional phenomena experimentally. In this Section, we overview the effect of periodic modulation on beam propagation in two-dimensional waveguide arrays, highlighting their similarity and difference with one-dimensional lattices.

### 2.4.1. Bloch oscillations and Zener tunneling

New effects are associated with Bloch oscillations and Zener tunneling in lattices of higher dimensionality. In particular, the process of Zener tunneling becomes nontrivial as the band gap structure can cause an enhanced tunneling in preferred directions determined by the lattice symmetries.

The first experimental observation of Bloch oscillations and Zener tunneling in two-dimensional periodic systems was reported by Trompeter and collaborators [300] using optically induced photonic lattices. The lattice was created by interfering four mutually coherent ordinary-polarized broad beams in a biased photorefractive crystal. The periodic light intensity distribution inside the crystal had the form of a square lattice as shown in Fig. 18(a). Such a periodic light pattern induces a 2D modulation of the refractive index for the extraordinary polarized probe beam. The ordinary-polarized lattice beams, however, remain stationary along the whole length of the crystal. In order to create a transverse refractive index gradient, the photorefractive crystal was illuminated from the top with an incoherent white light, which intensity was modulated transversely, and was constant along the crystal length. Propagation of waves in a homogeneous lattice (for a constant background illumination $I_m$) is characterized by the dispersion of Bloch waves. In Fig. 18(b), dispersion curves characterizing the band gap spectrum are shown together with a diagram of the first Brillouin zone (inset). When the transverse index gradient is applied, Zener tunneling from the initial $\Gamma_1$ point in Fig. 18(b) occurs, according to the adiabatic theory, when the effective propagation constant reaches the gap edge $M_1$ and then tunnels through the gap to the point $M_2$. Because of the square symmetry of the lattice, this process results in the splitting of the initial beam into four parts [see Fig. 18(c)]. Respective images of the signal beam at the exit facet of the crystal are displayed in Fig. 18(c), top row. As it was not possible to follow the evolution of the beam inside the crystal, the details of its behavior were inferred by varying the incident angle $a$ of the probe beam [see Fig. 18(c)]. For angles below the Bragg resonance, the excitation by the probe with a different transverse wave number is equivalent to different starting points in the Brillouin zone. Therefore, when the beam is launched at different input angles, the field evolution is scanned through the Bloch oscillation at the output facet of the crystal. In this way, different stages of the Bloch oscillations were monitored [300], which would normally require crystals of different lengths. In Fig. 18(c), top row, we can clearly see the predicted reshaping of the beam and the tunneling of the beam energy into three different channels.

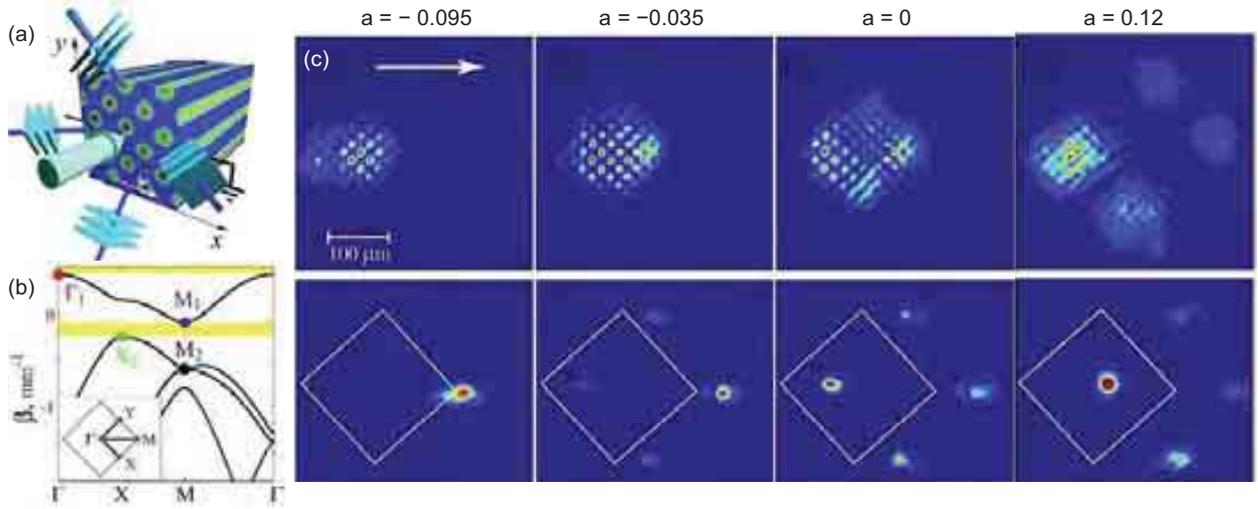

Figure 18: (a) Schematic of an optically induced square lattice. (b) Calculated band structure of the lattice for $I_m = const$. The inset shows symmetry points in the first Brillouin zone. (c) Light intensity profiles (top row) and corresponding Fourier spectra (bottom) measured at the crystal output for different initial inclinations of the input beam $a$ (in degrees). The arrow in (c) indicates the direction of the index gradient. White lines mark the first Brillouin zone. After Ref. [300].

A fully conclusive picture of the beam evolution during a Bloch oscillation and the related Zener tunneling was obtained by monitoring the Fourier spectrum of the output field [see Fig 18(c), bottom row]. For the initial beam tilt of −0.095° (negative angle corresponds to the initial motion of the beam against the force produced by the gradient), the beam intensity profile at the output facet of the crystal [Fig. 18(c), top row] is strongly modulated with adjacent maxima being approximately out-of-phase. Corresponding intensity distribution in the Fourier domain shows that the beam propagates at the Bragg angle in the M-symmetry point of the lattice. Increasing the input angle allows to scan through different positions inside the Bloch period and to monitor the light tunneling into the higher-order bands. Evidently, only one peak lies inside the first Brillouin zone for all the incident angles, and it corresponds to the field which undergoes Bloch oscillations [Fig. 18(c), bottom row]. The other three peaks representing tunneled beams lie outside the boundaries of the first Brillouin zone; thus, they belong to the second transmission band. For an angle of −0.035°, the beam just experiences its first Bragg reflection and the three tunneled beams emerge from all M-symmetry points of the lattice. At larger angles, all the beams get accelerated again by the index gradient, and the central part of the beam completes a full Bloch oscillation. same.

Zener tunneling accompanying Bloch oscillations of a broad beam that covers several sites of the lattice occurs only when the beam approaches the edge of the Brillouin zone. Similar to the one-dimensional case (Sec. 2.1), a drastically different scenario was observed by Trompeter and collaborators [300] for an input beam which is initially very narrow, comparable in size with a single site of the 2D lattice. Such a beam excites simultaneously waves with wave vectors distributed over the whole Brillouin zone. This results in a symmetric breathing of the beam in the first band as it periodically diffracts and refocuses in propagation. Because the narrow beam excites modes with the wave vectors distributed over the whole Brillouin zone, there always exists a component which propagates in the vicinity of the band edge. Hence, the tunneling occurs in this case continuously over the whole Bloch period.

It should be mentioned that Bloch oscillations in general waveguide lattices with an arbitrary diagonal interaction were recently considered by Szameit and collaborators [296]. The eigenfunctions (Wannier-Stark states) of the lattice Schrödinger equation with a linear potential were derived analytically, and a 2D propagator was presented to describe the evolution of arbitrary initial excitations, including single site and broad beam excitation. Using these results, Szameit and collaborators [296] predicted that in the general lattices a complete delocalization along any equipotential line that is parallel to an interaction force should take place, along with the occurrence of additional beating frequencies.

Recently, the effects of lattice symmetry on the two-dimensional Bloch dynamics were analyzed by Shchesnovich and collaborators [258]. As the illustrative example, the hexagonal lattice was considered [258], which is distinct from the square lattice in that it cannot be represented as a sum of two one-dimensional potentials. For lattices without transverse gradient, the possibility of the Rabi oscillations (Sec. 2.3.3) between the resonant Fourier amplitudes was demonstrated. In a two-dimensional periodic potential with an additional linear tilt, it was demonstrated [258] that the direction of the tilt together with the Fourier coefficients of the lattice potential determine how many of the lowest-order Bloch bands of the photonic band-gap spectrum are involved in the interband transitions. Three general regimes of the Zener tunneling were identified in the hexagonal photonic lattice: (i) quasi one-dimensional Zener tunneling (or, equivalently, simple Bragg resonance involving only two Bloch bands) which occurs when the Bloch index crosses the Bragg planes far from one of the high-symmetry points; (ii) three-fold Bragg resonance at the high-symmetry M-point with the Zener transitions between the three Bloch bands; and (iii) the six-fold Bragg resonance at the high-symmetry Γ-point with the Zener tunneling involving, in general, six Bloch bands. In special symmetric cases a new effect was found which can be called Zener tunneling of Rabi oscillations [258]. It was also shown that tunneling of phase dislocations (or optical vortices) results in the output waves carrying the same phase dislocations.

The occurrence of Bloch oscillations in zig-zag quasi-two-dimensional waveguide arrays was recently considered by Wang and collaborators [315]. The special topological configuration of the lattice leads to the appearance of the the next-nearest-neighbor interaction [see also Sec. 2.4.2 below], which results in the new features in the BO patterns, including a double turning-back path which occurs when the beam approaches the band edge.

*2.4.2. Dynamic localization and diffraction management*

Effective coupling approach described in Sec. 2.2.1 for one-dimensional lattices [see Eq. (16)] can be generalized for the multi-dimensional systems. Following Ref. [75], we consider propagation of optical beams in a two-dimensional hexagonal array of coupled optical waveguides, where the waveguide axes are periodically curved in the longitudinal propagation direction [see an example in Fig. 19(b)]. In this case, the waveguide positions can be represented as

$$x_{n,m} = d(n + \frac{m}{2}), \quad y_{n,m} = dm\frac{\sqrt{3}}{2}, \tag{53}$$

where $d$ defines the lattice period, and $n$ and $m$ are the waveguide numbers along the tilted $n-$ and $m$-axes [see schematic in Fig. 19(a)]. When only the nearest-neighbor waveguide coupling is taken into account, the general two-dimensional tight-binding equations can be derived in the following form (see Appendix 7.1.1),

$$i\frac{d\Psi_{n,m}}{dz} + \widetilde{C}_1^*\Psi_{n-1,m} + \widetilde{C}_1\Psi_{n+1,m} + \widetilde{C}_2^*\Psi_{n,m-1} + \widetilde{C}_2\Psi_{n,m+1} + \widetilde{C}_3^*\Psi_{n-1,m+1} + \widetilde{C}_3\Psi_{n+1,m-1} = 0, \tag{54}$$

where

$$\widetilde{C}_1 \equiv \widetilde{C}_{n,n+1,m,m} = C_1 \exp[-i\omega \dot{x}_0(z)], \tag{55}$$

$$\widetilde{C}_2 \equiv \widetilde{C}_{n,n,m,m+1} = C_2 \exp[-i\omega \dot{x}_0(z)/2 - i\omega \dot{y}_0(z)\sqrt{3}/2], \tag{56}$$

$$\widetilde{C}_3 \equiv \widetilde{C}_{n,n+1,m,m-1} = C_3 \exp[-i\omega \dot{x}_0(z)/2 + i\omega \dot{y}_0(z)\sqrt{3}/2], \tag{57}$$

$\omega = 2\pi n_0 d/\lambda$ is the dimensionless optical frequency, functions $x_0$ and $y_0$ define the tow-dimensional waveguide bending profile, the asterisk stands for the complex conjugation, and the dot stands for the derivative. The real-valued coefficients $C_1$, $C_2$, and $C_3$ define the coupling strength between the neighboring waveguides along the different high-symmetry directions in a hexagonal lattice [119] [see Fig. 19(a)], and they characterize diffraction in a straight hexagonal waveguide array with $x_0 \equiv 0$ and $y_0 \equiv 0$ [295]. When the lattice is composed of circular rods all coupling coefficients are the same, $C_1 = C_2 = C_3$. However, in the general case of non-symmetric waveguides, the transverse cross-section coupling coefficients along different directions may be all different [281].

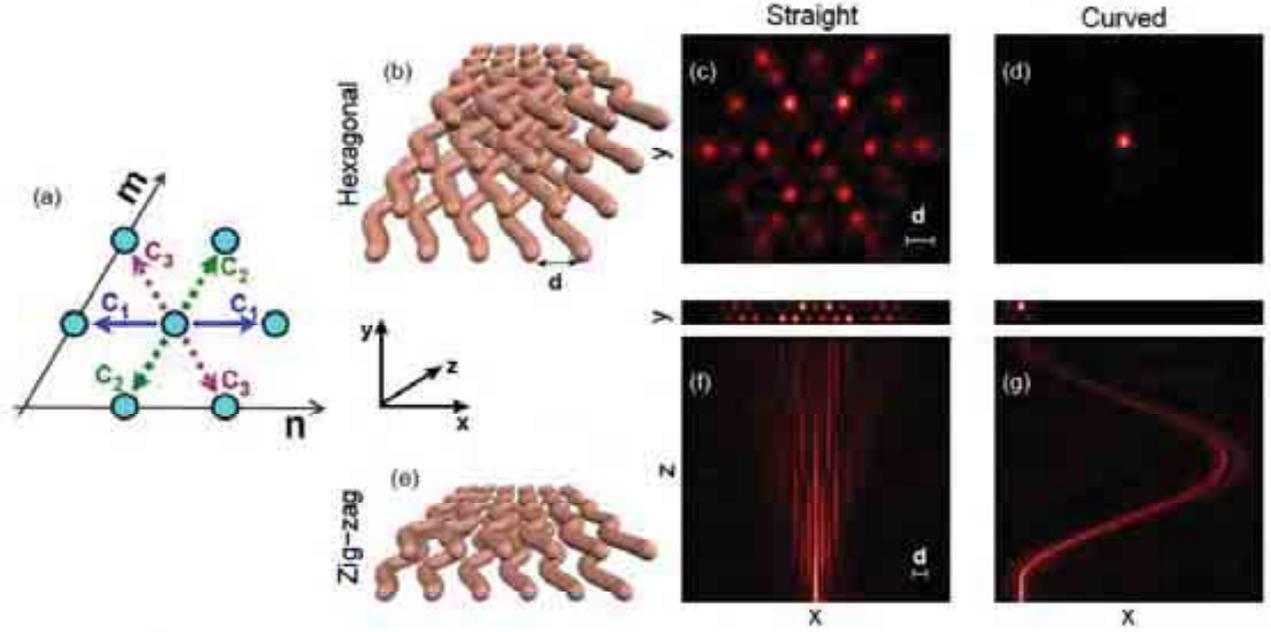

Figure 19: (a) Schematic of the couplings between the nearest neighbors in a hexagonal lattice. Lattice sites are numbered along the $n$ and $m$-axes. (b,e) Sketches of hexagonal and zig-zag periodically curved waveguide arrays. Insert shows orientation of the coordinate axes. (c,d) Experimentally measured output beam profiles for the straight and curved hexagonal lattices. Waveguide spacing is $d = 22$ $\mu$m. (f,g) Experimentally measured output beam profiles (top) and fluorescent images of beam propagation (bottom) for the straight and curved zig-zag arrays. Waveguide spacing is $d = 26$ $\mu$m. After Refs. [75, 291].

In order to specifically distinguish the effects due to diffraction management, we consider the light propagation in the waveguide arrays with symmetric bending profiles, since asymmetry may introduce other effects due to the modification of refraction, such as beam dragging and steering (Sec. 5.6). Specifically, we require that

$$x_0(z) = f_1(z - z_a) \text{ and } y_0(z) = f_2(z - z_a) \tag{58}$$

for some coordinate shift $z_a$, where functions $f_1(z)$ and $f_2(z)$ are symmetric,

$$f_1(z) = f_1(-z) \text{ and } f_2(z) = f_2(-z). \tag{59}$$

Then, by analyzing the plane-wave solutions of Eq. (54) it can be shown [75] that after the full bending period [$z \to z + L$, where $x_0(z) \equiv x_0(z + L)$ and $y_0(z) \equiv y_0(z + L)$] the beam diffraction in the periodically curved hexagonal waveguide array is the same as in a straight hexagonal array with the *effective coupling coefficients*,

$$C_1^{(\text{eff})} = \frac{C_1}{L} \int_0^L \cos\left[\omega \dot{x}_0(\zeta)\right] d\zeta, \tag{60}$$

$$C_2^{(\text{eff})} = \frac{C_2}{L} \int_0^L \cos\left[\frac{\omega}{2} \dot{x}_0(\zeta) + \frac{\sqrt{3}}{2} \omega \dot{y}_0(\zeta)\right] d\zeta, \tag{61}$$

$$C_3^{(\text{eff})} = \frac{C_3}{L} \int_0^L \cos\left[\frac{\omega}{2} \dot{x}_0(\zeta) - \frac{\sqrt{3}}{2} \omega \dot{y}_0(\zeta)\right] d\zeta. \tag{62}$$

Eqs. (60)-(62) show that the couplings along different high symmetry directions in a two-dimensional modulated lattice can be controlled *independently of each other* by designing appropriate bending profiles, providing new flexibility compared to the one-dimensional modulated lattices (Sec. 2.2.1). In particular, it becomes possible to engineer not only the strength and the sign of the diffraction in periodically curved two dimensional lattices, but to control also the effective lattice *geometry* and even the *dimensionality* of the lattice, as will be discussed in more detail in Sec. 2.4.3.

First, we explore possibility for the two-dimensional dynamic localization (2DDL). We follow Ref. [75] and consider curved hexagonal lattices where the waveguides are periodically bent in the $(x - z)$-plane along the propagation direction with a harmonic waveguide bending profile, which is expressed mathematically through the $z$-dependent transverse shift of the waveguides $x_0(z)$,

$$x_0(z) = \begin{cases} 0, & \text{if } 0 \leq z \leq z_0 \\ A\{\cos\left[\frac{2\pi(z-z_0)}{L-z_0}\right] - 1\}, & \text{if } z_0 \leq z \leq L \end{cases} \tag{63}$$

It was predicted in Ref. [75] that all three effective couplings (60)-(62) vanish, $C_1^{(\text{eff})} = C_2^{(\text{eff})} = C_3^{(\text{eff})} \equiv 0$, and a complete two-dimensional DL for both transverse dimensions can be realized for a special class of bending profiles, consisting of alternating straight and sinusoidal segments in each bending period $L$, provided that the bending parameters satisfy the conditions, $z_0 = [1 - 1/J_0(\xi)]^{-1}L$, and $A = -z_0\xi/\pi\omega J_0(\xi)$, where $\xi \approx 2.61$ is defined by the equation $J_0(\xi) = J_0(2\xi)$. Here $J_0$ is the Bessel function.

This setting for 2DDL was tested experimentally by Szameit and collaborators [291], who fabricated two-dimensional hexagonal waveguide arrays using the direct femtosecond laser-writing technique in fused silica glass. The experimental samples contained one full bending period with $L = 25$ mm, and the spacing between neighboring waveguides was $d = 22$ $\mu$m. The samples had hexagon-shaped boundaries with 5 waveguides at each facet [see Fig. 19(c)]. To characterize the arrays, cw laser light at the wavelength $\lambda = 633$ nm was launched into the central waveguide of the samples. Experimentally measured output beam profile for the straight sample is shown in Fig. 19(c), where strong beam diffraction is recorded. One can notice that light already hits the sample boundaries because of the finite number of the waveguides in the array. In contrast, in the curved sample with the bending profile (63) with the parameters tuned to the DL regime, the full suppression of the output beam diffraction in all transverse directions is clearly visible [Fig. 19(d)]. Numerical simulations performed by Szameit and collaborators [291] indicate that the diffraction is strongly reduced in the curved samples, such that the light does not hit the sample boundaries and DL is not affected by the finite array size in the experiments. Remarkably, it turns out that a simple one-dimensional lattice modulation in $x$-direction (63) is sufficient to completely suppress the *two-dimensional* diffraction in the whole transverse $(x - y)$-plane [Fig. 19(d)].

A similar DL effect was also observed [291] in the experiment with modulated zig-zag shaped lattices [see Fig. 19(e)], which can-be described by Eq. (54) with $m$ taking values $m = 1, 2$ [see schematic in Fig. 19(a)]. Interestingly, if we number all lattice sites with a single index $n$ according to their position along $n$-axis [Fig. 19(a)], such zig-zag lattices can be considered as quasi one-dimensional lattices with the next-nearest neighbor lattice site interaction [43]. The DL regime can be realized in the periodically curved zig-zag lattices with exactly the same waveguide bending profile (63) as was used for the hexagonal lattice. The equivalence between the two-dimensional diffraction suppression in the hexagonal and quasi one-dimensional zig-zag lattices can be ascribed to the topological similarity of the unit cells in both cases [cf. Fig. 19(b) and Fig. 19(e)]. To confirm these predictions experimentally, Szameit and collaborators [291] fabricated waveguide arrays with the zig-zag geometry and required waveguide bending [Fig. 19(e)]. The samples contained 31 waveguides (16 in the bottom row and 15 in the top row). In the experiments [291], light was coupled to the central waveguide in the top row of the zigz-ag array. Samples length was $L = 100$ mm and the waveguide spacing was $d = 26$ $\mu$m. The curved sample contained one full bending period. In order to visualize the light evolution which occurs inside the zig-zag lattice between the bending periods, a special fluorescence microscopy technique was employed [291]. In Figs. 19(f) and 19(g) experimentally recorded fluorescent images of the beam propagation inside the arrays (bottom) are shown together with the measured output intensity distributions (top). Clearly, in the straight array the light field broadens significantly during propagation [Fig. 19(f)]. In contrast, in the curved sample in the DL regime, the light field initially broadens but then it refocuses again into the excited waveguide at the output of the array after the propagation over one full bending period [Fig. 19(g)].

Recently, binary zig-zag arrays, which consist of two interleaved arrays of single-mode waveguides with detuned propagation constants, a primary array $A_n$, and an auxiliary array $B_n$, arranged in the zig-zag geometry shown in Fig. 19(e) with the bottom row composed of the waveguides $A_n$, and the top row composed of the waveguides $B_n$, were used by S. Longhi [176] to suggest *rectification of light refraction*. It was shown in Ref. [176] that if both arrays are periodically curved with a bending profile $x_0(z)$ made of a sequence of circular arcs with a constant curvature and alternating sign, light dynamics can be effectively decoupled, such that during propagation in the first segment with the positive curvature waveguide $A_n$ turns out to be coupled solely with waveguide $B_n$, and in the second segment with the negative curvature waveguide $B_n$ turns out to be coupled solely with waveguide $A_{n+1}$. This leads to a *ratchet effect,* i.e., a rectified transport in the absence of a net bias force, when after a full modulation cycle an arbitrary light distribution in the primary array $A_n$ is shifted, without distortion, by one unit from the left to the right. The result is a net drift of refracted light, with a locked to a specific refraction angle independent of the initial incidence angle and beam shape, and a suppression of discrete diffraction.

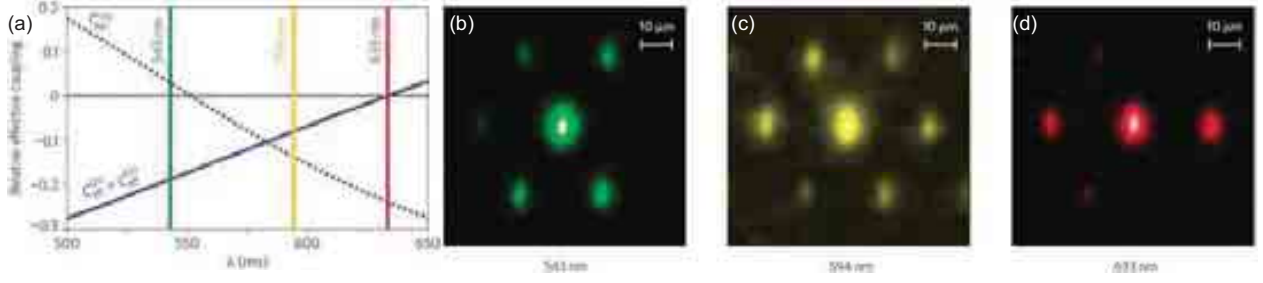

Figure 20: (a) Wavelength dependence of the effective couplings. Output diffraction profiles of light beams measured experimentally in the same modulated hexagonal lattice at three different wavelengths (b) 543 nm, (c) 594 nm, and (d) 633 nm. After Ref. [289].

It should be mentioned, that propagation of light waves in photonic lattices made of uniformly twisted *helical arrays* of evanescently coupled optical waveguides was theoretically investigated in Ref. [160]. It has been shown that a waveguide array with a helical structure can be used to mimic the effect of a uniform *magnetic field,* superimposed to a repulsive harmonic electrostatic force. The magnetic and electric forces are provided by the Coriolis and centrifugal forces experienced by optical rays in the non-inertial reference frame rotating with the twisted array [160]. In case of a one-dimensional helical waveguide array, the magnetic force does not play any role, and discrete diffraction in the array exactly mimics the dynamics of a quantum harmonic oscillator on a lattice. In two-dimensional helical arrays, light beam propagation is strongly influenced by the additional effective magnetic force, which leads to the appearance of a flowerlike trajectory of the optical beam in the transverse plane according to the semiclassical motion of a Bloch electron in the combined electric and magnetic fields.

*2.4.3. Spatial-spectral beam shaping*

In this Section, we examine diffraction of multi-color light beams in modulated two-dimensional photonic lattices. In Eqs. (60)-(62) the value of the effective coupling coefficients depends not only on the specific bending profile $x_0(z)$ and $y_0(z)$, but also on the optical frequency $\omega$, similar to the bending-induced dispersion which appears in one-dimensional periodically curved waveguide arrays (Sec. 2.2.4). This means that different frequency components may experience very different types of diffraction in *the same* physical structure. This feature provides unique opportunities for the control and reshaping of polychromatic light beams in two-dimensional photonic lattices.

To illustrate this effect, we follow Ref. [75] and consider the propagation of light beams of different wavelengths in the same modulated hexagonal lattice with a simple harmonic waveguide bending profile in the $x - z$ plane:

$$x_0(z) = A \cos\left(\frac{2\pi z}{L}\right), \tag{64}$$

where $A$ and $L$ are the bending amplitude and period, respectively, and $x_0(z)$ is the lattice transverse shift along $x$-axis. Although the waveguides are modulated in one plane, this affects coupling between the neighboring waveguides along different directions, defined by the effective coupling coefficients [75],

$$C_1^{(\text{eff})} = C(\lambda) J_0\left(\frac{\kappa}{\lambda}\right), \quad C_2^{(\text{eff})} = C_3^{(\text{eff})} = C(\lambda) J_0\left(\frac{\kappa}{2\lambda}\right), \tag{65}$$

where the parameter $\kappa = 4\pi n_0 dA/L$ depends only on the geometry of the lattice. Wavelength-dependencies of the effective coupling coefficients is presented in Fig. 20(a). One can see that the horizontal and diagonal coupling coefficients may vanish at different wavelengths, corresponding to partial DL along particular lattice symmetry directions.

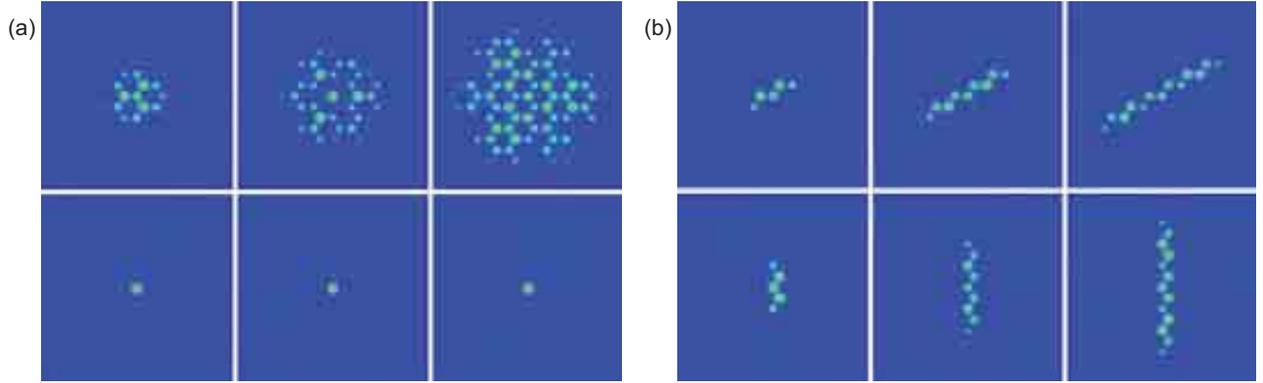

Figure 21: Optical field distributions at different propagation distances [propagation distance increases from the left to the right in (a) and (b)]. (a) Tunneling inhibition. Top row corresponds to an unmodulated array, while bottom row corresponds to a modulated array. (b) Anisotropic diffraction engineering. Top row: the waveguides are modulated in-phase along the diagonal of the array. Bottom row: the waveguides are modulated in-phase along the vertical axis. After Ref. [108].

Wavelength-controlled beam shaping in such structure was observed experimentally by Szameit and collaborators [289], who hexagonal array with harmonic bending (64) with bending period $L = 105$ mm and bending amplitude $A = 215$ $\mu$m. In the fabricated lattice, the diagonal couplings were suppressed at the wavelength $\lambda = 633$ nm [see red line in Fig. 20(a)]. In this spectral region light effectively experiences *one-dimensional* diffraction, see Fig. 20(d). On the other hand, for $\lambda = 583$ nm, all three couplings are reduced simultaneously by the same factor $-0.1$ [see yellow line in Fig. 20(a)], and the diffraction symmetry of the original hexagonal lattice is preserved. Output diffraction profiles at the closely tuned wavelength $\lambda = 594$ nm are shown in Fig. 20(c), where pronounced *hexagonal* diffraction pattern is visible. Finally, at $\lambda = 550$ nm [marked with green line in Fig. 20(a)] the horizontal coupling is canceled ($C_1^{(\mathrm{eff})} = 0$). At this wavelength, the beam can still spread across the whole lattice, yet the diffraction pattern is similar to those of *rectangular* lattices where each lattice site is coupled to its four immediate neighbors [see Fig. 20(b)]. The observed localization-induced transformation of lattice geometry suggests new approaches for flexible shaping of polychromatic light with ultra-broadband or supercontinuum spectra [251, 48].

*2.4.4. Tunneling control*

Control of light tunneling based on the self-collimation effect (see Sec. 2.2.3) can also be generalized to the two-dimensional geometries, provided that the out-of-phase refractive index modulation condition (29) can be satisfied. This is, however, not possible for two-dimensional lattices with arbitrary symmetries (for example, the out-of-phase modulation cannot be achieved for all lattice sites in hexagonal lattices).

It was suggested by Kartashov and collaborators [108] that two-dimensional control of light tunneling, can be achieved in honeycomb waveguide arrays, where one can realize a configuration when each waveguide is completely surrounded by neighbors with the out-of-phase longitudinal refractive index modulation. In this specific configuration it is possible to realize an out-of-phase modulation which results in the suppression of light tunneling. Light tunneling is inhibited almost completely for modulated honeycomb arrays under appropriate resonant conditions, see Fig. 21(a), lower row. This is in a sharp contrast to the unmodulated array [Fig. 21(a), upper row].

One can also engineer the diffraction by properly selecting clusters of out-of-phase or in-phase modulated waveguides. A nontrivial diffraction control might be realized by dividing the entire array into clusters where in each cluster the refractive index of adjacent guides oscillates in-phase, but in waveguides belonging to different clusters it oscillates out-of-phase [108]. In honeycomb arrays featuring three principal axes one can have waveguides oscillate in-phase in the direction parallel to the principal axis, but out-of-phase in the direction perpendicular to it. In this case light beams will diffract along the selected principal axis, while in the perpendicular direction the diffraction will be inhibited. This results in essentially one-dimensional anisotropic diffraction in an intrinsically two-dimensional array [see Fig. 21(b)].

In addition, it was shown in Ref. [108] that tunneling inhibition in honeycomb arrays is possible not only for simplest excitation of a single channel but also for multiple-channel excitations. This enables simultaneous diffrac-tionless transmission of several beams launched in different locations in the array. In particular, tunneling inhibition was demonstrated for optical vortices [108].

Recently, coherent destruction of tunneling and diffraction management was realized experimentally in square lattices by Zhang and collaborators [331], who used optical induction technique to create square lattices with out-of-phase index modulation. By changing the modulation amplitude, strength and sign of the diffraction was controlled. When the effective coupling coefficient is negative, it was observed [331] that the initially focused Gaussian beam diverges, and the initially upward-tilted beam bends downward, exhibiting typical behavior of anomalous diffraction and negative refraction. In the self-collimation regime, transmission of a two-dimensional image was demonstrated.

## 3. Modulated waveguides and couplers

*3.1. Preliminary remarks*

Axis bending or longitudinal modulation of the refractive index in a set of few coupled waveguides, such as in optical directional couplers, can deeply affect the propagation properties of light. The effect of modulation on light dynamics strongly depends on the spatial scale $\Lambda$ of the modulation as compared to the typical spatial scale $s$ of light evolution in the non-modulated optical structure. The latter is usually defined by the modal diffraction length for a single mono-mode waveguide, by the beating length of different optical modes in multi-mode waveguides, or by the coupling length for an optical directional coupler. Generally speaking, three different modulation regimes can be found depending on the ratio $\epsilon = \Lambda/s$. The first one corresponds to $\epsilon \ll 1$, i.e. to a modulation of axis bending or longitudinal index modulation on a spatial scale $\Lambda$ much shorter than $s$. In this regime, light waves that propagate along the dielectric structure can not follow the fast changes of the refractive index profile introduced by the modulation, and at leading order the dielectric medium behaves like a non-modulated medium with an effective refractive index obtained by averaging, over the spatial scale $\Lambda$, the rapidly-varying index profile. The second regime corresponds to the case $\epsilon \sim 1$, i.e. to a characteristic spatial scale $\Lambda$ of axis bending or index modulation which is of the same order as $s$. In this regime, resonance phenomena in the process of light transport can be envisaged. Finally, the third regime corresponds to the case $\epsilon \gg 1$, where light propagation is expected to adiabatically follow the slow changes of the refractive index profile. In all cases, modulation amplitudes and periods of axis bending should be chosen to avoid significant radiation losses in the guiding device. In this section we review some of the main results related to linear and nonlinear light dynamics in modulated or bent optical waveguides and couplers, highlighting some potential applications to monochromatic and polychromatic light transfer control and to optical switching. As light dynamics in modulated photonic lattices show strong similarities with coherent dynamics of quantum particles in periodic potentials, such as electrons in crystalline solids or semiconductor superlattices or Bose-Einstein condensates in optical lattices, light propagation and control in modulated waveguides and couplers discussed in this section shear strong similarities with coherent quantum control methods encountered in atomic and molecular physics contexts [175].

*3.2. Modulated single waveguides*

The first example of light control in a modulated single-mode waveguide is provided by beam splitting and adiabatic stabilization of light in a periodically curved optical waveguide. As axis bending in waveguide theory is generally associated to the excitation of radiation modes and it is thus responsible for the appearance of radiation losses [68, 199], in modulated waveguides radiation losses can be mostly suppressed under proper conditions and light can be still guided in the structure. Such a curious phenomenon, which was theoretically predicted and experimentally observed by Longhi and collaborators [181, 182], is the optical analogue of adiabatic stabilization (i.e. the reduction or even suppression of ionization) and wave packet dichotomy of atoms in intense and high-frequency laser fields predicted more than 20 years ago in the atomic physics context [235, 234, 76]. Suppression of radiation losses and adiabatic stabilization of light in a modulated optical waveguide with a high spatial modulation frequency can be simply captured within an averaged model of the scalar and paraxial wave equation, as discussed in [181, 182]. Let us consider light propagation in an optical waveguide with a curved axis profile $x_0(z) = A(z)\cos(2\pi z/\Lambda)$ of spatial period $\Lambda$ shorter than the characteristic diffraction length of the waveguide mode and with the profile shown in Fig. 22(a). The scalar wave equation that describes propagation of light waves at wavelength $\lambda$ in the modulated waveguide is given by

$$i\frac{\partial E}{\partial z} = -\frac{\lambda}{4\pi n_s}\frac{\partial^2 E}{\partial x^2} + \frac{2\pi}{\lambda}\Delta n(x - x_0(z))E, \qquad (66)$$

where $n_s$ is the substrate refractive index and $\Delta n(x) = n_s - n(x)$ the refractive index change of the waveguide core of the non-modulated (straight) waveguide. The amplitude $A(z)$ of the modulation is assumed to slowly vary over one spatial period $\Lambda$ and to adiabatically increase from zero to a steady-state value $A_0$. If the final amplitude $A_0$ is too small, the optical beam intensity $|E|^2$ spreads far away from the waveguide core, as shown in Fig. 22(b) which depicts a typical evolution of the beam intensity $|E(x, z)|^2$ as obtained by numerical integration of Eq.(66). This case corresponds to strong bending-induced radiation losses of the periodically-curved waveguide, i.e. to atomic ionization in the quantum mechanical analogy. As the amplitude $A_0$ is further increased, adiabatic stabilization, associated to a splitting of the beam, is observed, as shown in Fig. 22(c). In such a regime, light propagation can be safely described by a cycle-averaged equation. Indeed, for a short modulation period $\Lambda$ of waveguide bending, the light field cannot follow the rapid longitudinal variation of the refractive index and beam dynamics is governed, at leading order, by a cycled-averaged refractive index potential. The high-frequency limit thus leads to the cycle-averaged scalar wave equation

$$i\frac{\partial E}{\partial z} = -\frac{\lambda}{4\pi n_s}\frac{\partial^2 E}{\partial x^2} + \frac{2\pi}{\lambda}\Delta n_{av}(x,z)E, \tag{67}$$

where

$$\Delta n_{av}(x,z) = \frac{1}{\Lambda}\int_0^\Lambda dz\,[n_s - n(x - x_0(z))] \tag{68}$$

is the cycle-averaged refractive index change of the waveguide. The behavior of $\Delta n_{av}(x,z)$, which slowly varies with $z$ owing to the adiabatic change of the amplitude $A(z)$, is shown in Fig. 22(d). Note that $\Delta n_{av}(x,z)$ corresponds to the refractive index profile of a Y splitter, i.e. the curved waveguide with short bending period $\Lambda$ is equivalent, at leading order, to a Y adiabatic splitter. Suppression of radiation losses and wave packet dichotomy are thus simply explained as due to the appearance of an adiabatic splitter in the cycle-averaged limit. In the previous analysis, the waveguide has been assumed to be single-mode and bent in the $(x,z)$ plane. Adiabatic stabilization can occur also in three-dimensional waveguides with non-planar axis bending [153]. In particular, a helicoidal axis bending [Fig. 23(a)] realizes the optical analogue of adiabatic stabilization of a two-dimensional atom in a circularly-polarized laser field [153]. In this case, the cycle-average effective waveguide acts as an annular channel, and the launched beam adiabatically evolves into a ring-shaped pattern, as shown in Fig. 23(b).

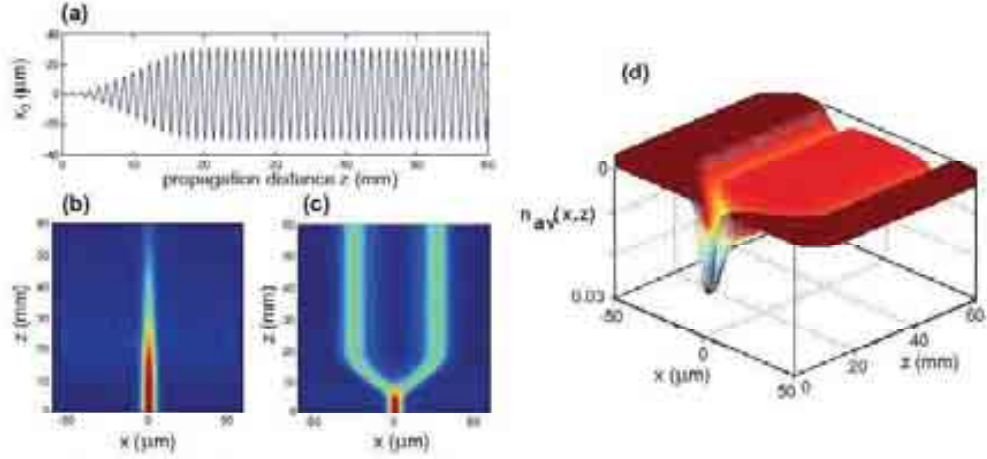

Figure 22: Adiabatic stabilization of light in periodically-curved waveguides. (a) Typical behavior of the waveguide axis bending profile $x_0(z)$, corresponding to a sinusoidal bending with a maximum amplitude $A_0 = 30$ $\mu$m (after the adiabatic region) and with period $\Lambda = 1200$ $\mu$m. (b) Evolution of beam intensity $|E(x,z)|^2$ along the waveguide in the lossy regime, corresponding to $\Lambda = 120$ $\mu$m and $A_0 = 2$ $\mu$m. (c) Same as in (b), but in the adiabatic stabilization regime ($\Lambda = 120$ $\mu$m and $A_0 = 30$ $\mu$m). The behavior of the cycle-averaged index change $n_{av}(x,z)$, corresponding to the stabilization regime of (c), is depicted in (d). Parameter values are: $\lambda = 1.55$ $\mu$m, $n_s = 2.138$, $n(x) = \Delta n_0 \{\text{erf}[(x+w)/D_x] - \text{erf}[(x-w)/D_x]\}/2$, $\Delta n_0 = 0.0074$, $w = 5$ $\mu$m, and $D_x = 2.375$ $\mu$m. After Ref. [175].

The simple analysis so far discussed assumes that the straight waveguide sustains a single (bound) mode, however a more complex scenario can be observed for beam propagation in periodically curved multimode waveguides. If the spatial frequency of modulation is shorter than the propagation constant separation of the various guided modes, the average model can be still adopted, and adiabatic stabilization (i.e. suppression of radiation losses) can still be observed, however splitting of the light wave packet may not occur in the multimode regime [198]. On the other hand, if the spatial frequency of the waveguide axis modulation matches the propagation constant mismatch between two guided modes, resonant power oscillations between the different guided modes can be observed [117]. In some sense, the waveguide axis modulation acts in this case like a grating that couples the two guided modes. This phenomenon

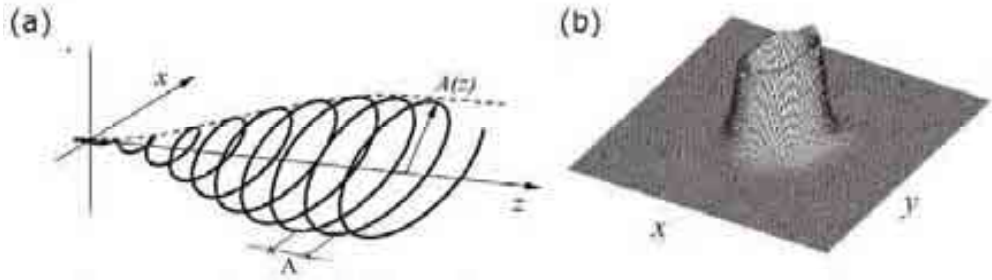

Figure 23: Adiabatic stabilization of light in a twisted channel optical waveguide. (a) Schematic of the waveguide axis bending. (b) Example of an adiabatically-stablized annular mode guided by the cycle-average effective waveguide after the adiabatic region. After Ref. [153].

is analogous to Rabi oscillations in a two-level atom model driven by a resonant laser field.

*3.3. Modulated optical couplers*

The waveguide optical directional coupler is a device which exploits light tunneling between two optical waveguides placed in close proximity to each other [97, 189]. It is a basic component of optical networks, which is used to combine and split optical signals. A monochromatic light beam injected into one of the two waveguides of the coupler oscillates back and forth between the two waveguides as it propagates in the device [97]. The period is defined by the coupling length, which is inversely proportional to the coupling rate between the two waveguides. Several works, briefly reviewed in this section, have shown that the flow of monochromatic or polychromatic beams in the coupler can be strongly modified by periodic axis bending or by the introduction of a longitudinal modulation of the refractive index change or channel width of waveguides.

*3.3.1. Coherent destruction of photonic tunneling*

Periodic axis bending of the optical coupler enables to effectively control the coupling length. In particular, for a sinusoidal axis bending and at special values of amplitudes and spatial frequencies of modulation, the effective coupling coefficient vanishes, i.e. an effective suppression of light tunneling between the two waveguides can be realized [149, 34]. Such a phenomenon is the optical analogue of the coherent destruction of tunneling of a quantum particle in a bistable potential driven by an external ac field, originally proposed by P. Hänggi and collaborators [83, 82] 20 years ago but observed experimentally for matter waves solely very recently using cold atoms trapped in optical lattices [120]. The optical potential $V_e(x) = n_s - n(x)$ of a planar optical directional coupler, with effective refractive index profile $n(x)$ and substrate index $n_s$, acts for light waves like a double-well potential for matter waves. The double-well optical potential $V_e(x)$ generally supports a set of guided modes for a given wavelength $\lambda$, with energies $E_1 < E_2 < E_3 < ...$ [see Fig. 24(b)]. In the optical context, the two lowest-energy guided modes are generally referred to as the coupler supermodes [320], and $\Delta\beta = 2\pi(E_2 - E_1)/\lambda$ gives the propagation constant detuning of such supermodes. In the framework of the coupled-mode theory, the propagation constant detuning is related to the coupling rate $C$ between the two waveguides by the simple relation $\Delta\beta = 2C$. For a symmetric coupler, the two supermodes correspond to the symmetric and antisymmetric superpositions of the single guided modes of the two waveguides. A light wave initially launched into one of the two waveguides periodically tunnel back and forth between the two waveguides owing to the propagation constant shift $\Delta\beta$ of symmetric and antisymmetric supermodes. Sinusoidal bending of the coupler axis, with spatial period $\Lambda = 2\pi/\omega$ and amplitude $A$ [see Fig. 24(a)], acts for light waves like an external time-varying sinusoidal force for the quantum particle in the bistable potential $V_e(x)$ [34, 175]. The temporal evolution of the quantum particle in the driven bistable potential is thus mapped into light beam evolution along the spatial axial direction of the coupler. For certain parameter ratios between the amplitude and spatial period of the modulation, photon tunneling between the two waveguides of the coupler can be brought to a complete standstill [83], i.e. the optical analogue of coherent destruction of tunneling is realized. The experimental demonstration of coherent destruction of tunneling for light waves in sinusoidally-curved optical directional couplers was reported by Della Valle and collaborators in Ref.[34]. The energy level diagram of the double-well potential, in absence of the external driving field and corresponding to the optical coupler used in the experiment by Della Valle and collaborators, is shown in Fig. 24(b). The two lowest-energy levels $E_1$ and $E_2$ correspond to the symmetric and antisymmetric supermodes of the optical coupler, whereas the other energy levels correspond to higher-order modes. When the coupler is sinusoidally bent, the energy levels of the time-independent optical Hamiltonian $\mathcal{H} = -\lambda^2/(8\pi^2 n_s)\partial_x^2 + V_e(x)$ are replaced by the quasi-energy levels of a time-periodic Hamiltonian $\mathcal{H}' = -\lambda^2/(8\pi^2 n_s)\partial_x^2 + V_e(x) - F(z)x$, where the sinusoidally-varying homogeneous force $F(z)$ is given by $F(z) = (4\pi^2 A n_s/\Lambda^2)\cos(2\pi z/\Lambda)$ [34]. In particular, the two energies $E_1$ and $E_2$ of the symmetric and antisymmetric supermodes of the straight coupler are replaced by the quasi-energies $\epsilon_1$ and $\epsilon_2$ in the modulated coupler. Coherent destruction of tunneling occurs approximately for a spatial modulation frequency $\omega$ in the range $(E_2 - E_1) < (\lambda/2\pi)\omega < (E_3 - E_2)$ and for a modulation amplitude that corresponds to exact crossing between the quasi-energies $\epsilon_1$ and $\epsilon_2$, i.e. for $\epsilon_1 = \epsilon_2$ [82]. The manifold of quasi-energy level crossing $\epsilon_2 = \epsilon_1$ in the $(A, \Lambda)$ plane is shown in Fig. 24(c) by the solid line. An approximate expression of the quasi-energy splitting $\Delta\epsilon = \epsilon_2 - \epsilon_1$, which is valid in the high-frequency limit $\lambda/(2\pi)\omega \gg (E_2 - E_1)$, can be calculated by coupled-mode theory and reads [149]

$$\Delta\epsilon = (E_2 - E_1)J_0(\Gamma) \tag{69}$$

where

$$\Gamma = \frac{4\pi^2 n_s a A}{\lambda \Lambda} \tag{70}$$

and $a$ is the distance between the two waveguides. Note that, under such an approximation, the first manifold of quasi-energy crossing corresponds to first zero of $J_0$ Bessel function and is represented by the straight dashed line in Fig. 24(c). More generally, the modulated coupler behaves, at propagation distances integer multiplies of the modulation period, like a straight coupler with an effective coupling rate $C_{eff}$ given by $C_{eff} = CJ_0(\Gamma)$. Noticeably, modulation of the coupler can reverse the sign of the coupling rate, a property which can be of interest in diffraction management and control problems (see, for instance, [147, 54]). The experimental demonstration of coherent destruction of tunneling for light waves, reported in Ref.[34], was simply based on fluorescence imaging of light propagation in a series of sinusoidally-curved optical directional couplers manufactured by the ion-exchange technique in an active Er-doped phosphate glass. Single waveguide excitation at 980 nm is accomplished by fiber butt coupling, and spatial mapping of light propagation along the couplers is achieved from the top of the sample using a CCD camera connected to a microscope, as shown in Fig. 24(d). Light waves at 980 nm are partially absorbed by the $Er^{3+}$ ions, yielding a green upconversion fluorescence which is monitored by the CCD. Since the fluorescence is proportional to the local photon density, the recorded fluorescence patterns map the profile of $|E(x,z)|^2$ in the waveguide reference frame. The onset of CDT is clearly shown in Fig 24(e): as in the straight coupler light waves tunnel back and forth between the two waveguides, with a coupling length $\sim 1/6$ of the sample length, in the sinusoidally curved couplers, with period $\Lambda$ and amplitude $A$ chosen on the quasi-energy crossing manifold, optical tunneling is suppressed. In fact, the fluorescence images taken for the sinusoidally-curved couplers clearly show that light remains trapped in the initially-excited (upper) waveguide, following the sinusoidally-bent path of its optical axis without tunneling into the adjacent (bottom) waveguide.

It should be mentioned that the control of the effective coupling rate in an optical directional coupler can be achieved by keeping the two guiding channels straight, but modulating their refractive index or widths in a periodic and out-of-phase fashion along the longitudinal direction [184, 270, 294]. In the fast modulation regime, the two methods of tunneling control are essentially equivalent and a relation similar to Eq.(69) holds for the index modulation case as well. However, especially in case of more than two waveguides and at small modulation frequencies, the two methods give different results. In particular, in Ref.[184] it was shown that in a waveguide array exact suppression of light spreading can not be achieved in the index modulation case. Apart from such a difference, for both axis bending and index modulation methods coherent destruction of tunneling is a resonant effect, occurring only for specific modulation amplitude and frequency values. Recently, it was however suggested by X. Luo and collaborators [186] that nonlinearity may cause a broadening of the resonance in directional couplers. Such a broadening effect, related to the so-called nonlinear coherent destruction of tunneling, yields a tunneling inhibition also in structures with a frequency slightly detuned from the resonance one and for powers well below the power required for the formation of a soliton in the unmodulated system. The phenomenon of nonlinear coherent destruction of tunneling has been observed in a recent experiment by A. Szameit and collaborators [294], based on a femtosecond-laser-written optical waveguide coupler in fused silica. The experiment by Szameit and collaborators confirmed two salient features of the theoretical prediction (see Fig. 25): (1) nonlinearity extends the destruction of tunneling to a finite parameter range; (2) periodic driving reduces the threshold power for the tunneling suppression.

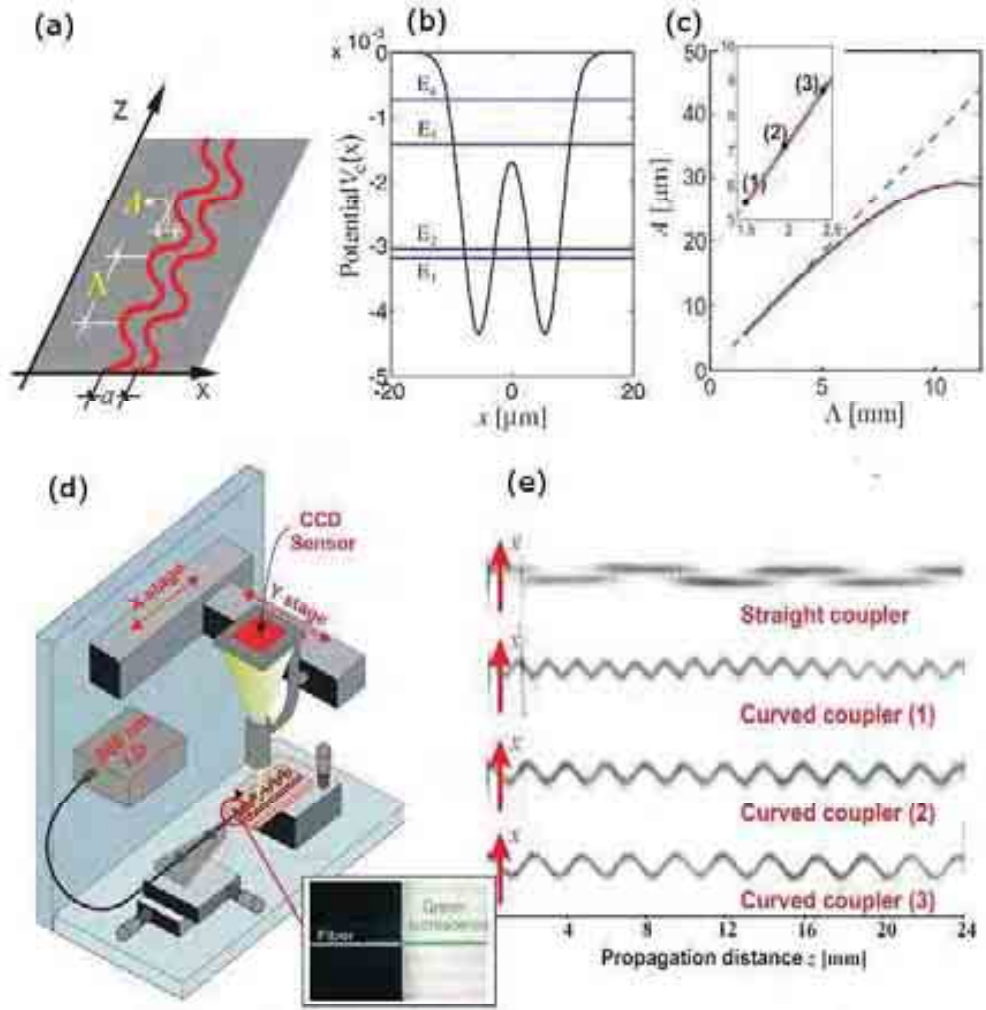

Figure 24: Coherent destruction of photonic tunneling in a sinusoidally-curved synchronous directional coupler. (a) Schematic of the optical coupler. The axis bending profile is sinusoidal with amplitude $A$ and period $\Lambda$. (b) Energy level diagram of the double-well optical potential for the directional coupler used in the experiment of Ref.[34]. (c) Manifold of quasi-energy crossing in the $(A, \Lambda)$ plane. (d) Experimental set-up for fluorescence light imaging: the green fluorescence emitted by Er ions is collected by a microscope objective and imaged onto a CCD sensor. (e) Measured fluorescence patterns in a straight directional coupler (upper figure) and in three sinusoidally curved couplers (bottom figures) for geometric parameters corresponding to points (1), (2) and (3) of Fig. 24(c). After Refs.[34, 175].

An all-optical switch based on a nonlinear directional coupler with periodic modulation, which shows a much lower switching threshold power and sharper switching width as compared to a traditional nonlinear directional coupler switch, has been discussed in Ref.[317]. Extensions of photonic tunneling control in longitudinally-modulated optical couplers include light tunneling inhibition in arrays of optical couplers [112] and switching management in couplers with biharmonic longitudinal modulation of refractive index [111].

*3.3.2. The polychromatic coupler*

As discussed in Sec. 3.3.1, in a straight directional coupler [Fig. 26(a)] a monochromatic beam of wavelength $\lambda$ initially injected into one waveguide of the coupler undergoes a periodic beating back and forth between the two

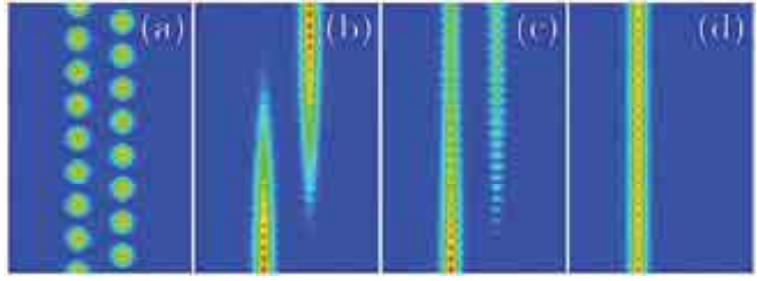

Figure 25: Light propagation in a nonlinear waveguide coupler: (a) without periodic modulation; (b) with periodic modulation and weak laser power; (c) with periodic modulation and strong laser power, corresponding to nonlinear destruction of tunneling. After Ref.[294].

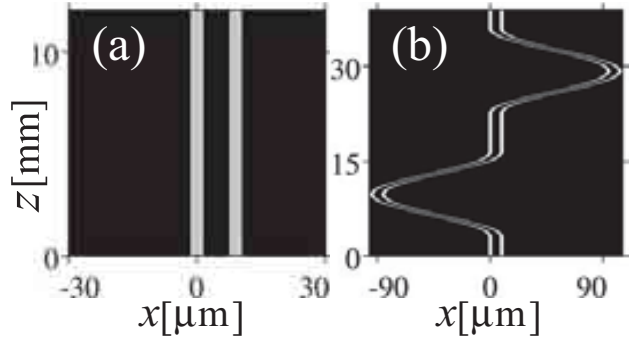

Figure 26: (a) Conventional directional coupler composed of two evanescently coupled straight waveguides. (b) Polychromatic light coupler with specially designed bending of the waveguide axes. Waveguide width and separation between waveguide axes are 3 $\mu$m and 9 $\mu$m, respectively. Refractive index contrast is $\Delta n = 8 \times 10^{-4}$, substrate index $n_0 = 2.35$. After Ref. [70].

waveguides with a period defined by the coupling length

$$Z_c = \frac{\pi}{2C(\lambda)}, \qquad (71)$$

where $C(\lambda)$ is the coupling coefficient at wavelength $\lambda$. Then, in the linear regime, signal switching between output coupler ports is realized by choosing the device length as an odd number of coupling lengths. Since the first experimental demonstration of a subpicosecond switching in a dual-core fiber coupler [66], various aspects of switching in different coupler configurations has been extensively analyzed [140, 5, 104, 264, 14].

However, conventional couplers can only perform switching of signals with rather limited spectral bandwidth, because the coupling length depends on optical frequency [203, 97] and tends to increase at the red spectral edge, resulting in strong separation of different frequency components between the waveguides, as shown in Figs. 27(a) and 27(c).

In Ref. [70] a new configuration of directional coupler was designed for switching of polychromatic light, such as light with supercontinuum frequency spectrum generated in photonic-crystal fibers and fiber tapers [245, 311]. The spectral bandwidth of the suggested device is five times wider compared to conventional straight couplers, making it possible to collectively switch wavelengths covering almost the entire visible region.

Following Ref. [70], the operating bandwidth of conventional coupler consisting of straight parallel waveguides [Fig. 26(a)] can be improved by introducing special bending of waveguide axes in the propagation direction as illustrated in Fig. 26(b). The effect of axes bending on light propagation in two coupled waveguides can be described in terms of the effective coupling coefficient $C_{\text{eff}}$. It was shown [149] that, in the limit when bending period ($\Lambda$) is much smaller than the coupling length for straight waveguides ($Z_c$), the light distribution at the output of the curved coupler is the same as for straight structure with the coupling $C_{\text{eff}}$ between the waveguides. Numerical simulations indicate that the effective coupling accurately describes the device operation for a broad class of bending profiles even when $L$, $\Lambda$, and $Z_c$ are of the same order, where $L$ is the length of the device. As shown in Ref. [70], wavelength-insensitive effective coupling can be realized in a hybrid coupler consisting of alternating straight and sinusoidal segments [see

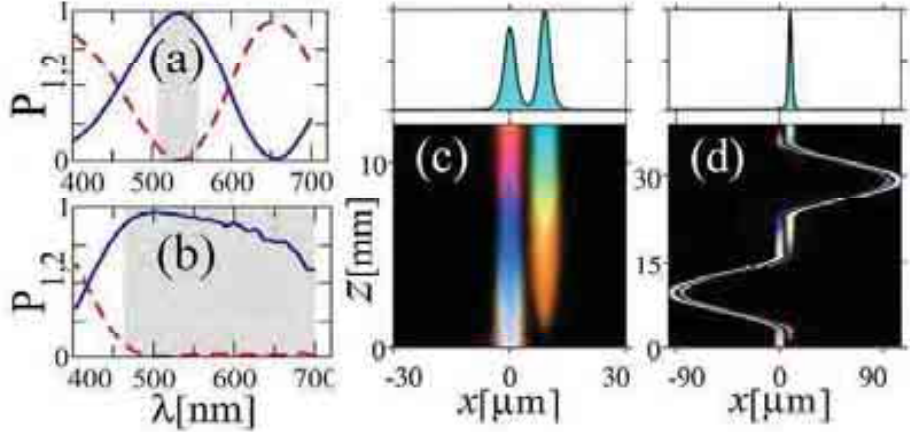

Figure 27: (a,b) Wavelength dependence of linear transmission of straight and optimized curved couplers, respectively. The figure shows the output powers in the left (dashed curve, $P_1$) and right (solid curve, $P_2$) coupler arms when light is launched into the left coupler arm at the input. Shading marks spectral regions where the switching ratio $P_2/P_1$ is larger than 10. (c,d) Evolution of polychromatic light with flat spectrum covering $450 - 700$ nm in the straight and in the optimized curved structures, respectively. Top panels in (c) and (d) show the total intensity distributions at the output. After Ref. [70].

Fig. 26(b)]. Such optimized curved coupler can be used to collectively switch all spectral components around the central wavelength $\lambda_0$ from one input waveguide to the other waveguide at the output if the device length is matched to the effective coupling length at the central wavelength $\lambda_0$, i.e. $L = \pi/[2C_{\text{eff}}(\lambda_0)]$. We note that the broadband behavior of such coupler is based on the generic coupled-mode analysis, and does not depend on a specific device realization, unlike e.g., broadband twin-core photonic crystal fiber couplers which require a sophisticated refractive index profile in the core regions for their operation [130]

Numerical simulations based on full continuous model confirmed [70] that the proposed coupler structure indeed exhibits extremely efficient switching into the crossed state simultaneously in a very broad spectral region of about $450 - 700$ nm, which covers almost the entire visible range, see Figs. 27(b) and 27(d). This is in a sharp contrast to the conventional straight coupler [Figs. 27(a) and 27(c)] that can only operate in the spectral region of $\sim 510 - 560$ nm, which is about five time less than for the proposed curved coupler. We note that slight decrease of the output power at the red edge of the spectrum for the curved coupler [Fig. 27(b)] is caused by the radiation at the waveguide bends [149], but such losses do not affect the broadband switching behavior.

### 3.4. Tunneling control and diffraction management in twisted fibers

In the previous section it has been shown that a periodic in-plane bending of the optical axis of a directional coupler modifies the effective tunneling rate between adjacent waveguides. It should be mentioned that control of photonic tunneling in coupled waveguide fibers or fiber arrays can be also achieved by twisting (rather than bending) the optical axis. The geometric twist of the optical structure along the propagation direction is responsible for the appearance, in the reference frame of the waveguides, of noninertial Coriolis and centrifugal forces for photons, which are analogous the former to a magnetic (Lorentz) force and the latter to an electrostatic (repulsive) harmonic force (see, for instance, [178, 223, 160]). In particular, in case of annular fibers [223] or coupled waveguide fibers arranged along a ring [162], the Coriolis force is the most important one. It is responsible, for example, of the optical analogue of the Aharonov-Bohm effect [223] and can be exploited to suppress the tunneling between two communicating defects [223, 162] (the so-called topological suppression of photonic tunneling). Interestingly, diffraction management and self-imaging phenomena can be obtained in a circular fiber waveguide array by a suitable geometric twist. Following Ref.[162], let us consider a circular array of $N$ single-mode fibers in the geometrical setting of Fig.28(b), an let us indicate by $\Psi_k$ the amplitudes of the light modes in the various waveguides ($k = 1, 2, 3, ..., N$). The optical structure is twisted along the propagation direction $z$ by a twist rate $\epsilon(z)$. Coupled-mode equations describing light transport in the twisted array, as derived in Ref.[162], read explicitly

$$i\frac{d\Psi_k}{dz} = -C\left[\exp(-i\delta)\Psi_{k+1} + \exp(i\delta)\Psi_{k-1}\right], \tag{72}$$

where $C$ is the coupling constant of adjacent fibers (in the absence of the twist), $k = 1, 2, 3, ..., N$, $\Psi_0 = \Psi_N$, and $\Psi_{N+1} = \Psi_1$. In Eq.(72), the additional phase term $\delta(z)$ in the coupling constant arises because of the geometric twist and is given by

$$\delta(z) = \frac{4\pi^2 \epsilon(z) n_s r_0^2}{N\lambda} \tag{73}$$

where $\lambda$ is the light wavelength in vacuum, $n_s$ is the substrate refractive index, and $r_0$ is an effective radius of the the circular array, which is close to its geometrical radius $R_0$ (for a precise definition of the effective radius $r_0$ we refer the reader to Ref. [162]). For an arbitrary $z$-dependence of $\epsilon$, the most general solution to the coupled-mode equations (72) is given by an arbitrary superposition of the following $N$ fundamental solutions

$$\Psi_s^{(k)}(z) = \exp[iQ_k s + i\theta(z, Q_k)] \tag{74}$$

with $k = 0, 1, 2, ..., N - 1$, $Q_k = 2\pi k/N$, and

$$\theta(z, Q_k) = 2C \int_0^z d\xi \cos[Q_k - \delta(\xi)]. \tag{75}$$

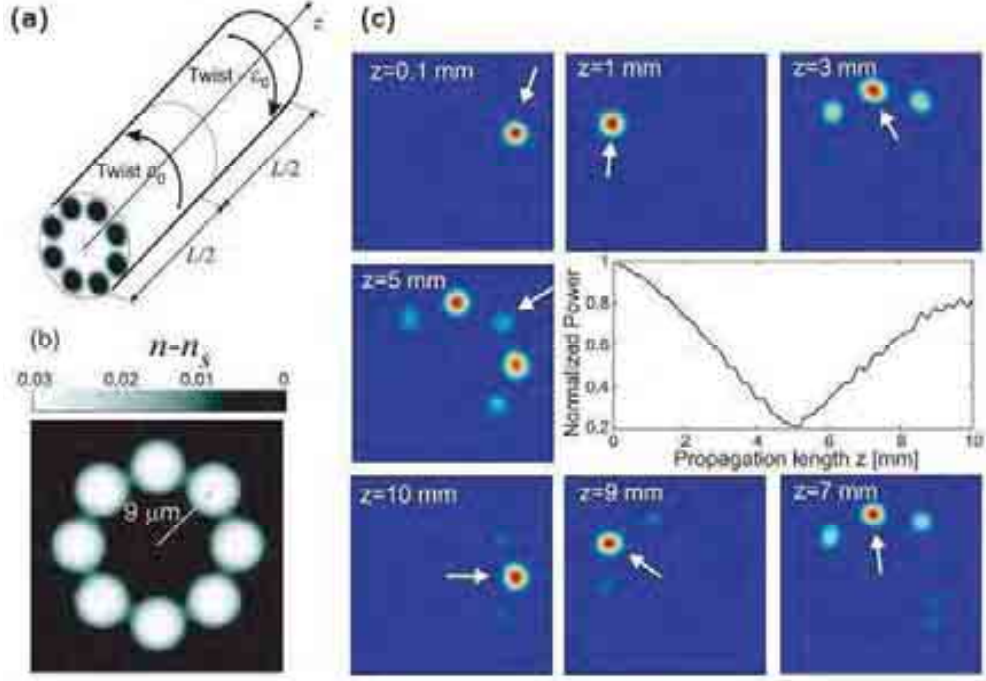

Figure 28: Self-imaging in a twisted circular fiber array with alternating clockwise and counter-clockwise twists $\pm\epsilon_0$. (a) Schematic of the fiber array with $N = 8$ cores and with alternating twist geometry. (b) Refractive index profile $n(x,y) - n_s$ of the fiber array. (c) Beam propagation evolution along the fiber array ($\lambda = 980$ nm). The arrows in the figures locate the initially excited fiber of the array, whereas the central plot shows the fractional beam power trapped in such fiber versus propagation distance. After Ref. [162].

An important case is the self-imaging effect, which was discussed for periodically curved arrays in the previous chapter and related to the phenomenon of dynamic localization. The condition for self-imaging for a twisted fiber array, from the input plane $z = 0$ to the output plane $z = L$, requires $\Psi_s^{(k)}(L) = \Psi_s^{(k)}(0)$ for any $k = 0, 1, 2, ..., N - 1$, a condition that can be fulfilled provided that

$$\int_0^L d\xi \exp[-i\delta(\xi)] = 0. \quad (76)$$

Note that such a condition for self-imaging is similar to the one encountered in a planar array of bent optical waveguides [183]. The simplest case of twist management that realizes optical self-imaging is that corresponding to the sequence of two twisted regions, of equal length $d = L/2$, with uniform twist rates $\epsilon = \epsilon_1$ and $\epsilon = \epsilon_2$ in the two regions. In such a case, the self-imaging condition yields $\exp(-i\delta_1) + \exp(-i\delta_2) = 0$, which can be satisfied by assuming either $\delta_1 = 0, \delta_2 = \pi$ or $\delta_1 = \pi/2, \delta_2 = -\pi/2$, i.e. for either

$$\epsilon_1 = 0, \ \epsilon_2 = \frac{N\lambda}{4\pi n_s r_0^2} \quad (77)$$

or

$$\epsilon_1 = -\epsilon_2 = \frac{N\lambda}{8\pi n_s r_0^2}. \quad (78)$$

Note that the condition for self-imaging is independent of the length $d$ of each region. Note also that Eq.(77) means that discrete diffraction in an untwisted fiber array after an arbitrary propagation length $d$ can be always cancelled adding a successive twisted region of the same length provided that the twist rate is chosen according to Eq.(77). Conversely, the conditions expressed by Eq.(78) correspond to propagation in a twisted circular fiber array with alternating sign for the twist, as shown in Fig.28(a). An example of diffraction cancellation in a $N = 8$ $L = 1$-cm-long circular fiber array via twist management, based on alternating twist regions, is shown in Fig.28. As an initial condition, one of the fiber in the array has been excited in its fundamental mode, and the arrows in the various panels of Fig.28(c) indicate the position of the initially-excited fiber at a few propagation lengths. Note that, after a full twist cycle. i.e. at the output plane $z = 10$ mm, the light is almost completely returned in the initially excited fiber, thus realizing optical self-imaging for the finite array of circular fibers.

*3.5. Adiabatic light transfer*
*3.5.1. The Landau-Zener adiabatic coupler*

As discussed in Sec. 3.3.2, in a synchronous optical coupler with straight axis of a given length the transfer efficiency from one channel to the other one is strongly sensitive to the wavelength $\lambda$. A possible method to realize directional coupling which is rather insensitive to the wavelength, at least within a given range, is to adopt the so-called Landau-Zener adiabatic coupler [150] . In this device, light tunneling between the two waveguides mimics the Landau-Zener process governing the transition dynamics of a 2-level quantum mechanical system, with a time-dependent Hamiltonian varying such that the energy separation of the two states is a linear function of time [133, 327]. The Landau-Zener directional coupler can be realized by two closely-spaced waveguides (waveguide spacing $d$) with a cubically-bent axis with bending profile given by

$$x_0(z) = \frac{A}{L_1^3}(z - L_1)^3 - A \quad (79)$$

($0 < z < L$), where $2A$ ($A \ll L$) is the full lateral shift of the waveguides between input ($z = 0$) and output ($z = L$) planes and $L_1$ is the position at which the axis curvature $\ddot{x}_0$ vanishes. Light transfer between the waveguides is governed by the following coupled-mode equations for the amplitudes $\Psi_1(z')$ and $\Psi_2(z')$ of light waves trapped in the two waveguides [150]

$$i\frac{d}{dz'}\begin{pmatrix} \Psi_1 \\ \Psi_2 \end{pmatrix} = \begin{pmatrix} \eta^2 z' & C \\ C & -\eta^2 z' \end{pmatrix} \begin{pmatrix} \Psi_1 \\ \Psi_2 \end{pmatrix} \quad (80)$$

where $z' = z - L_1$ ($-L_1 < z' < L - L_1$), $C$ is the coupling strength between the two waveguides of the coupler, and the parameter $\eta^2$ governs the adiabaticity of level crossing and reads explicitly

$$\eta^2 = \frac{3dAn_s}{L_1^3} \quad (81)$$

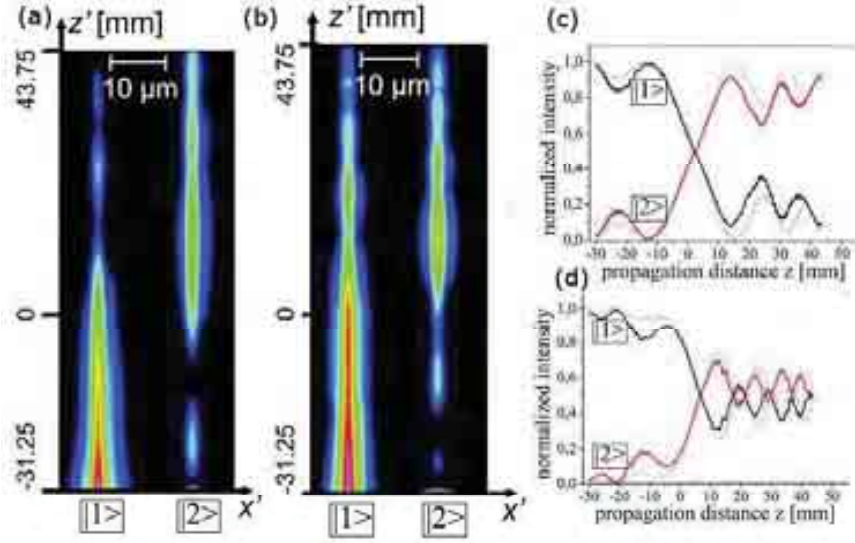

Figure 29: Light transfer in a Landau-Zener optical directional coupler. Measured fluorescence images of light tunneling in cubically-bent waveguide couplers [(a) and (b)], and corresponding behavior of light power evolution in the two waveguides [(c) and (d), theoretical and experimental curves] for different values of the adiabatic parameter. In (a) and (c) $d = 17$ $\mu$m, $A = 300$ $\mu$m and $L_1 = 31.25$ mm, whereas in (b) and (d) $d = 17$ $\mu$m, $A = 500$ $\mu$m and $L_1 = 31.25$ mm, resulting in a faster level crossing and near 50% splitting. After Ref.[42].

In their present form, Eqs.(80) describe Landau-Zener tunneling with linear crossing of energy levels, at a rate $\eta^2$, and with constant coupling $C$ of finite duration $L$. The solution to Eq.(80) can be expressed in terms of parabolic cylinder functions (see Ref.[309] for details). The overall transfer efficiency of the coupler turns out to strongly depend on the value of the adiabaticity parameter $\eta^2$ according to the Landau-Zener formula [327]. In particular, complete and nearly wavelength-insensitive light transfer from one waveguide (at the input plane) to the other waveguide (at the output plane) occurs in the adiabatic ($\eta \to 0$) limit. The experimental demonstration of the Landau-Zener optical coupler has been reported by F. Dreisow and collaborators in Ref.[42]. In the experiment, a set of cubically-curved waveguide couplers were manufactured by fs laser microstructuring of fused silica. The use of fused silica glass with a high content of silanol leads to massive formation of nonbridging oxygen-hole color centers, which emit fluorescence light around 650 nm when excited with a He-Ne laser at 633 nm wavelength. Typical experimental results of light transfer in curved couplers are shown in Figs.29 and compared to the theoretical predictions. The figures clearly indicate how the overall transfer efficiency is controlled by the adiabatic parameter $\eta^2$. In particular, full light transfer is attained as $\eta \to 0$. The dependence of the adiabatic parameter on the wavelength $\lambda$ is smooth as the wavelength is tuned, at around a central values $\lambda_0$, by $\Delta\lambda \sim 0.1 - 0.2\lambda_0$, and therefore the transfer efficiency of the Landau-Zener coupler is much less sensitivity to wavelength changes than an ordinary synchronous coupler.

*3.5.2. Coherent tunneling by adiabatic passage*

Light transport in few coupled waveguide systems can benefit from several control schemes developed in the contexts of atomic and quantum physics which can be transferred to optics [175]. For example, coherent transport in space of a quantum particle among a chain of quantum wells is an interesting example of quantum tunneling control which has been widely studied in the quantum physics context [51, 81, 259, 52, 80, 88]. The transport scheme, which is referred to as "coherent tunneling adiabatic passage" (CTAP), is based on dynamical tuning of the tunneling rates between adjacent quantum units by changing either the distance or the height of the neighboring potential wells following a counterintuitive scheme which is reminiscent of the celebrated stimulated Raman adiabatic passage (STIRAP) technique [10, 310], originally developed for transferring population between two long-lived atomic or molecular energy levels optically connected to a third auxiliary state. The application of CTAP to light transfer management among evanescently-coupled optical waveguides has been theoretically proposed and experimentally demonstrated in a series of recent works [224, 156, 179, 35, 132, 253]. To understand the physical principle underly-ing photonic CTAP, it is worth recalling the problem of coherent population transfer in a three-level atomic system by means of two delayed optical pulses [10]. Let us consider three energy levels $|1\rangle$, $|2\rangle$ and $|3\rangle$ of an atomic or molecular system, where $|1\rangle$ and $|2\rangle$ are two metastable states optically connected to an auxiliary excited state $|3\rangle$. The optical transition $|1\rangle \leftrightarrow |2\rangle$ is assumed to be electric-dipole forbidden. Assume that initially the population occupies level $|1\rangle$, and we like to completely transfer the population to level $|2\rangle$ by use of two laser pulses, the pump (P) and Stokes (S) pulses, which are in resonance with the two electric-dipole allowed transitions $|1\rangle \leftrightarrow |3\rangle$ and $|3\rangle \leftrightarrow |2\rangle$, respectively, as shown in Fig.30. The condition of resonance can be partially relaxed, however it is fundamental that the two-photon resonance condition is met [10]. If $\Omega_P(t)$ and $\Omega_S(t)$ denote the Rabi frequencies of pump and Stokes pulses, which are assumed to be not chirped, in the rotating-wave approximation the evolution equations for the amplitude probabilities $c_1$, $c_2$ and $c_3$ of the three levels read [10, 310]

$$i\frac{dc_1}{dt} = -\frac{\Omega_P(t)}{2}c_3 \tag{82}$$

$$i\frac{dc_2}{dt} = -\frac{\Omega_S(t)}{2}c_3 \tag{83}$$

$$i\frac{dc_3}{dt} = -\frac{\Omega_P(t)}{2}c_1 - \frac{\Omega_S(t)}{2}c_2. \tag{84}$$

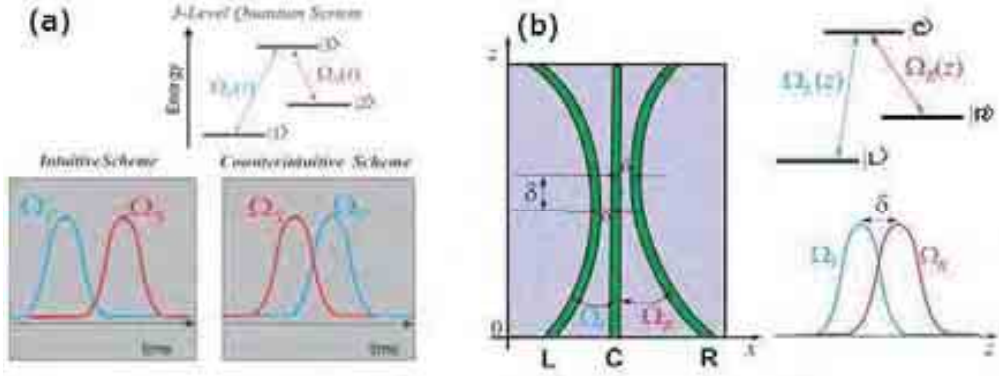

Figure 30: (a) Coherent population transfer in a three-level atomic system via coherent interaction with two resonant optical pulses (the pump and Stokes pulses) with Rabi frequencies $\Omega_P(t)$ and $\Omega_S(t)$. STIRAP corresponds to the counterintuitive pulse sequence order shown in the right panel. (b) Coherent tunneling by adiabatic passage in a triplet optical waveguide structure that mimics atomic STIRAP. $\Omega_L$ and $\Omega_R$ are the coupling rates between adjacent waveguides, which play the same role as the pump and Stokes laser pulses of atomic STIRAP. Time is replaced by the spatial propagation distance $z$. After Ref.[175].

In the above equations, we assumed coherent interaction between the atom and the laser fields, neglecting dephasing effects and decay of level $|3\rangle$. The latter effect may be simply included, if needed, by adding a phenomenological term $-(\gamma/2)c_3$ on the right hand side of Eq.(84), where $\gamma$ is the decay rate of the level $|3\rangle$ to other levels of the atom. Initial conditions, corresponding to a population in level $|1\rangle$, are $c_1(0) = 1$ and $c_2(0) = c_3(0) = 0$. A possible scheme to transfer population from level $|1\rangle$ to level $|2\rangle$ consists of sending first a pump pulse $\Omega_P(t)$ of area $\pi$ (i.e. $\int dt \Omega_P(t) = \pi$), which transfers population from level $|1\rangle$ to the intermediate level $|3\rangle$, and then a Stokes pulse $\Omega_S(t)$ of area $\pi$ which transfers population from the intermediate state $|3\rangle$ to level $|2\rangle$ [left panel in Fig.30(a)]. Such a pulse scheme, which is referred to as the *intuitive* pulse sequence scheme, suffers from several drawbacks, such as the need to carefully control the area of the pulses and the detrimental effect of possible decay of intermediate level $|3\rangle$, which is populated during the transfer process [10, 310]. A powerful technique to overcome such limitations is represented by STIRAP, in which the temporal sequence of pump and Stokes pulses is reversed, as indicated in the right panel of Fig.30(a). In such a pulse sequence scheme, which is referred to as the *counterintuitive* pulse scheme, population transfer is robust against changes in the shape or area of the pulses and even against decay of the intermediate level $|3\rangle$, because such a level is not populated during the transfer process. STIRAP requires a certain temporal overlapping between pump and Stokes pulses and, very important, the evolution of system (82-84) must be adiabatic. Population transfer via STIRAP is based on the existence of an adiabatic dark state (also called trapped state) for Eqs.(82-84), given by

$$c_1 = \frac{\Omega_S}{\sqrt{\Omega_P^2 + \Omega_S^2}}, \quad c_2 = \frac{-\Omega_P}{\sqrt{\Omega_P^2 + \Omega_S^2}}, \quad c_3 = 0. \tag{85}$$

For the counterintuitive pulse scheme, at $t \to -\infty$ one has $\Omega_P(t)/\Omega_S(t) \to 0$, and therefore the dark state [Eq.(85)] corresponds to $c_1 = 1$, $c_2 = c_3 = 0$, i.e. to the population on level $|1\rangle$. At $t \to \infty$, one has conversely $\Omega_S(t)/\Omega_P(t) \to 0$, and therefore the dark state [Eq.(85)] corresponds to $c_1 = c_3 = 0$, $c_2 = -1$, i.e. to the population on level $|2\rangle$. If the conditions for adiabatic evolution are satisfied, during the sequence of Stokes and pump pulses the atomic system evolves remaining in its dark state, and therefore population is fully transferred from level $|1\rangle$ to level $|2\rangle$ without never exciting the intermediate level $|3\rangle$. The condition for adiabatic evolution may be roughly expressed in saying that the temporal overlap interval $\Delta\tau$ of pump and Stokes pulses should be much larger that the maximum value of $\Omega_{eff} = \sqrt{\Omega_P^2 + \Omega_S^2}$ [10]. This simplest STIRAP protocol of atomic physics can be applied to photonic tunneling by adiabatic passage in a triplet waveguide system, which was demonstrated in the experiment by S. Longhi and collaborators in Ref.[179]. The optical system here simply consists of three evanescently-coupled optical waveguides L, C and R in the geometry depicted on the left panel of Fig.30(b). A central straight waveguide C, which plays the role of the intermediate atomic state $|3\rangle$, is side coupled to two circularly and oppositely curved left L and right R waveguides, which are displaced along the longitudinal $z$ axis by a distance $\delta$. Therefore, the minimum distance $\rho$ between waveguides L and C, and between waveguides R and C, are reached at propagation distances $z$ which are displaced by $\delta$, as shown on the right panel of Fig.30(b). Complete light transfer between the outer waveguides L and R can be achieved following the previously discussed atomic STIRAP transfer technique. Indeed, coupled mode equations describing light transfer among the three waveguides are analogous to the STIRAP equations (82-84), where $z$ plays the role of time and the Rabi frequencies of pump and Stokes pulses are played by the coupling amplitudes $\Omega_L(z)$, $\Omega_R(z)$ between adjacent waveguides [see Fig.30(b), right panel]. Note that, as the distances between waveguides L-C and C-R change along the propagation distance $z$, the coupling amplitudes $\Omega_L(z)$ and $\Omega_R(z)$ correspondingly describe two bell-shaped and equal waveforms, shifted each other by $\delta$, which mimic pump and Stokes laser pulses. If waveguide R is excited in its fundamental mode at the input plane $z = 0$, as $\Omega_L$ precedes $\Omega_R$ the equivalent sequence of pump and Stokes pulses follows a counterintuitive scheme, and light is fully transferred to waveguide L with negligible excitation of the central waveguide C, as shown on the left panel of Fig.31. Conversely, if waveguide L is excited in its fundamental mode at the input plane $z = 0$, the intuitive sequence of pump and Stokes pulses is now mimicked, and transfer among the three waveguides follows a rather involved path which is shown in the right panel of Fig.31. The light transfer mechanism shown in the left panel of Fig.31, in which light transfer between two outer waveguides is possible via an intermediate (but never excited) waveguide, is the optical analogue of atomic STIRAP, and it is therefore relatively tolerant against moderate changes in the waveguide geometry.

Extensions of photonic tunneling control by adiabatic passage schemes in coupled optical waveguides have been investigated in several subsequent works. Among others, we mention the optical analogues of multilevel population adiabatic passage [156], straddle STIRAP [35], Raman chirped adiabatic passage [163], population transfer via a continuum [169, 173, 172, 41], nonlinear optical STIRAP [132], and fractional STIRAP [40]. In particular, in a recent experiment by F. Dreisow and collaborators [40] the technique of fractional (or interrupted) STIRAP was applied to the realization of a broadband adiabatic splitter. In the fractional STIRAP, the central waveguide is interrupted at the point where $\Omega_P = \Omega_S$, and no more power transfer between the waveguides is possible after waveguide interruption. According to Eq.(85), in the adiabatic limit 50% beam splitting is achieved after waveguide interruption. The splitting turns out to be broadband because, in spite of the wavelength dependence of the coupling strengths, the splitting ratio is independent of wavelength. In the experiment of Ref.[40], the three-waveguide beam splitter was realized by femtosecond laser inscription of curved micro-optical channels with engineered separation distance, and broadband $\sim 50\%$ beam splitting was demonstrated in the whole wavelength range from 500 nm to 1100 nm. Multicolor fluorescence measurements, showing the light transfer evolution in the interrupted STIRAP waveguide structure, is shown in Fig.32, whereas the measured splitting ratio of the device in the 400-1100 nm spectral range is depicted in Fig.33. The condition for the wavelength independence to be valid is ultimately limited by the adiabaticity criterion of light transfer. While for the blue light the evolution is limited by strongly confined modes and the resulting negligible coupling between the waveguides, the limit in the red spectral region is the growing coupling strength, which arises from the high degree of the waveguides mode overlap and makes the STIRAP process nonadiabatic. Indeed, the fractional STIRAP device benefits from the fact that both limits are well separated, which yields the large spectral operating range.

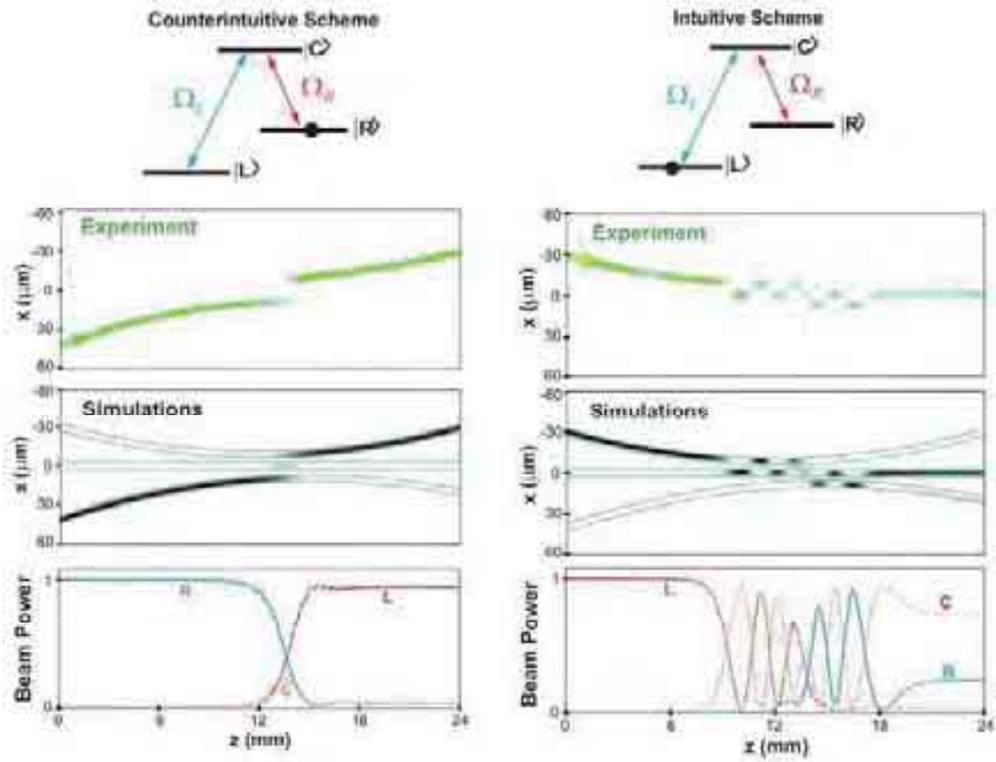

Figure 31: Experimental demonstration of coherent photonic tunneling by adiabatic passage in a triplet waveguide system. Beam propagation simulations and measured fluorescence images are depicted that show light transfer under beam excitation mimicking a counterintuitive (STIRAP) pulse sequence (left panel) and an intuitive pulse sequence (right panel). After Ref.[179].

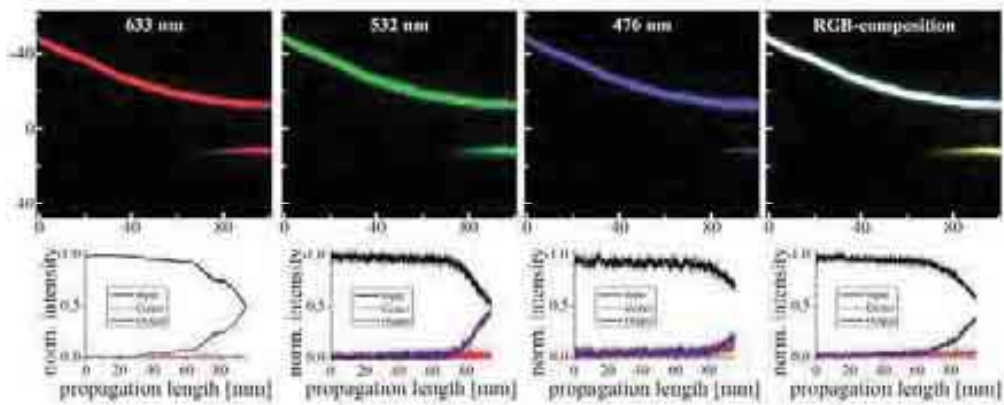

Figure 32: Multicolor fluorescence measurements, showing light transfer in the fractional STIRAP device. Upper row: fluorescence images for three wavelengths and merged RGB-image. Lower row: measured intensities of input (l), center (c), and output (r) waveguides, and the incoherent sum of 633 nm, 532 nm, and 476 nm (fourth column). After Ref.[40].

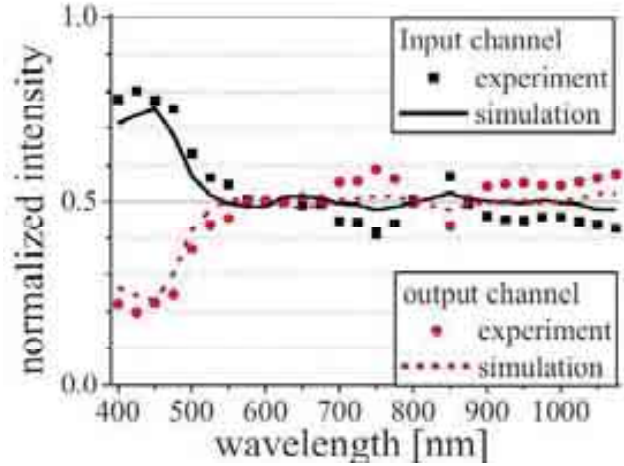

Figure 33: Broad band beam splitter in a three-waveguide structure based on fractional STIRAP. The figure shows the experimentally measured and theoretical predicted splitting ratio as a function of wavelength . After Ref.[40].

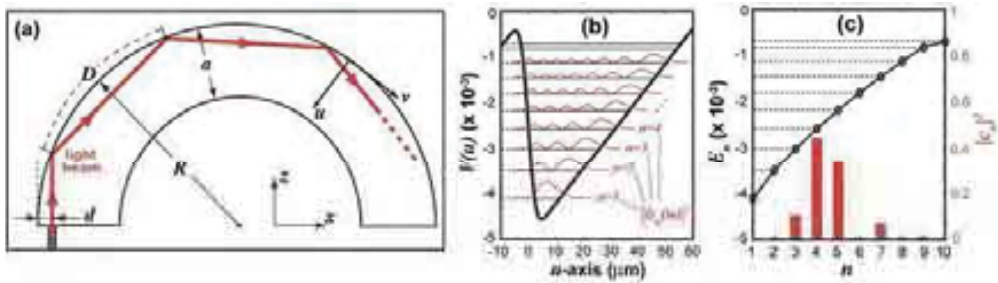

Figure 34: (a) Schematic of the photonic bouncer, based on a curved wide channel waveguide. (b) Effective one-dimensional potential well $V(u)$. Numerically computed energy eigenvalues $E_n$ and corresponding eigenstates $\phi_n(u)$ of the well are also shown. (c) Wave packet spectrum $|c_n|^2$ for an initial Gaussian wave packet. After Ref.[36].

### 3.6. Photon bouncing ball

Light propagation in a circularly-curved wide channel waveguide or, similarly, in a circularly-curved semi-infinite waveguide array show several interesting interference phenomena, like the collapse and revival of a wave packet, providing an optical realization of a quantum bouncer [167, 36]. The motion of a quantum particle falling in a constant gravitational field, eventually reflected by a hard surface [77], is of great relevance in different areas of physics, mainly because of the possibility of observing quantum effects of gravity [220, 221], testing the equivalence principle [308], and observing freely accelerating particles (Airy wave packets [13]). Experimental realizations of a quantum bouncer, i.e. of a quantum particle bouncing on a hard surface under the influence of gravity [77], have been reported using ultracold atoms and neutrons. Owing to the formation of a gravitational quantum well, the energy levels of the bouncing atoms or neutrons are quantized. Since the gravitational well is anharmonic, the quantum motion of an initially localized particle strongly deviates from the classical one at long observation times, and in the full quantum regime wave packet collapses and revivals should be observed as a result of quantum interference [246]. Unfortunately, the energy levels in a gravitational well are very closely spaced, making the classical limit ubiquitous in most practical cases and quantum interference effects hardly observable. Since collapse and revival effects are in their essence a manifestation of the wave nature of matter and because of the similarity between quantum and classical interference [12], they can be observed for optical waves as well. The simplest realization of a photon bouncer, i.e. of an optical analogue of a quantum bouncing ball, was proposed by G. Della Valle and collaborators in an experiment that uses a curved wide channel waveguide [36]. The optical analogue of the quantum bouncer is based on trapping of a light beam near the outer edge of a circularly-curved wide channel waveguide of radius $R$ and channel size $a$, as schematically shown in Fig. 34(a). A light beam injected near the outer edge of the waveguide undergoes a sequence of bounces due to total internal reflection, which mimics the motion of a quantum bouncing ball. The trapping mechanism of light near the waveguide edge relies on the existence of Airy bound states similar to the whispering gallery waves found in spherical or toroidal microresonators [306]. In the curvilinear reference frame $(u, v)$ shown in Fig. 34(a), which is related to the physical $(x, z)$ frame by a conformal mapping, propagation of a monochromatic and scalar light beam at wavelength $\lambda$, localized near the outer boundary $u \sim 0$ of the waveguide at a distance much smaller than $R$, is described by the following effective Schrödinger-type wave equation [36]

$$i\frac{\partial E}{\partial v} = -\frac{\lambda}{4\pi n_s}\frac{\partial^2 E}{\partial u^2} + \frac{2\pi}{\lambda}V(u)E \qquad (86)$$

where $E(u, v)$ is the complex electric field envelope, $n_s$ is the substrate refractive index, $\lambda$ is the photon wavelength, $V(u) = n_s - n(u) + n_s u/R$, and $n(u)$ is the effective refractive index profile of the straight channel waveguide. Near $u = 0$ the potential $V(u)$ behaves like a gravitational well, in which the hard surface is provided by the outer waveguide edge whereas the gravitational potential is played by the fictitious transverse refractive index gradient perceived by light in the curved reference frame. The behavior of $V(u)$ for the experiment of Ref.[36] is shown in Fig. 34(b), together with the numerically-computed (dimensionless) energy eigenvalues $E_n$ and corresponding eigenstates $\phi_n(u)$ of the well. Such eigenstates closely resemble the set of shifted Airy functions found in the quantum bouncer problem [77]. In the experiment, the waveguide was excited by a Gaussian beam with a full width at $1/e^2$ of $\sim 9.5$ $\mu$m, which excites a few eigenstates $\phi_n(u)$ of the optical well with an estimated spectrum $|c_n|^2$ centered at around $n_0 = 4$ as shown in Fig. 34(c), where $c_n = \int du \phi_n(u) E(u, 0)$. For the values of $R$ and $d$ used in the experiment, the beam makes about $N \sim 28$ classical bounces over the full curved waveguide path [see Fig. 34(a)]. To study wave packet collapse and revival phenomena, let us expand the eigenvalues $E_n$ at around $n = n_0$ up to second order; the propagated wave packet $E(u, v)$ at the arc length $v$ is then given by

$$\begin{aligned}E(u, v) &= \exp\left(-i\frac{E_{n_0}v}{\lambda}\right) \sum_{l=0,\pm 1,\ldots} c_{n_0+l}\phi_{n_0+l}(u) \times \\ &\times \exp\left[-2\pi i v(l/T_1 + l^2/T_2)\right]\end{aligned} \qquad (87)$$

where the spatial scales $T_1$ and $T_2$ are defined by

$$T_1 = 2\pi\lambda/(dE_n/dn)_{n_0} \ , \ T_2 = 4\pi\lambda/(d^2E_n/dn^2)_{n_0}. \qquad (88)$$

and $\lambda = \lambda/(2\pi)$ is the reduced wavelength of photons. For the spectrum shown in Fig. 34(c), at $\lambda = 980$ nm one obtains $T_1 \simeq 2.3$ mm and $T_2 \simeq 49$ mm, i.e. the two spatial scales are well separated. The beam evolution over the shortest spatial scale $\sim T_1$ reproduces the classical (ray optics) result. Here, the quadratic term in $l$ entering in the exponent on the right hand side of Eq.(87) can be neglected and the wave packet undergoes a periodic motion with a spatial periodicity $T_1$ very close to the classical bouncing period $D \simeq 2.2$ mm predicted by the ray optics analysis (semiclassical regime). In this regime, $E(u, v) \simeq E_{cl}(u, v)$, where $E_{cl}(u, v)$ is the 'classical' component of the wave packet [246] which is obtained by letting $T_2 \to \infty$ in Eq.(87). The time scale $T_2$ at the next order in the expansion is analogous to the quantum revival time scale of the quantum bouncer and it is responsible for the appearance of collapsed and revival states for propagation distances in the spatial scale $T_2$ [77, 246]. In particular, when $v$ varies near $v \simeq T_2/2$, the quadratic phase term in Eq.(87) is close to $\pi l^2/2$, and one has $|E(u, v)|^2 \simeq |E_{cl}(u, v - T_1/2)|^2$, i.e. the semiclassical wave packet dynamics is retrieved, however the motion is out of phase as compared to the classical prediction by half a period. This corresponds to a fractional revival of the input wave packet. Similarly, at $v \simeq T_2$, one obtains $|E(u, v)|^2 \sim |E_{cl}(u, v)|^2$, i.e. the classical bouncing motion is retrieved (first integer revival). Between subsequent revivals, a collapsed state, corresponding to a kind of 'incoherent' superposition of the various wave packet components and leading to a fully delocalized field over a distance $d$ from the waveguide boundary, is expect on the basis of general considerations [246]. The propagation distances at which the collapsed states occur depend on the precise shape of the input beam, and can be determined by direct numerical analysis of beam propagation. Figures 35(a) and 35(b) show the measured and predicted evolution of the beam center of mass, as reported in the experiment of Ref.[36], together with the paraboliclike trajectory of the beam undergoing successive bounces, expected by a ray optics analysis (the classical limit of the quantum bouncer). Figure 36 also shows detailed measured and predicted beam intensity distributions at a few propagation distances along the curved waveguide, clearly revealing the regimes of classical, collapsed, fractional revival and integer revival states.

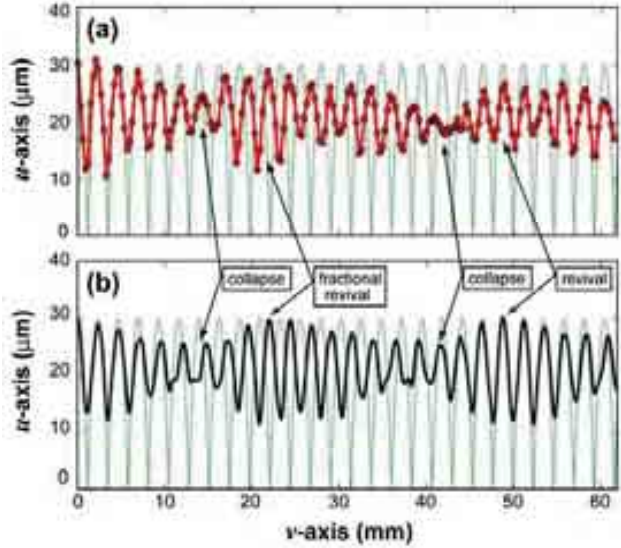

Figure 35: (a) Experimental and (b) numerically computed path followed by the beam center of mass of the photon bouncing ball versus the curvilinear propagation distance $v$. The classical trajectory (dotted line) is also shown for comparison. After Ref.[36].

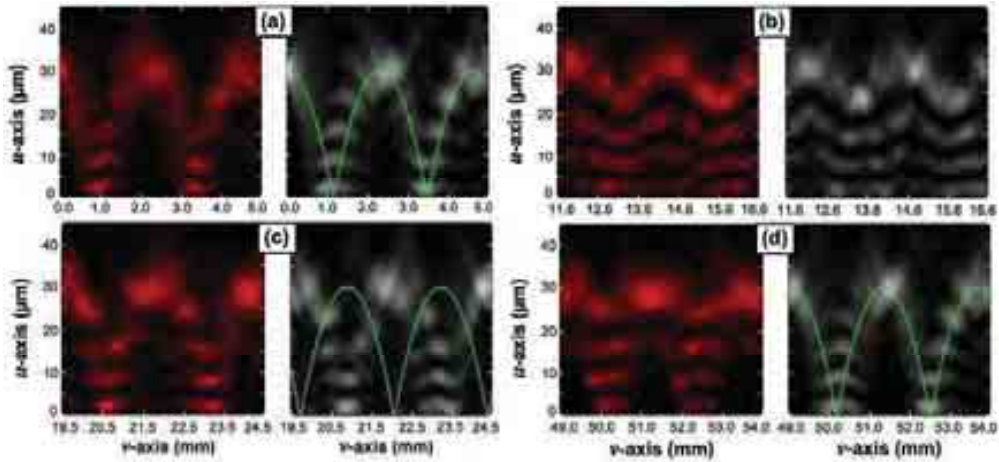

Figure 36: Measured (left panels) and numerically computed (right panels) light intensity map $|E(u,v)|^2$ of beam evolution in the photonic bouncer during (a) the first two bounces, (b) the first collapse, (c) the fractional revival, and (d) the first integer revival. The dashed line in (a), (c) and (d) is the classical trajectory. After Ref.[36].

A modified version of the quantum bouncing ball problem, referred to as the "quantum bouncing ball on a lattice", was theoretically proposed by S. Longhi [167], along with an optical realization based on discrete light diffraction near the edge of circularly curved coupled waveguide arrays. As compared to the continuous quantum bouncing ball problem previously discussed, in the lattice system the effective ball mass varies during the motion in a similar fashion of a driven Bloch particle in a period lattice. Though the classical orbits are correspondingly modified, the genuine quantum phenomena of wave-packet collapses and revivals persist in the lattice model and can be easily visualized in the proposed optical structure. Finally, it should be mentioned that an interesting extension of the photon bouncing ball experiment by Della Valle and collaborators could be the use of a circularly-curved wide waveguide with an undulating outer edge. The undulation of the outer edge corresponds to the introduction of a periodic vibration of the reflecting hard surface for the bouncing photons. Such a structure would provide an optical realization of the so-called Fermi-Pustyl'nikov quantum accelerator, which has served as a testing ground for the study of classical and quantum chaos phenomena (see, for instance, [252] and references therein).

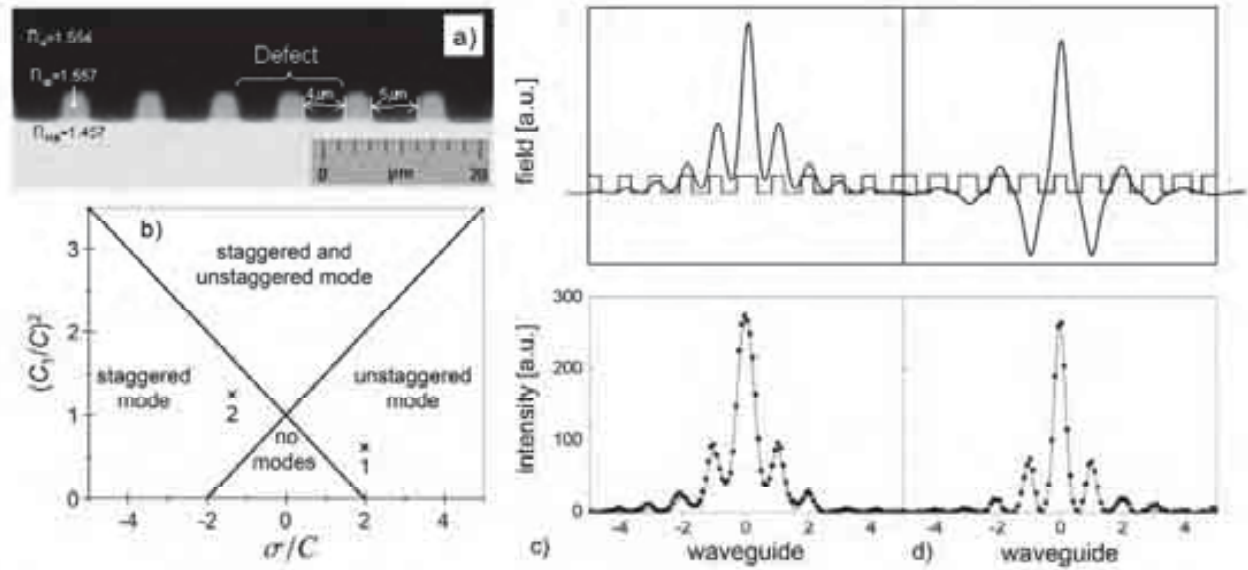

Figure 37: (a) Cross section of a polymer waveguide array with a defect. The defect is introduced by reducing the waveguide spacing compared with the homogeneous array. (b) Regions of existence for symmetric staggered and unstaggered modes in the plane of defect parameters. (c,d) Field (top) and intensity (bottom) profiles of localized modes corresponding to crosses in figure (b) marked 1 and 2, respectively. Solid line: theory, dots: experiment. After Ref. [304].

## 4. Interfaces and lattices with defects

### 4.1. Localization and transmission through defects

The study of propagation, scattering, and localization of linear waves in discrete lattices with inhomogeneities is a subject of continuous interest. We will first summarize the key properties of defect modes in unmodulated waveguide arrays, and then discuss how the modes can be tailored by introducing periodic lattice modulation.

Following Ref. [304], we consider a periodic lattice composed of an array of optical waveguides, with a single symmetric defect created by changing the width of a central waveguide and/or its separation from the neighboring waveguides on its sides, see an example in Fig. 37(a). In the framework of coupled-mode description, the defect located at the lattice site $n = 0$ is characterized by the mode detuning $\sigma$, and modified coupling $C_1$ with neighboring sites:

$$i\frac{d\Psi_0}{dz} + \sigma \Psi_0 + C_1(\Psi_1 + \Psi_{-1}) = 0,$$
$$i\frac{d\Psi_{\pm 1}}{dz} + C_1 \Psi_0 + C\Psi_{\pm 2} = 0, \quad (89)$$
$$i\frac{d\Psi_n}{dz} + C(\Psi_{n-1} + \Psi_{n+1}) = 0, \quad n \neq 0, \pm 1,$$

where $\Psi_n$ are the mode amplitudes at individual waveguides, $z$ is the propagation distance and $C$ is the coupling coefficient away from the defect location.

Whereas in homogeneous lattices beams experience diffraction, a defect can support localized modes in the form $\Psi_n(z) = \psi_n \exp(i\beta z)$, where a mode profile $\psi_n$ has an exponential dependence away from the defect, $\psi_{\pm n} = \psi_{\pm 1} \mu^{|n|-1}$ for $|n| \geq 2$. Solutions for localized modes correspond to the value of the decay rate coefficient $|\mu| < 1$, and depending on its sign the modes can be unstaggered ($\mu > 0$) or staggered ($\mu < 0$). The existence regions of these modes are summarized in Fig. 37(a), and characteristic mode profiles are shown in Figs. 37(b),(c). The features and dynamics of defect modes have been further explored theoretically and experimentally in the nonlinear regime, demonstrating possibilities for all-optical beam switching [128, 231, 213].

The properties of defect modes can be strongly modified in periodically curved lattices. Let us consider a simplest case of a single defect created at a lattice site number $n = 0$, assuming that the coupling between the neighbouring waveguides is not modified as illustrated in Fig. 38(a). Then, the coupled-mode equations can be formulated as [158]:

$$i\frac{d\Psi_n}{\partial z} + C(\Psi_{n-1} + \Psi_{n+1}) + \sigma \Psi_0 \delta_{n,0} = f(z)n\Psi_n, \tag{90}$$

where $\Psi_n$ are the mode amplitudes, $C$ is the coupling coefficient, $\sigma$ is the defect strength, $\delta_{n,0}$ denotes the Kronecker's delta, and $f(z)$ defines the lattice modulation. In the absence of lattice modulation (for $f \equiv 0$), Eq. (90) reduces to Eqs. (89) if we put $C_1 = C$. We see that in this case the defect always supports exactly one mode. We now consider a periodically modulated lattice with $f(z) = A\cos(\omega z)$, where $A$ is proportional to the bending amplitude and $\omega$ is the spatial modulation frequency. It was shown in Ref. [158] that if the modulation period is smaller than the coupling length, which corresponds to the condition $C \ll \omega$, and $\omega \sim A \sim \sigma$, then one can derive approximate equations for the slowly varying amplitudes $u_n$ which describe light dynamics averaged over a modulation period,

$$i\frac{du_n}{\partial z} + C_{n-1}u_{n-1} + C_n u_{n+1} + (\sigma - m\omega)u_0 \delta_{n,0} = 0. \tag{91}$$

Here integer parameter $m$ defines the order of resonance when $(\sigma - m\omega)/\sigma \simeq C/\omega$, and the averaged coupling coefficients are $C_n = C_{eff} = J_0(A/\omega)C$ for $n \neq 0, -1$, $C_{-1} = J_n(A/\omega)C$, and $C_0 = J_{-n}(A/\omega)C$. The lattice modulation not only modifies the effective defect strength, but also the effective coupling coefficients at the defect location are different than in the rest of the lattice, as illustrated in Fig. 38(b). In case of a resonance of even order, Eq. (91) reduces to Eqs. (89) by substituting $u_n \rightarrow \Psi_n$, $J_0(A/\omega)C \rightarrow C$, $C_{-1} \rightarrow C_1$, and $(\sigma - m\omega) \rightarrow \sigma$. For odd-order resonances, such reduction is not possible since $C_{-1} \neq C_0$. The existence regions for the localized modes calculated for the first-order resonance ($m = 1$) are presented in Fig. 38(c), and the characteristic mode profiles are shown in Fig. 38(d). We see that the amplitude and frequency of the lattice modulation can determine the disappearance or the appearance of one or even two distinct localized modes at the impurity site [158], whereas exactly one mode exists in the absence of modulation. The size of the localized modes can be flexibly controlled through the lattice modulation [92, 90]. It is interesting to note that in the dynamic localization regime, when $J_0(A/\omega) = 0$, the central lattice sites may still be effectively coupled with the neighboring sites since $C_{-1}$ and $C_0$ would be non-zero. When additionally the ratio of the defect strength and the lattice modulation frequency is close to an integer number $m$, i.e. the effective defect strength $(\sigma - m\omega)$ is small, then there appear resonant oscillations between the defect site and its two nearest neighbor sites [329, 330]. This is in sharp contrast to a case of unmodulated lattices, where for strong lattice defects ($\sigma \gg C$) the modes becomes trapped at a single defect site.

The lattice modulation also strongly modifies the transmission properties of waves through the defect. In a periodic lattice away from the defect, the plane wave solutions are found as $u_n = u_0 \exp[iQn + 2iC_{eff}\cos(Q)]$, where $Q$ is the wavenumber. A characteristic dependence of the transmission coefficient on the lattice modulation for all possible wavenumbers of incident waves is shown in Fig. 38(e). We see that the transmission can increase dramatically in modulated lattices, and an example of almost complete beam transmission through a defect is presented in Fig. 38(f), whereas the beam would experience strong reflection in the absence of modulation.

Modulated lattices offer many possibilities to control wave transport and localization in presence of multiple defects. It was predicted that the effective coupling between two defects can be varied by orders of magnitude by adjusting the modulation parameters [316]. In case of disordered lattices with random distribution of defects, the modification of the effective coupling coefficient can be used to control Anderson localization [91, 332].

*4.2. Interfaces and surface modes*

Wave propagation can be strongly modified in the vicinity of lattice boundaries and interfaces between different structures, where the ideal periodicity is broken. Under certain conditions, the interfaces can support a special type of localized modes known as surface waves [30]. In particular, such waves can exist at the interface separating periodic and homogeneous linear dielectric optical media [323, 322]. Optical surface waves are particularly attractive for optical sensing, measurement and characterization applications. We will first review the optical beam dynamics at the edge of unmodulated lattices, and then discuss a range of effects which can arise due to periodic modulation.

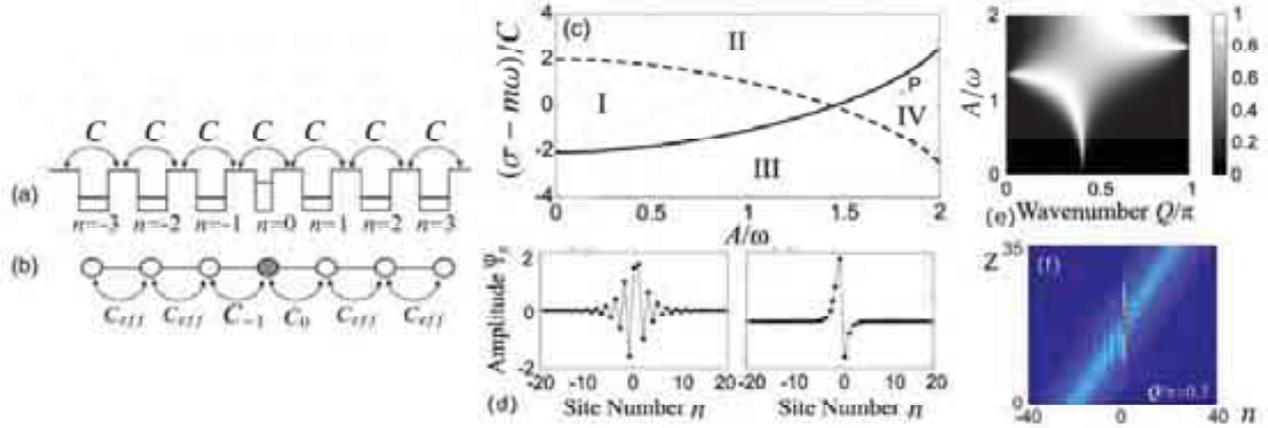

Figure 38: Localization and transmission properties of a defect in a modulated lattice. (a) Schematic of a lattice with a defect at site $l = 0$. (b) Effective averaged coupling between the lattice sites due to lattice modulation. (c) Existence domain of localized defect modes around the first resonance ($n = 1$). Region I, no localized defect states; region II and III, one localized defect mode; region III, two localized modes. (d) Amplitude profiles of localized modes, parameters correspond to marked point P in plot (c). (e) Spectral transmission $T(Q)$ of propagating plane waves through the defect versus the normalized lattice bending amplitude for $n = 1$, $\sigma/C = 8$, $\omega/C = 7.5$. (f) Transmission of a Gaussian wave packet through the defect, parameters correspond to plot (e) with $A/\omega = 0.9$. After Ref. [158].

#### 4.2.1. Surface waves in straight waveguide arrays with edge defects

In unmodulated optical lattices composed of identical straight optical waveguides, surface waves can exist only when the effective refractive index of the boundary waveguide or the mode coupling to the surface waveguide are modified, as illustrated in Fig. 39(a). The corresponding coupled-mode equations can be formulated as [158]:

$$i\frac{d\Psi_1}{dz} + C_0\Psi_2 + \sigma\Psi_1 = 0, \quad i\frac{d\Psi_2}{dz} + C\Psi_3 + C_0\Psi_1 = 0, \quad i\frac{d\Psi_n}{dz} + C(\Psi_{n-1} + \Psi_{n+1}) = 0 \, (n \geq 2) \quad (92)$$

where $\Psi_n$ are the mode amplitudes, $C$ is the coupling constant between identical neighboring waveguides, $C_0$ is the coupling constant between the boundary waveguide ($n = 1$) and its adjacent waveguide ($n = 2$), and $\sigma$ is the normalized difference between the propagation constants of the boundary waveguide and the other waveguides of the array. Solutions in the form of localized surface modes have the form:

$$\Psi_1 = \psi_1 e^{i\beta z}, \quad \Psi_n = \psi_2 \mu^{-(n-2)} e^{i\beta z} \, (n \geq 2) \quad (93)$$

where $\psi_1$ is an arbitrary amplitude, $\psi_2 = \psi_1(\beta - \sigma)/C_0$, $\beta = C(\mu + 1/\mu)$, and $\mu$ can take one of two possible values:

$$\mu_{1,2} = \frac{\sigma}{2C} \pm \sqrt{\left(\frac{\sigma}{2C}\right)^2 + \left(\frac{C_0}{C}\right)^2 - 1}. \quad (94)$$

The mode profiles are localized if $|\mu| < 1$, and according to this condition there can be two, one, or zero surface modes depending on the structural parameters as summarized in Fig. 39(b). In the absence of surface defects (for $C_0 = C$ and $\sigma = 0$), surface modes do not exist and the beam experiences strong reflection from the boundary [191], see Fig. 39(c). In the absence of surface modes, the light intensity at the edge waveguide always goes to zero at long propagation distances, although the decay dynamics can be non-exponential [155, 158, 15]. We note that the surface defects supporting localized modes can be induced by the optical beam itself in nonlinear media, leading to a formation of discrete surface solitons [191, 211, 280, 116, 263, 265, 249, 278, 137], see an example in Fig. 39(d). These waves can be considered as an optical analogue of electronic Tamm surface states [298], underlying the universality of surface localization mechanisms.

#### 4.2.2. Defect-free surface waves in curved waveguide arrays

In modulated lattices, the effect of boundaries was first revealed in numerical simulations which identified a presence of eigenmodes associated with the boundaries, which propagation constants are shifted outside the wave spectrum thus preventing complete band collapse in the dynamic localization regime [89]. It was further predicted that if the surface waveguide is modified, then the critical defect strength required to support a Tamm-like mode can be reduced when the effective coupling between the waveguides is lowered due to the lattice modulation [92]. Most remarkably, it was also found that, in a sharp contrast with arrays of straight waveguides, the surface localization is possible in the absence of edge defects when all periodically curved waveguides are exactly identical [74]. The corresponding coupled-mode equations can be represented in the following normalized form [74],

$$i\frac{d\Psi_n}{dz} + C\exp[-i\kappa_0(z)]\Psi_{n+1} + C\exp[i\kappa_0(z)]\Psi_{n-1} = 0, \tag{95}$$

where $\Psi_n(z)$ is the field amplitude in the n-th waveguide, $n = 1, \ldots,$ and $\Psi_{n\leq 0} \equiv 0$ due to the structure termination (i.e. the waveguide with $n = 1$ corresponds to the surface of the semi-infinite lattice). Transverse shift $x_0(z) = x_0(z+L)$ defines the periodic longitudinal lattice modulation. Coefficient C defines the coupling strength between the neighboring waveguides, it characterizes diffraction in a straight waveguide array with $x_0 \equiv 0$ [266]. Expression (95) shows that the effect of periodic lattice modulation appears through the modifications of phases of the coupling coefficients along the propagation direction $z$. Following Ref. [74], we consider the waveguide arrays with symmetric bending profiles, requiring that $x_0(z) = f(z - \bar{z})$ for a given coordinate shift $\bar{z}$, where function f(z) is symmetric, $f(z) \equiv f(-z)$. Note that asymmetry, such as introduced by superimposing modulations with different periodicity [103], may introduce other effects due to the modification of refraction including beam dragging and steering which will be discussed in Sec. 5.6 below.

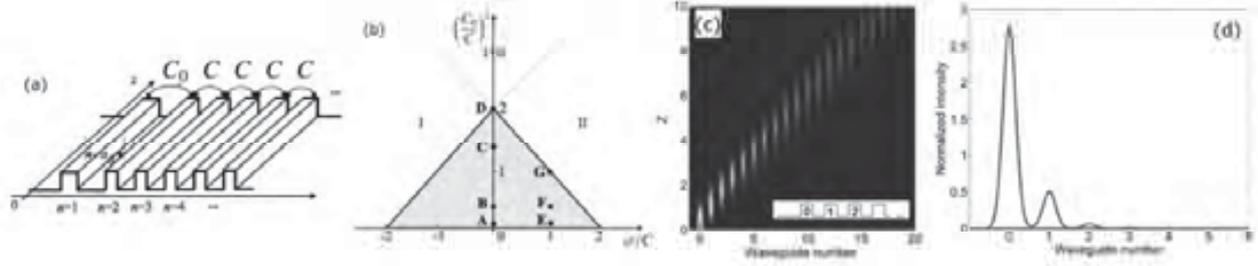

Figure 39: Beam dynamics at the boundary of straight waveguide arrays. (a) Schematic of a semi-infinite array of tunneling-coupled optical waveguides with a defect boundary waveguide ($n = 1$). (b) Existence domain of bound modes of the semi-infinite array in the plane. Regions I and II support one bound mode corresponding to either one of the two values $\mu_{1,2}$ given by Eq. (94); region I+II supports both bound modes. In the triangular shaded region no bound modes are supported by the array. (c) Diffraction pattern in a semi-infinite array when only the first waveguide is excited. The inset depicts a semi-infinite waveguide array. (d) Characteristic intensity profile of a localized surface mode (discrete surface soliton), supported by a self-induced nonlinear defect. After Refs. [159] (a,b) and [191] (c,d).

If the bending period L is small, such that the parameter $\kappa = 2\pi/L$ is large, $\kappa \gg 1$, then it is possible to derive equations for averaged values of the mode amplitudes [74], $u_n = <\Psi_n>$,

$$i\frac{du_n}{dz} + C_0 u_{n+1} + \bar{C}_0 u_{n-1} + \delta_{1,n}(\Delta_1 u_1 + \Delta_2 u_2) + \delta_{2,n}\bar{\Delta}_2 u_1 = 0. \tag{96}$$

Here $\delta$ is the Kronecker delta,

$$\Delta_1 = -\kappa^{-1}\sum_{m\neq 0}|C_m|^2 m^{-1}, \quad \Delta_2 = \kappa^{-2}\sum_{m\neq 0}\sum_{j\neq 0,-m} C_j C_m C_{j+m} j^{-1} m^{-1}, \tag{97}$$

the bar stands for the complex conjugation, $u_{n\leq 0} \equiv 0$, $C_m$ are the coefficients of Fourier expansion, $C\exp[-i\kappa_0(z)] = \sum_m C_m \exp(im\kappa z)$, and it is assumed that the modulation parameters are tuned close to the dynamic localization condition such that $C_0$ is a small value, $|C_0| \sim O(\kappa^{-1})$. Quite interestingly, the averaged Eqs. (96) have the same form as Eqs. (92) if we make a substitution $C_0 \to C$, $C_0 + \Delta_2 \to C_0$, $\Delta_1 \to \sigma$, provided all coefficients being substituted are real. This demonstrates that the effect of periodic modulation is to introduce the "virtual" defects (which values are defined by $\Delta_1$ and $\Delta_2$) at the lattice boundary. Accordingly, Eqs. (96) can admit solutions for localized surface modes in the form $u_n(z) = u_n(0)\exp(ikz/L)$, where $k$ is the propagation constant. Specifically, there exists one surface state

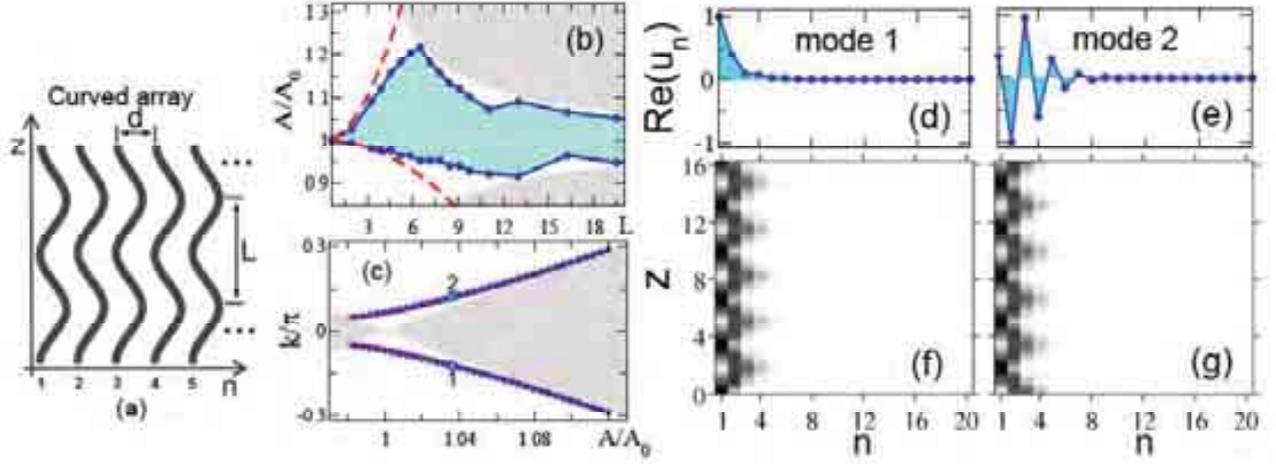

Figure 40: Properties of defect-free surface modes in periodically curved waveguide arrays. (a) Sketch of a periodically curved waveguide array. (b) Solid lines bound domain (hatched) of existence of defect-free surface modes in the sinusoidal modulated lattice calculated numerically. Dashed lines show modes cut-offs obtained from the asymptotic expansion. Solid shading marks the region where localized modes cannot exist. (c) Defect-free surface modes in sinusoidal modulated lattice with modulation period L = 3.25. Circles and solid lines show modes Bloch wave numbers k calculated numerically and using asymptotic expansion, respectively. Shading marks transmission band of the lattice. (d-e) Numerically calculated modes profiles at the input, and (f-g) their propagation dynamics are shown for the two complementary modes marked 1 and 2 in (c), respectively. After Ref. [74].

if
$$\alpha_2 - \alpha_1 \leq 2 \leq \alpha_1 + \alpha_2, \quad (98)$$

and two surface states emerge if
$$\alpha_2 - \alpha_1 > 2, \quad (99)$$

where $\alpha_1 = |\Delta_1/C_0|$ and $\alpha_2 = |1 + \Delta_2/C_0|^2$. If the modulation is symmetric, such that $x_0(z + L/2) = -x_0(z)$, then $|C_m| = |C_{-m}|$ and accordingly $\Delta_1 = 0$, meaning that the modes should always appear in pairs. Moreover, this conclusion is valid even beyond the applicability of the asymptotic expansion, since there is an exact symmetry of the model Eq. (95) in case of symmetric modulations: for each solution $\Psi_n(z)$,

$$\widetilde{\Psi}_n(z) = (-1)^n \Psi_n\left(z + \frac{L}{2}\right) \quad (100)$$

is also a solution. Therefore, in symmetric structures surface modes always appear in pairs with the Bloch wavenumbers of the opposite sign. This property is fundamentally different from the existence features of Shockley surface states, which cannot appear on both sides of a single band due to the different symmetries of modes in odd and even gaps [326].

As an example, we consider a harmonic modulation function in the form
$$x_0(z) = A \cos\left(\frac{2\pi z}{L}\right). \quad (101)$$

In this case, the Fourier coefficients can be calculated analytically, $C_m = C J_m[\xi A/A_0]$, where $J_m$ is the Bessel function of the first kind of order m. The modulation amplitude $A_0$ corresponds to the dynamic localization condition, $A_0 = \frac{\xi}{2\pi}L$, where $\xi \simeq 2.405$ is the first root of the Bessel function $J_0$. Since the sinusoidal modulation is symmetric, $\Delta_1 = 0$, and the virtual surface defect is defined by a single coefficient $\Delta_2 = -\frac{1}{2\pi^2}C^3 L^2 J_1^2(\xi A/A_0) J_2(\xi A/A_0)$. Then for each modulation amplitude A such that $A_{crit}^- < A < A_{crit}^+$, this virtual defect supports (at least) a pair of surface modes, where the left and right mode cut-offs can be estimated according to Eq. (99),

$$\frac{A_{crit}^2}{A_0} = 1 \pm \frac{2C^2 J_1[\xi] J_2[\xi]}{\xi \kappa^2 (\sqrt{2} \mp 1)}. \quad (102)$$

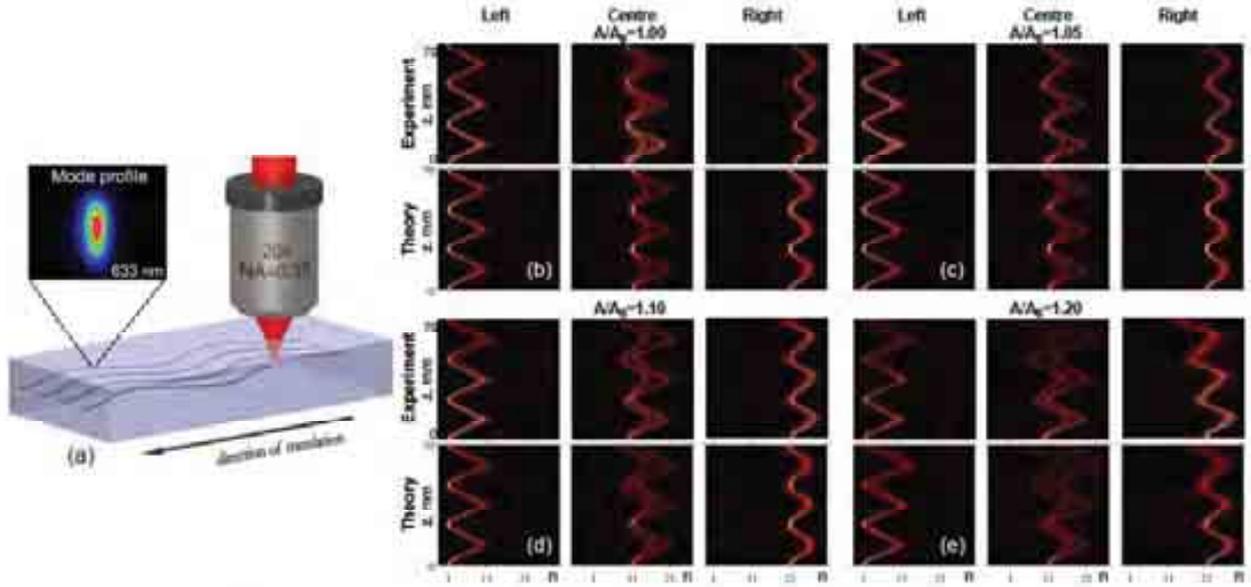

Figure 41: (a) Schematic of the femtosecond laser direct-writing setup. Insert shows waveguide mode profile measured at 633 nm wavelength. (b-e) Light propagation in the four curved arrays with different bending amplitude A. In each of the four blocks fluorescent images are shown on the top, and corresponding numerical simulations are shown at the bottom. Light is launched into the left edge waveguide of the array (left), centre waveguide (centre), and right edge waveguide (right). After Ref. [290].

For large modulation periods the averaged Eqs. (96) are no longer valid, however numerical simulations confirm the existence of defect-free surface modes for a broad range of modulation parameters. These results are summarized in Fig. 40(a), where hatched is the domain of the existence of the defect-free surface modes on the (L, A) parameter plane for C = 1. Note that the results can be mapped to the other coupling values using a simple transformation $\Psi_n(z, C, L, A) = \Psi_n(Cz, 1, CL, CA)$. For small modulation periods, the analytical expressions of Eq. (102) provide a good estimate for the surface modes cut-offs (dashed lines). For the large modulation periods the domain of the existence of the defect-free surface modes is limited by the region where lattice transmission band extends to the whole Brillouin zone ($-\pi \leq k \leq -\pi$), and therefore localized states cannot exist. This region is defined as $L \geq \pi(2|C_0|) = \pi/(CJ_0[\varepsilon A/A_0])$, and it is shown with solid shading in Fig. 40(a). The dependence of the mode wavenumber on modulation amplitude for a fixed modulation period (L = 3.25) is presented in Fig. 40(b), which demonstrates the existence of a pair of symmetric surface modes outside the lattice transmission band, and the wave numbers of surface modes calculated using asymptotic expansion are in excellent agreement with those calculated numerically. At the cross-section $z = 0$, one mode has unstaggered input profile [Fig. 40(c)], while the other one exhibits staggered structure [Fig. 40(d)]. The number of defect-free surface modes can be more than two for larger modulation periods.

The defect-free surface modes have been observed experimentally in periodically-curved waveguide arrays fabricated with femtosecond laser direct-writing technique in fused silica samples [290], see Fig. 41(a) for a schematic illustration of the laser-writing method. The samples were 70 mm long and contained 21 waveguide each with a bending profile according to Eq. (101). The bending period was L = 23.33 mm, such that each curved sample contained three full bending periods. For the the spacing between the centres of the adjacent waveguides $d = 14$ $\mu$m, the laser wavelength $\lambda = 633$ nm, the refractive index of the silica glass $n_0 = 1.45$, the bending amplitude required for the dynamic localization is $A_0 = 44.32$ $\mu$m. In order to study the generation of the defect-free waves, curved waveguide arrays with different bending amplitudes were fabricated: $A/A_0 = 1.00$, $A/A_0 = 1.05$, $A/A_0 = 1.10$, and $A/A_0 = 1.20$. First, we consider the light propagation in the curved sample with $A/A_0 = 1.00$, which is tuned to the exact dynamic localization. When the beam is launched into the left edge waveguide of the array, we observe the formation of a localized surface mode [Fig. 41(b), top left image] in the excellent agreement with the corresponding numerical simulation [Fig. 41(b), bottom left image]. When the beam is launched into the array centre, far away from the boundaries, the dynamic localization is observed [Fig. 41(b), centre], similar to the previous experiments [183].

The experimental measurements also demonstrate that surface modes appear in pairs in agreement with theoretical predictions. When the laser beam is launched into the left edge waveguide of the array, the one of the two surface modes is generated which has a higher excitation coefficient at the input $z = 0$ [i.e. the generated mode has most of its power concentrated in the first waveguide ($n = 1$) at the input], which is shown in Fig. 41(b), left. According to the symmetry relation formulated in Eq. (100), the second mode would have most of its power concentrated in the first waveguide at the distance $z = L/2$. However exciting the left edge waveguide at the distance $z = L/2$ is equivalent to exciting the right edge waveguide ($n = 21$) at the distance $z = 0$ because of the symmetry of our structure [see the sketch in Fig. 40(a)]. Accordingly, the second mode was excited by launching the laser beam into the right edge waveguide, as shown in Fig. 41(b), right.

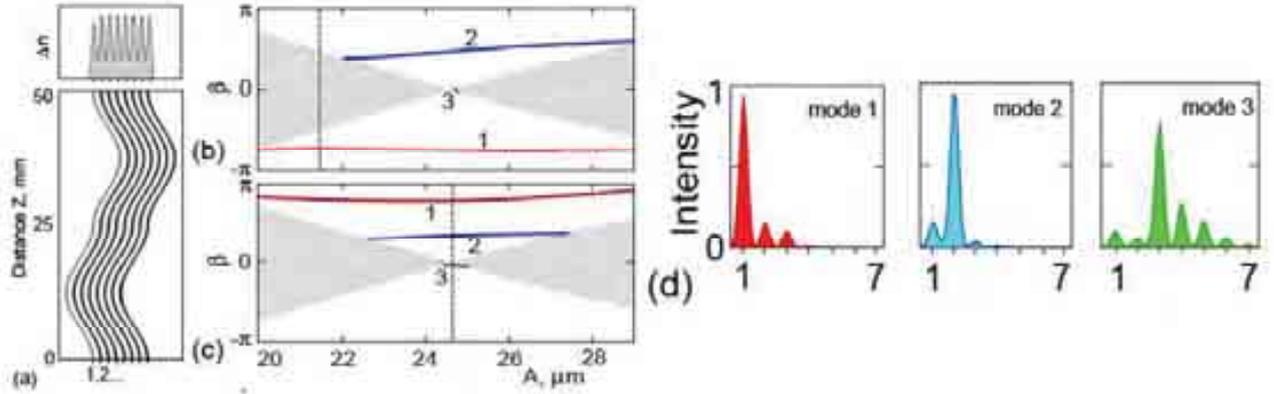

Figure 42: Linear surface modes in curved waveguide array with a surface defect. (a) Schematic of the array (bottom) and index change (top). (b,c) Propagation constants of linear surface modes vs. waveguide bending amplitude $A$ for surface defects (b) $\sigma = -0.52$ and (c) $\sigma = -0.56$. Grey shading marks the array transmission band and the vertical dashed line corresponds to (b) $A = 21.5\,\mu m$ and (c) $A = 24.5\,\mu m$. (d) Intensity profiles at $z = 0$, $L$ of three different surface modes for $A = 24.5\,\mu m$, corresponding to mode dispersion lines marked 1–3 in (c). After Ref. [240]

The surface waves were also observed in a curved array which bending amplitude is slightly detuned from the exact dynamic localization value, $A/A_0 = 1.05$. When the beam is launched into the left and right array edges, the generation of two complementary surface modes is registered [see Fig. 41(c), left; Fig. 41(b), right]. In this case the generated surface waves are slightly less localized compared to the surface waves in Fig. 41(b), left, and Fig. 41(b), right, due to the increased initial radiation caused by the mismatch between the mode input profile and the single-site excitation which we use at the input. In contrast, when we the beam is launched into the array centre, significant beam diffraction is observed [Fig. 41(c), centre], as the dynamic localization condition is not fulfilled for this bending amplitude.

For the third sample with the bending amplitude $A/A_0 = 1.10$, excitation of the edge waveguides still yields the formation of defect-free surface waves [see Fig. 41(d), left; Fig. 41(d), right]. In this case, initial radiation is observed, as the generated surface modes are quite close to the cut-off. When array centre is excited, strong diffraction is observed [Fig. 41(d), centre].

The bending amplitude $A/A_0 = 1.20$ of the fourth sample is beyond the surface modes cut-off, and strong beam diffraction is observed in all the cases, whether the beam is launched into the array centre [Fig. 41(e), centre] or into the edge waveguides [Figs. 41(e), left, and 41(e), right], in good agreement with the theoretical predictions.

### 4.2.3. Surface waves in curved waveguide arrays with edge defects

We now consider a more general situation, when the edge waveguide is modified in a curved waveguide arrays, such as illustrated in Fig. 42(a). In this case, there can appear an interplay between different mechanisms of surface localization due to (i) waveguide bending and (ii) fabricated surface defect [240, 239]. The coupled-mode equations describing the beam dynamics in such structures can be written as

$$i\frac{d\Psi_n}{dz} + C\exp[-ix_0(z)]\Psi_{n+1} + C\exp[ix_0(z)]\Psi_{n-1} + \delta_{n,1}\sigma\Psi_n = 0, \qquad (103)$$

where $\Psi_n(z)$ are the normalized mode amplitudes, $n = 1, 2\ldots$ is the waveguide number, and $\Psi_{n\leq 0} \equiv 0$ due to the structure termination, i.e. the waveguide with $n = 1$ corresponds to the surface of the semi-infinite array. The value of $\sigma$ defines the detuning of the surface waveguide. Characteristic dependencies of the mode propagation constant $\beta$ (defined as the phase accumulated over one modulation period) on the bending amplitude for different values of the surface defect are presented in Figs. 42(b,c). The profiles of the three fundamental modes (marked 1–3) are similar for $\sigma = -0.56$ [shown in Fig. 42(d)] and $\sigma = -0.52$ due to their common physical origin. Mode 1 is supported by the surface defect, similar to Tamm-like modes discussed in Sec. 4.2.1 above. It exists for a wide range of bending amplitudes and has a profile with intensity maximum at the first waveguide [Fig. 42(d), left]. Modes 2 and 3 exist due to waveguide bending, similar to the defect-free surface states discussed in Sec. 4.2.2. The input and output mode profiles (at $z = 0, L$) have intensity maxima at the second and third waveguides, respectively [Fig. 42(d), middle and right]. Experimental measurements performed in $LiNbO_3$ waveguide arrays confirmed the existence of these modes, and additionally demonstrated all-optical beam switching due to mode coupling through nonlinear self-action, which is discussed in Sec. 5.4 below.

*4.2.4. Light tunneling control between a curved boundary waveguide and straight array*

We now discuss light dynamics in a photonic structure composed of a periodically curved boundary waveguide and a straight waveguide array, as schematically illustrated in Fig. 43(a). In addition to the periodic bending of the boundary waveguide, it can also have a different refractive index profile compared to the straight waveguides (similar to the configuration discussed in Sec. 4.2.3 above), see an example in Fig. 43(b). The coupled-mode equations for the evolution of the mode amplitudes ($\Psi_n$) can be written as [161]:

$$i\frac{d\Psi_0}{dz} + \sigma\Psi_0 + C_0(z)\Psi_1 = 0, \quad i\frac{d\Psi_1}{dz} + C_0(z)\Psi_0 + C\Psi_2 = 0, \quad i\frac{d\Psi_n}{dz} + C(\Psi_{n-1} + \Psi_{n+1}) = 0 \ (n \geq 2), \quad (104)$$

where $n$ is the waveguide number ($n = 0$ for the boundary waveguide), $\sigma$ is the detuning of the propagation constant of the boundary waveguide, $C$ is the constant coupling coefficient between the straight waveguides, $C_0(z) = C_0(z + \Lambda)$ is the coupling coefficient between the curved boundary waveguide and the first straight waveguide, and $\Lambda$ is the modulation period of the boundary waveguide. Note that, in contrast to the models for curved waveguides discussed above, it is assumed in Eqs. (104) that the coupling coefficient $C_0(z)$ is real-valued. Since the imaginary part of the coupling coefficient is related to the waveguide bending [see, e.g., Eq. (103) above], the imaginary part of $C_0(z)$ can be made arbitrarily small by considering waveguide bending with a long modulation period.

Before performing a detailed analysis of Eqs. (104), it is useful to first consider the spectral properties in the case of separated boundary waveguide and the straight waveguide array, when $C_0 = 0$. For an isolated boundary waveguide, the mode is $\Psi_0(z) = \Psi_0(z = 0) \exp(i\beta z)$, where the propagation constant value is $\beta = \sigma$. For a semi-infinite straight waveguide array, extended modes correspond to a range of propagation constants $-2C \leq \beta \leq 2C$. If the value of $C_0(z)$ is non-vanishing, then there will appear a coupling between a discrete level of the boundary waveguide and a continuous spectral band of the semi-infinite array, as illustrated in Fig. 43(c).

The effect of coupling between the discrete and continuous spectral regions can be analyzed analytically in the limit of small coupling, $C_0(z) \ll C$. Under this condition, the decay rate at large propagation distances (when $z$ is much larger than the modulation period $\Lambda$) of the light intensity at the boundary waveguide can be found as [161],

$$|\Psi_0(z)|^2 \simeq e^{-R(z)Q(z)}, \ R(z) \simeq \frac{2\pi \sum_m |\epsilon_m|^2 G(\sigma + m\Omega)}{\sum_m |\epsilon_m|^2}, \ Q(z) = \int_0^z dz' \frac{C_0(z')}{C_{0m}}, \ G(\omega) = \frac{C_{0\max}^2}{\pi C}\left[1 - \left(\frac{\omega}{2C}\right)^2\right]^{1/2}, \quad (105)$$

where the value of $G(\omega)$ is taken to be zero for $|\omega| > 2C$, $C_{0\max} = \max_z C_0(z)$, $\Omega = 2\pi/\Lambda$, and $\epsilon_n$ are the Fourier coefficients of $C_0(z)/C_{0\max}$ decomposition, i.e. $C_0(z)/C_{0\max} = \sum_m \epsilon_m \exp(im\Omega z)$. The meaning of $R(z)$ is the effective decay rate, and $Q(z)$ is the effective (interaction) length. The interaction length $Q(z)$ is introduced to facilitate a comparison between the decay rates for curved [labeled $W_C$ in Fig. 43(a)] and straight [labeled $W_S$ in Fig. 43(a)] waveguides. Note that for a straight boundary waveguide $Q(z) = z$, i.e. the effective interaction length is equal to physical length, whereas $Q(z) < z$ for curved waveguides.

Using Eq. (105), it is possible to inhibit or accelerate the decay [161] depending on the value of $\sigma$. Such decay control is analogous to the scheme of quantum mechanical decay control based on modulation of the coupling to the continuum (a generalization of the concept of frequent observation of a system) suggested by Kofman and Kurizki [124, 7]. The predicted decay regimes were observed experimentally in femtosecond laser written waveguide arrays [44]. In experiments, samples with different strengths of the effective refractive index for a boundary waveguide were fabricated by varying the writing velocity, corresponding to different values of the detuning parameter $\sigma$.

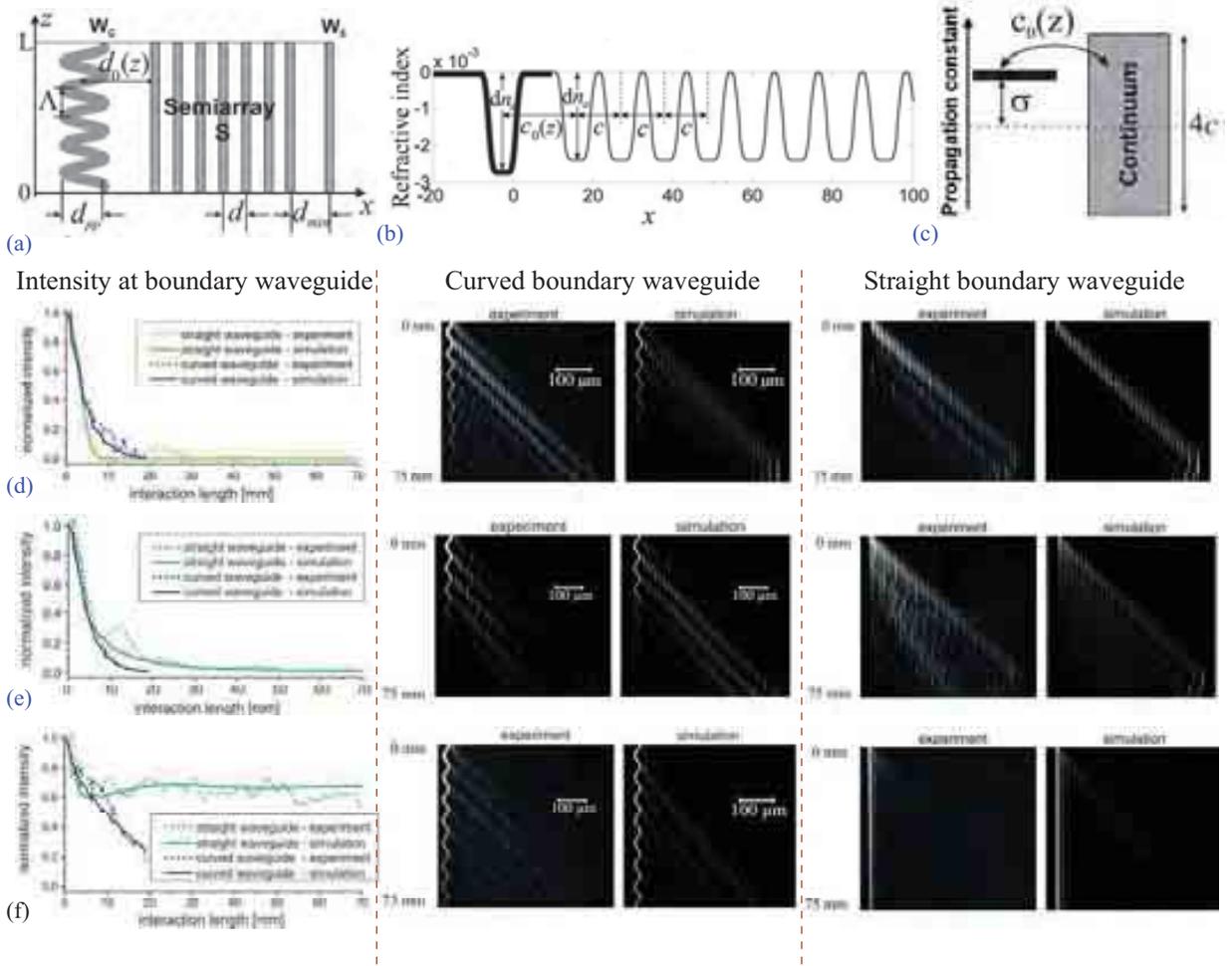

Figure 43: (a) Schematic of the photonic lattice for optical tunneling control between a curved ($W_C$) or straight ($W_S$) boundary waveguide and a straight waveguide array; (b) Characteristic refractive index modulation at a particular distance $z$: array –thin solid curve, the boundary waveguide – thick solid curve. (c) The diagram of propagation constants in an isolated boundary waveguide and the photonic band in a waveguide array; (d-f) Left column: Measured decay of light power versus interaction length $Q$ (dotted curves) and the corresponding curves predicted by numerical simulations (solid curves) for both straight and periodically curved boundary waveguides. Right and middle columns: Measured fluorescence images representing light intensity profile inside the arrays, and the corresponding numerical simulations for the curved and straight boundary waveguides as indicated by labels. The boundary waveguide detuning is (d) $\sigma = 0$, (e) $\sigma = 1.2C$, and (f) $\sigma = 2.2C$. After Refs. [44] (a),(c-f) and [161] (b).

For $\sigma = 0$, the modulation of the discrete-to-continuum coupling $C_0(z)$ leads to decay deceleration, as shown in Fig. 43(d), since the decay rate for the nonmodulated waveguide $W_S$ is approximately given by result $R = 2\pi G(\sigma)$, which is largest at $\sigma = 0$. For the modulated waveguide $W_C$, the decay rate is diminished owing to averaging over the spectral components of the modulation profile according to Eq. (105). Figure 43(d), left, shows the measured light power trapped in either straigh or curved boundary waveguides versus the interaction length $Q(z)$ and corresponding theoretical predictions based on the coupled-mode Eqs. (104) (solid lines). The measured fluorescence patterns, showing light leakage to the semiarray $S$, together with the corresponding theoretical predictions, are also shown in Fig. 43(d) as indicated by labels.

As the detuning $\sigma$ is increased, but remains bounded as $\sigma < 2C$ (note that under such conditions straight boundary waveguide cannot support localized surface modes as discussed in Sec. 4.2.1 above), then the decay rate for the straight waveguide slows down considerably (but remains finite), since the discrete-to-continuum coupling spectrum $G(\omega)$ decreases with $\sigma^2$. On the contrary, the modulation of the discrete-to-continuum coupling for a curved waveguide leads to acceleration of the decay, which is the optical analogue of quantum anti-Zeno effect, see Fig. 43(e).

A further shift of detuning above the energy band ($\sigma > 2C$) leads to the appearance of a localized surface mode at the straight boundary waveguide. If $\sigma$ is not shifted considerably above the energy band edge, only a part of the excited energy can be trapped at the surface, and one observes a fractional (or limited) decay. Most importantly, modulation of the discrete-to-continuum interaction yields a suppression of the bound state, and the decay fades from fractional to complete, as illustrated in Fig. 43(f).

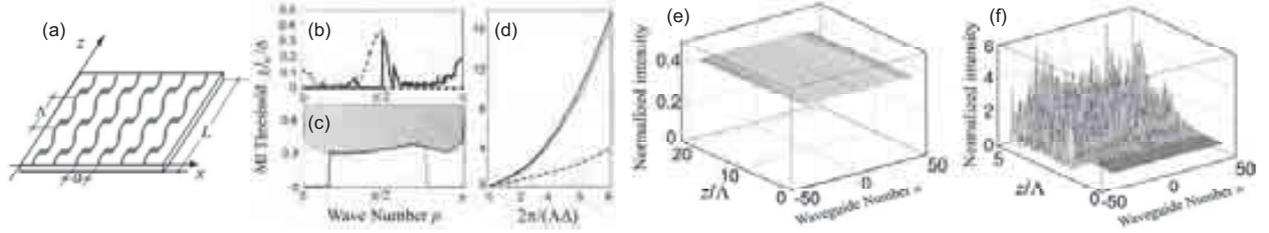

Figure 44: (a) Array of periodically curved optical waveguides. Threshold value $\chi I_0/\Delta$ for MI in a sinusoidally curved array [self-focusing, $2\pi/(\Lambda\Delta) = 1$] vs. wave number $p$ of the nonlinear Bloch wave: (b) far from self-imaging (solid curve, $\Gamma = 1$; dashed curve, $\Gamma = 3$) and (c) close to self-imaging [solid curve, $\Gamma = 2.395$; dashed curve, $\Gamma = 2.420$]. In (c) the shaded area is the modulational instability domain at exact self-imaging. (d) MI threshold at $\Gamma = 2.405$ vs. $2\pi/(\Lambda\Delta)$ for p=0 (solid curve), $p = \pi/2$ (dashed curve), and $p = \pi$ (dotted curve). MI in the self-imaging condition for (e) $\chi I_0/\Delta = 0.4$ and (f) $\chi I_0/\Delta = 1$ and for $2\pi/(\Lambda\Delta) = 1$. MI threshold is $\chi I_0/\Delta = 0.692$ [see plot in (c) at $p = 0$]. After Ref. [152].

## 5. Nonlinear modulated lattices

*5.1. Modulation instability*

In the nonlinear propagation regime, one of the important problems is the effects of periodic bending on the modulational instability (MI) of nonlinear Bloch waves. For an array of straight waveguides, the phenomenon of discrete MI is well known [123]: for, e.g., a self-focusing medium, the nonlinear Bloch wave is unstable at any intensity level for $|p| < \pi/2$ (where $p$ is the wavenumber) and stable at any intensity level otherwise.

To study discrete modulational instability in the curved array [see Fig. 44(a)], Longhi [152] performed standard linear stability analysis by setting $\Psi_n(z) = \bar{\Psi}_n(z)[1+u(z)\exp(iqn)+v^*(z)\exp(-iqn)]$, where $\bar{\Psi}_n(z)$ is the nonlinear Bloch function, $q$ is the wave number of transverse discrete perturbation, and $u$ and $v$ are small perturbation amplitudes, which satisfy the corresponding linearized and periodic system. Applying again a Floquet theory, we can determine the Floquet exponents $\mu_\pm(q)$ by standard numerical methods; an instability at wave number $q$ occurs whenever the real part of $\mu_\pm(q)$ becomes positive.

Figures 44(b) and 44(c) show the intensity threshold for the occurrence of MI as a function of the wave number $p$ in a sinusoidally modulated array of waveguides for a few values of the dimensionless parameter $\Gamma = 4\pi^2 n_s aA/\Lambda\Delta)$, highlighting inhibition of MI as the self-imaging condition is approached. For $\Gamma$ far from and smaller than the self-imaging value of $\Gamma = 2.405$ (the first zero of Bessel function $J_0$), a thresholdless MI occurs, as for a straight array, in the region $|p| < \pi/2$, whereas a MI with a finite threshold and irregular boundary can be observed for the other wave numbers [solid curve in Fig. 44(b)]. This situation is reversed when $\Gamma$ lies between the first two zeros of $J_0$ [dashed curve in Fig. 44(b)]. As $\Gamma$ approaches the first zero of $J_0$ from both sides, the interval of thresholdless MI shrinks, as shown in Fig. 44(c). At $\Gamma = 2.405$, i.e., in the self-imaging condition, a finite threshold for MI is observed in the entire spectrum of nonlinear Bloch waves [Fig. 44(c)]. As a general rule, the threshold intensity for the occurrence of MI increases as the bending period $\Lambda$ decreases [Fig. 44(d)].

As an example, the evolution of normalized intensity of waveguide channels is shown for an intensity below [Fig. 44(e)] and above [Fig. 44(f)] the MI threshold corresponding to a nonlinear Bloch waves with $p = 0$, initially perturbed by an added random noise to seed the instability.

*5.2. Diffraction-managed solitons*

In this section we discuss the effect of nonlinearity on the beam propagation in arrays of periodically curved (or modulated) waveguides. Nonlinearity changes the refractive index which modifies the beam propagation and enables all-optical functionalities and active control of light propagation.

Propagation of light in nonlinear dielectric media with a periodically-varying refractive index is known to exhibit many novel features, which do not occur in homogeneous nonlinear materials [28]. The underlying periodicity can strongly modify the physics of nonlinear beam self-action, in both self-focusing and self-defocusing nonlinear regimes, and can lead to the beam self-trapping in the form of *discrete spatial solitons* [26, 57].

The combination of tailored diffraction characteristics and light self-action opens new possibilities for the power-controlled beam shaping and switching in nonlinear photonic structures. The recent theoretical studies of the nonlinear beam propagation in the lattices with longitudinally modified linear diffraction predicted that solitons can be generated in various types of *diffraction-managed nonlinear lattices*, including periodically-curved waveguide arrays [1] and other types of modulated one and two-dimensional photonic structures [228, 269, 125, 20], and such solitons are reminiscent of dispersion-managed temporal solitons [67, 273, 193]. On the other hand, many properties of discrete diffraction-managed solitons may be completely different.

An especially intriguing problem is the nonlinear beam self-action under the condition of linear dynamic localization, where diffraction is suppressed in all orders. In the latter case, the modulational instability is suppressed [152], suggesting that discrete soliton formation may demonstrate unusual features. In particular, numerical simulations indicate that narrow beams propagating in arrays of curved waveguides with reduced diffraction should exhibit nonlinear self-trapping to discrete solitons at increased powers [228], similar to the dynamics of a particle in a nonlinear chain under the action of dc field [29]. At the intermediate power levels, nonlinearity may instead lead to beam broadening due to the destruction of periodic linear beam refocusing.

We follow Ref. [73] and study the propagation of light beams in a one-dimensional array of coupled nonlinear optical waveguides with the transverse period $d$ in the $x$ direction, where the waveguide axes are periodically curved in the propagation direction $z$ with the period $L \gg d$. When the tilt of beams and waveguides at the input facet is less than the Bragg angle, the beam propagation is primarily characterized by coupling between the fundamental modes of the individual waveguides, and it can be described by the nonlinear tight-binding equations taking into account the periodic waveguide bending, $i(d\Psi_n/dz) + C(\omega)\exp(-i\omega\dot{x}_0(z))\Psi_{n+1} + C(\omega)\exp(i\omega\dot{x}_0(z))\Psi_{n-1} + \gamma|\Psi_n|^2\Psi_n = 0$, where $\gamma$ is an effective nonlinear coefficient which accounts for the Kerr-type nonlinear response of the waveguide material. Then, the total electric field envelope $E(x,z)$ is represented as a superposition of the modes $E_0(x)$ of the individual waveguides.

The distinctive features of discrete beam dynamics become most evident when only one waveguide is excited at the input. Then the light evolution for both positive and negative nonlinearities is fully equivalent in the framework of the tight-binding model [152, 183]. In the simulations presented below, we use the following values which are typical for the experiments with optical waveguide arrays: $d = 9$ $\mu$m, $n_0 = 2.35$, $\lambda = 532$ nm, $C_0 = 0.13$ mm$^{-1}$, $\gamma = 1.9$. Normalization chosen is such that $x$ is measured in $\mu$m and $z$ is measured in mm. We use the discrete model for the calculations presented in this paper, however we have confirmed the validity of our results by simulating the full parabolic equations for the continuous electric field envelopes. We note that in case of the strong coupling between the waveguides, the long range coupling between non-nearest neighbors becomes important. In this case, exact dynamical localization can also be realized in arrays of curved waveguides [95], and effects similar to the presented in this paper may be expected to take place in the nonlinear regime.

Following Ref. [73], we consider as a specific example curved waveguide array with the harmonic bending for which the effective coupling coefficient is

$$C_{\text{eff}} = C J_0\left(\frac{2\pi\omega A}{L}\right). \tag{106}$$

The effective diffraction can be made either normal, zero, or anomalous depending on the value of the bending amplitude. Cancelation of the effective coupling and periodic beam dynamic localization takes place at low powers when

$$A = A_{\text{sc}} = \frac{\xi L}{2\pi\omega}, \tag{107}$$

where $\xi \simeq 2.405$ is the first root of the function $J_0$. For example, for the bending period $L = 15$ mm dynamic localization occurs when $A_{\text{sc}} = 23.0$ $\mu$m.

The beam width is determined as the width of the transverse cross-section function centred at the current centre of mass of the beam where 75% of the beam power is concentrated. When the bending amplitude differs from the dynamic localization value, the beam experiences discrete diffraction at low powers, similar to the effect observed in straight waveguide arrays [28].

When the power of the input beam increases, nonlinearity destroys the dynamic localization condition by changing the refractive index of the waveguide material. Initially, we observe that the beam shape experiences irregular distortion, such that the periodicity of the dynamic localization is lost [see Fig. 45(a)]. However, the beam does not broaden significantly, and it still experiences approximate self-restoration at some points.

When the input power increases further, the beam no longer experiences self-restoration, and the nonlinear broadening takes place, where the beam experiences significant broadening and self-defocusing, as shown in Fig. 45(b).

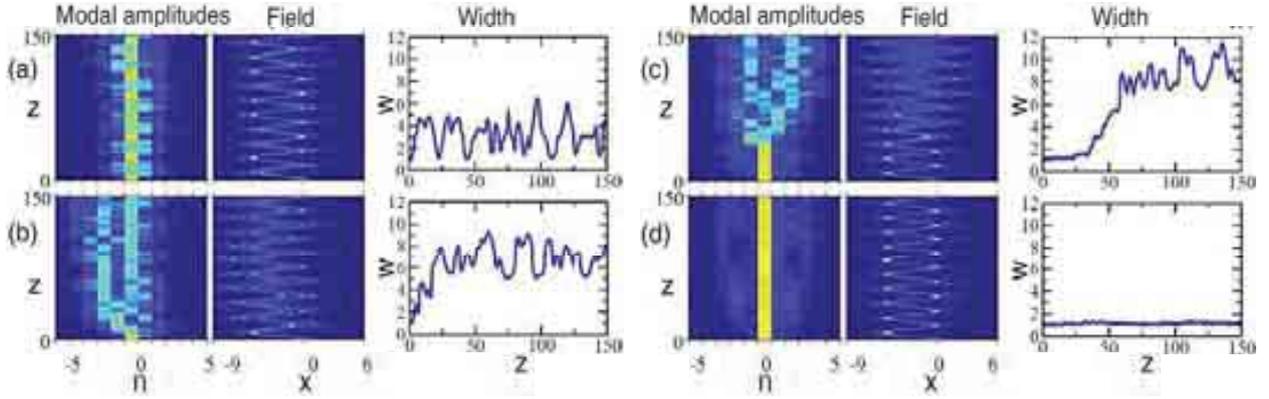

Figure 45: Nonlinear propagation of light in an array of periodically curved waveguides. Shown are the absolute values of the amplitudes of the modes of individual waveguides (left) and corresponding optical field patterns (centre). Right: beam width $w$ [normalized to the input width] as a function of the propagation distance. The input power is (a) $P/P_0 = 0.70$, (b) $P/P_0 = 1.7$, (c) $P/P_0 = 2.7$, and (d) $P/P_0 = 3.4$, where $P_0$ is the power required for the formation of one-site discrete soliton in the straight array. After Ref. [73].

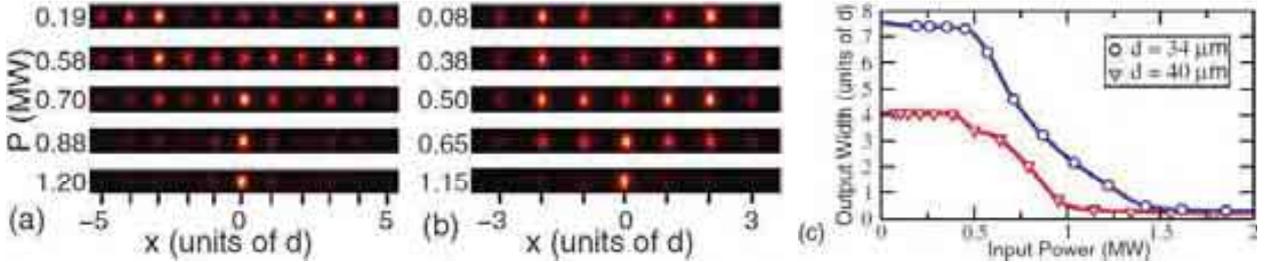

Figure 46: (a,b) Output beam profiles as a function of input peak power, measured in curved laser written waveguide arrays with waveguide spacing (a) $d = 34\ \mu$m and (b) $d = 40\ \mu$m. (c) Output beam width vs. input power. Circles correspond to (a), triangles to (b). After Ref. [287].

This self-defocusing is intrinsically limited due to the diffraction cancelation in the waveguide array. After propagation over some distance the beam broadens and its intensity is reduced accordingly. Therefore, the further beam spreading stops when the average beam width achieves a certain value. Such a peculiar nonlinear beam dynamics has no analogies in bulk media [122] or discrete systems [28] analyzed before.

At even higher input powers, transitional self-trapping of the beam is observed. The beam initially becomes self-trapped upon the launch into the array, but after propagation for some distance (which somewhat depends on the input power), it broadens rapidly and experiences again nonlinear broadening. Finally, at some critical power we observe a sharp transition from the nonlinear broadening to the discrete self-trapping over the whole length of the array, and the discrete lattice soliton is formed [see Fig. 45(d)].

In Ref. [287] these theoretical predictions were verified experimentally. In the experiments [287], waveguide arrays were created with a sinusoidal axis bending profile. Following Ref. [287], we study light propagation in the curved waveguides for different input powers. At low powers (linear regime), we observe dynamic localization of light in the curved waveguides, similar to [56, 183, 95]. Indeed, at the output facet of the arrays all the light is collected back into the same central waveguide in which it was coupled initially at the input [see Figs. 46(a) and 46(b), top].

When the input power is increased, transitional beam broadening is observed, in an agreement with the numerical simulations above. The beam experiences significant self-induced broadening [see Figs. 46(a) and 46(b), $P \sim 1$ MW] because the nonlinearity destroys the dynamic localization condition by changing the refractive index of the waveguide material. With propagation, the beam broadens and its intensity is reduced accordingly, such that the effect of the nonlinearity becomes weaker. Since the linear discrete diffraction is fully suppressed in the curved waveguide arrays, the beam broadening stops when the average beam width reaches a certain value.

At higher input powers, nonlinear self-trapping of the beam to a single lattice site occurs [see Figs. 46(a,b), bottom]. Whereas similar transition from nonlinear delocalization to self-trapping was predicted for Bloch oscilla-tions [228], it was not observed in previous experiments [216, 84]. We find, in good agreement with the theoretical predictions [73], that the power required for the formation of the diffraction-managed solitons in curved waveguide arrays is more than two times higher than the critical power of lattice solitons in exactly the same but straight waveguide arrays.

To confirm that the discrete diffraction-managed solitons can also form in periodically curved waveguide arrays with *defocusing* nonlinearity A. Szameit *et al.* [287] also performed experiments with curved LiNbO$_3$ waveguide arrays. They observed that the dynamic localization regime is destroyed and self-induced beam broadening takes place when the input power is increased. At higher powers, however, nonlinear beam self-trapping was observed.

### 5.3. Diffraction-managed polychromatic solitons

Nonlinear propagation of polychromatic light in curved waveguide arrays demonstrates many novel features [238]. At moderate light powers the nonlinear self-action breaks the left-right symmetry of the polychromatic beam, resulting in the separation of different spectral components due to the wavelength-dependent spatial shift. At high light powers a diffraction-managed polychromatic soliton is formed. These results demonstrate new possibilities for tunable demultiplexing and spatial filtering of supercontinuum light.

Following Ref. [238], first we discuss the general symmetry properties of polychromatic beams propagating in curved nonlinear waveguide arrays. The evolution of field amplitudes at individual waveguides, $a_{m,n}(z)$, can be approximately described by the coupled-mode equations [287]: $id\Psi_{m,n}/dz + C_m\Psi_{m,n+1} + C_m^*\Psi_{m,n-1} + \bar{\mathcal{G}}\Psi_{m,n} = 0$. Here $m$ is the number of the spectral component, and $n$ is the waveguide number, $C_m = C\exp[-2\pi i n_0 d\dot{x}_0(z)/\lambda_m]$ are the coupling coefficients depending on the local waveguide curvature, $C$ is the coupling coefficient in a straight array, and $d$ is the waveguide spacing. $\bar{\mathcal{G}}$ is a nonlinear term.

If $a_{m,n}$ is a solution of the coupled-mode equations then $\Phi_{m,n}(z) = (-1)^n\Psi_{m,-n}^*(z)e^{i\phi_m}$, is also a solution of the same equations but with the opposite sign of the nonlinearity $\gamma \to -\gamma$, where $\phi_m$ are arbitrary constant phase coefficients. This is a non-trivial generalization of the symmetry which is known to exist for optical Bloch oscillations [233] and between positive and negative nonlinearities in straight waveguide arrays [202]. In the linear propagation regime, $\gamma \equiv 0$, $\Psi_{m,n}$ and $\Phi_{m,n}$ satisfy exactly the same evolution equations. If only a single waveguide ($n = 0$) is excited at the input, the initial conditions for functions $\Psi_{m,n}$ and $\Phi_{m,n}$ can always be made the same by the proper choice of the constant phase coefficients, $\phi_m = 2\arg[\Psi_{m,0}(0)]$. Therefore $\Psi_{m,n}$ and $\Phi_{m,n}$ are identically equal at any propagation distance $\Psi_{m,n}(z) \equiv \Phi_{m,n}(z) \equiv \Psi_{m,-n}^*(z)e^{i\phi_m}$, and the beam intensity profile is symmetric since $|\Psi_{m,n}(z)|^2 \equiv |\Psi_{m,-n}(z)|^2$. In contrast, in the nonlinear regime the evolution equations for functions $\Psi_{m,n}$ and $\Phi_{m,n}$ are no longer the same, they differ in the sign of the nonlinear coefficient $\gamma$. Thus the left-right reflection symmetry of the intensity profile is immediately broken in the nonlinear regime. On the other hand, if one considers two identical arrays with the opposite sign of the nonlinearity and the same input single site excitation, then there exists exact reflection symmetry of the beam profiles between the two signs of the nonlinearity since $|a_{m,n}(z;+\gamma)|^2 \equiv |a_{m,-n}(z;-\gamma)|^2$.

To confirm the predictions based on the coupled-mode theory, the complete set of nonlinear equations was solved numerically using a finite-difference beam propagation method [238]. The waveguide bending profile was selected to be composed of two sinusoidal sections, and the transverse refractive index profile was taken as $\Delta n(x) = \Delta n_0 \cos^2(\pi x/d)$, with the refractive index change (fitted to the experimental value) $\Delta n_0 = 3.6\times10^{-4}$. The photosensitivity of LiNbO$_3$ was approximated as $\alpha(\lambda) = \exp[-\log(2)(\lambda - \lambda_b)^2/\lambda_w^2]$, where $\lambda_b = 400$ nm and $\lambda_w = 150$ nm. In the simulations, single input waveguide at $x = 0$ was excited with a Gaussian beam consisting of a large number of spectral components, $M = 60$, and a flat intensity spectrum $|E_m(0,0)|^2 = I$.

In numerical simulations, Qi *et al.* [238] observed that the output diffraction pattern in the curved array is symmetric with respect to the input position only at low input power (i.e., in the linear regime). This is the case for all spectral components despite their nontrivial evolution inside the array. In contrast, in the nonlinear regime symmetry of the output beam profile gets broken. For the defocusing nonlinearity ($\gamma = -1$), more blue components are concentrated on the left and more red spectral components are present on the right hand side of the beam. For the focusing nonlinearity ($\gamma = 1$), the output beam profile is also asymmetric and it is almost identical, up to the left-right reflection, to the output profile for the defocusing nonlinearity at the same input power level.

The spectral distribution between the output waveguides depends strongly on the input intensity, see Fig. 47. For high input powers noticeable difference starts to occur in the beam profiles for the defocusing [Fig. 47(a)] and focusing [Fig. 47(b)] nonlinearity. The coupled-mode analysis predicts an exact reflection symmetry of the beam profiles for both signs of the nonlinearity, however this approach starts to fail at high levels of nonlinearity. As the power is increased further, simultaneous nonlinear self-trapping of all spectral components to a single waveguide channel and formation of polychromatic diffraction-managed soliton occurs for both defocusing and focusing nonlinearities. Similar to monochromatic diffraction-managed solitons [287], the critical power required for formation of polychromatic diffraction-managed solitons in the curved arrays is several times higher than in straight arrays.

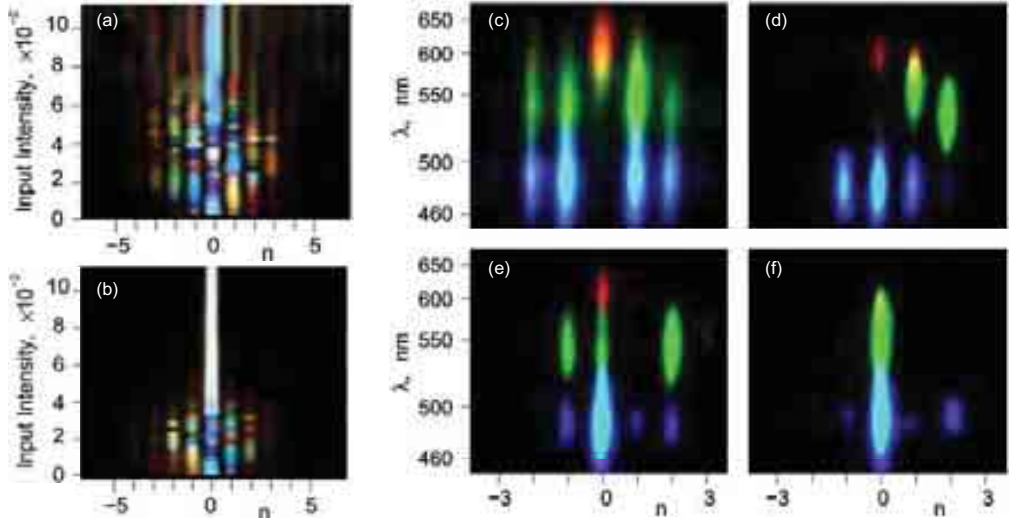

Figure 47: Numerically calculated output light intensity in the curved array as a function of the input power for (a) defocusing and (b) focusing nonlinearity. Experimental spectrally resolved output beam profiles: (b) Illumination time of 5 s, input power of $1\mu$ W; (c)-(f) illumination time of 35, 60, and 420 s, respectively; input power is 18.0 mW. After Ref. [238].

These theoretical predictions have been tested experimentally in Ref. [238], for the curved waveguide arrays fabricated by Titanium indiffusion in a 50 mm long X-cut niobate ($LiNbO_3$) mono-crystal, excited by a supercontinuum beam in the spectral range of $460-660$ nm, generated by femtosecond laser pulses coupled into highly nonlinear photonic crystal fiber. At low input power, in the essentially linear propagation regime, the measured output beam profile is symmetric with respect to the central waveguide ($n = 0$) [Fig. 47(c)]. At higher input power, shorter (blue) wavelengths are predominantly concentrated in the left waveguides and longer (green) wavelengthes are concentrated in the right waveguides [Figs. 47(d) and 47(e)], in a qualitative agreement with the theoretical predictions. The key result was the observation of *diffraction-managed polychromatic soliton* through self-localization of all spectral components to a single waveguide at long enough illumination times Fig. 47(f).

*5.4. Nonlinear surface waves*

The existence and properties of linear surface waves existing at the edge of an array of periodically curved optical waveguides has been discussed above in Sec. XXX. Here, we describe nonlinear surface waves generated at the edge of modulated waveguide arrays. Such surface modes possess a nontrivial dynamics due to an interplay between three mechanisms of surface localization: (i) waveguide bending, (ii) fabricated surface defect, and (iii) nonlinear beam self-action. We demonstrate experimentally that nonlinear beam self-action can provide an effective control of the output beam profile, including switching between different waveguides near the surface.

The experiments [240, 239] were performed with nonlinear modulated waveguides fabricated by titanium indiffusion in a 50 mm long X-cut $LiNbO_3$ crystal with defocusing photorefractive nonlinearity [202], featuring a transverse refractive index profile containing a negative surface defect (lower refractive index value at the first waveguide). The waveguide bending profile was selected as being composed of sinusoidal sections. Two curved arrays with different bending amplitudes $A = 21.5\,\mu$m and $A = 24.5\,\mu$m were fabricated. Both samples contain one full bending period of $L = 50$ mm.

The curved waveguide arrays can support linear surface modes even in the absence of surface defects [74]. In the nonlinear regime, the beam dynamics can be studied numerically, including the effect of nonlinear self-action [240].

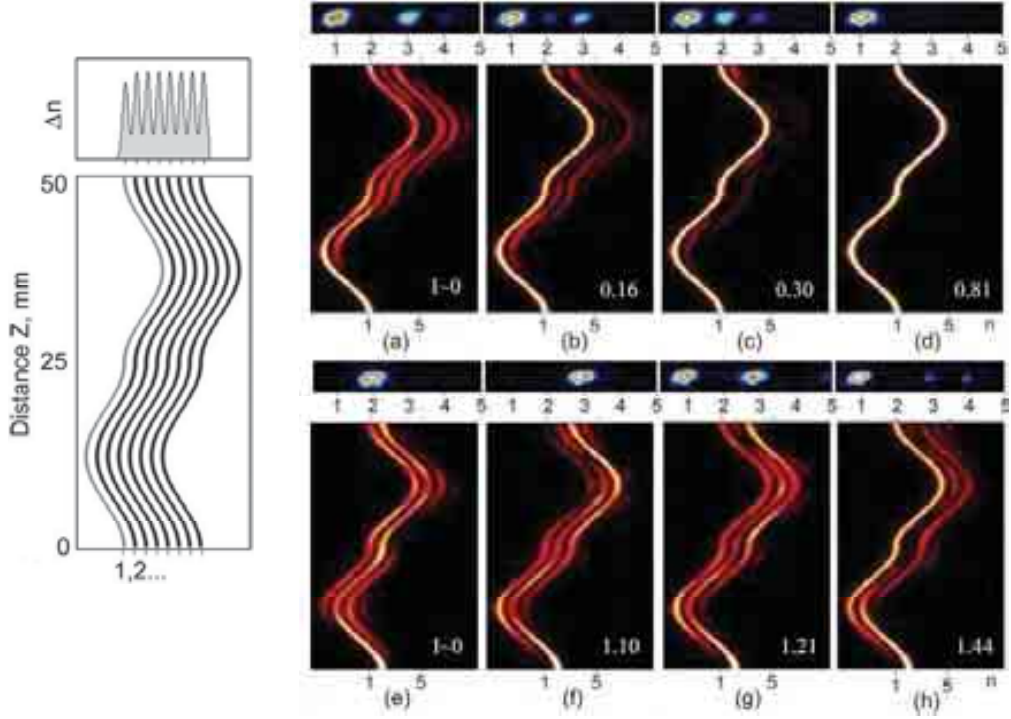

Figure 48: Surface waves in modulated waveguide array ($A = 24.5\,\mu$m). The insert shows schematic of the array (bottom) and corresponding refractive index profile (top). In each block, top images - experimental results for different illumination times, bottom images - numerical simulations. In (a-d) the beam is launched into the first waveguide, in (e-h) the second waveguide is excited. (a,e) Linear propagation - input power $1\,\mu$W, illumination time 5 s. In (b-d) illumination time is 0.50, 1.25, and 4.25 minutes; in (f-h) illumination time is 10.92, 17.67, and 18.42 minutes, respectively. Input power in (b-d) and (f-h) is 4 mW. In the numerical calculations the normalized input intensity $I$ is marked in the corners. After Ref. [240].

For light coupled to a single waveguide number $n$ at the input with the normalized intensity $I = |\Psi_n(0)|^2$, the simulated beam evolution in an array with $A = 24.5\,\mu$m in Fig. 48. In Figs. 48(a-d) we excite the first waveguide ($n = 1$). At low powers, in the essentially linear propagation regime, the first (defect) surface mode is excited. At higher input powers, the defocusing nonlinearity increases the strength of the negative defect in the first waveguide, and eventually the entire beam gets trapped to the first waveguide [see Figs. 48(c-d)].

In Figs. 48(e-h), the second waveguide ($n = 2$) is excited. At low light intensity, the second surface mode is excited. However, at higher intensity, nonlinear coupling and interaction between different linear modes is present. We note that even away from the surface, the nonlinear beam dynamics is highly nontrivial at these power levels [73, 288]. It is observed switching of the output beam position between the second, third, and first waveguides as we increase the intensity [Figs. 48(f,g)]. As the input intensity grows further, the beam becomes localized at the edge waveguide [Fig. 48(h)], indicating the formation of a self-trapped nonlinear surface wave. At even higher intensities (not shown) the nonlinearity completely detunes the input channel and the light gets trapped back to the second waveguide.

In our experiments, we use the beam from a cw laser ($\lambda = 532$ nm). The beam is $o$−polarized, perpendicularly to the plane of the array, to minimize bending losses and radiation. The nonlinear refractive index change in the photorefractive LiNbO$_3$ depends on the input power and slowly increases with illumination time. Therefore, it was monitored the output intensity distribution onto a CCD camera with increasing illumination, till a steady state is reached. We focus the beam to the first and second waveguides, and for each case show the output intensity profiles at four different illumination times in Fig. 48(a-h, top images). As the nonlinear response increases, the output beam switches between the different waveguides. There is a good agreement between the experiments and numerical simulations, except at very high input powers, when a slow oscillatory beam motion between the first two waveguides was registered, due to the charge dynamics in photorefractive crystal which is not described by the coupled-mode equations.

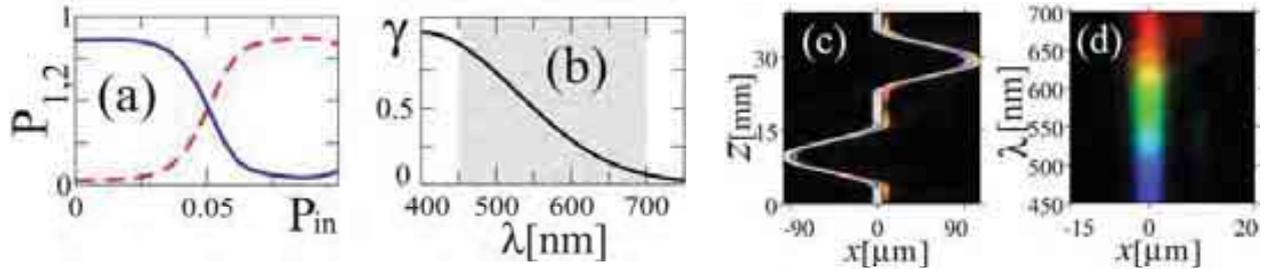

Figure 49: Nonlinear switching of polychromatic light. (a) Power distribution at the output ports of the coupler as a function of the input power. Solid and dashed curves show power in the left ($P_1$) and in the right ($P_2$) output coupler ports, respectively. At the input, polychromatic light with flat spectrum covering $450 - 700$ nm is fed into the left arm of the coupler. (b) Sensitivity function $\gamma$ describing wavelength-dispersion of the nonlinear response. (c,d) Propagation dynamics and output spectrum, respectively, in the nonlinear switched state realized at the total input power $P_{in} = 0.085$. Nonlinear coefficient is $\alpha = 10$. After Ref. [70].

The properties of two-dimensional surface solitons supported by an interface of a waveguide array whose nonlinearity is periodically modulated was analyzed theoretically by Kartashov *et al.* [114]. It was shown that, when the nonlinearity strength reaches its minima at the points where the linear refractive index attains its maxima, nonlinear surface waves exist and can be made stable only within a limited band of input energy flows and for lattice depths exceeding a lower threshold.

*5.5. Nonlinear switching of polychromatic light*

Periodic bending of the waveguide axes can be employed in a number of useful nonlinear optical devices, such as a nonlinear coupler, for realizing all-optical switching of polychromatic light with a very broad spectrum covering all of the visible region. The bandwidth of such a device can be enhanced five times or more compared with conventional structures composed of straight waveguides. This approach suggests novel opportunities for the creating all-optical logical gates and switches for polychromatic light with white-light and supercontinuum spectra.

Here we discuss nonlinear switching of polychromatic signals in a directional coupler composed of two coupled periodically modulated waveguides. Such a nonlinear coupler can be realized in media with slow nonlinear response, where the optically-induced refractive index change is defined by the time-averaged light intensity of different spectral components [210, 19].

The optimized curved coupler can be used to collectively switch all spectral components around the central wavelength $\lambda_0$ from one waveguide at the input to the other waveguide at the output [70]. At high input powers, nonlinear change of the refractive index modifies waveguide propagation constant and decouples waveguides from each other similar to other nonlinear coupler structures studied before [97, 203, 66]. This causes switching from crossed state into the parallel state as shown in Figs. 49(a), 49(c) and 49(d). Remarkably, nonlinear switching also takes place in a very broad spectral region ∼ 450 − 700nm, which enables the coupler to act as an all-optical digital switch for polychromatic light.

In the simulations of Ref. [70], the case of a photorefractive medium such as $LiNbO_3$ was considered, where optical waveguides of arbitrary configuration are fabricated by titanium indiffusion [21, 202]. In the $LiNbO_3$ waveguide arrays, the photovoltaic nonlinearity arises due to charge excitations by light absorption and corresponding separation of these charges due to diffusion.

The spectral response of this type of nonlinearity depends on the crystal doping and stoichiometry, and it may vary from crystal to crystal. In general light sensitivity appears in a wide spectral range with a maximum for the blue spectral components [256], but the sensitivity extends also in the near infra-red region [98]. The photosensitivity dependence can be approximated by a Gaussian function

$$\gamma(\lambda) = \exp\left[-\log(2)\frac{(\lambda - \lambda_b)^2}{\lambda_w^2}\right], \qquad (108)$$

where $\lambda_b = 400$ *nm* and $\lambda_w = 150$ *nm* [Fig. 49(b)]. It was checked [70] that the switching behavior of the coupler remains essentially the same for a range of other values of $\lambda_w$, which primarily affect the quantitative characteristics such as the switching power.

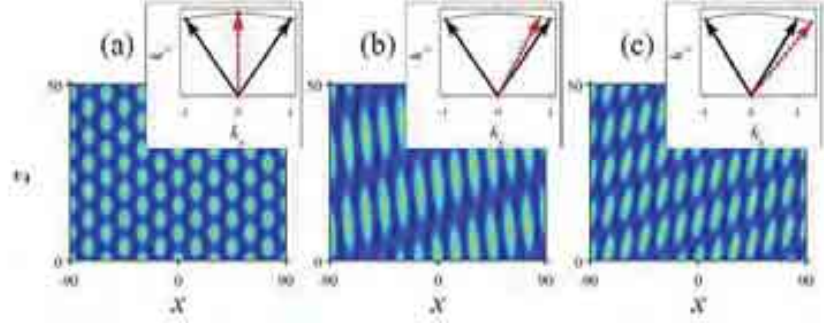

Figure 50: Examples of one-dimensional photonic lattices modulated by the third beam with the transverse wave number $k_{3x}$: (a) $k_{3x} = 0$, (b) $k_{3x} = 0.8k_{12x}$, and (c) $k_{3x} = 1.3k_{12x}$. Insets show the wave vectors of two input beams which form the lattice, and the wave vector of the third beam (red, dashed). Parameters are $A_{12} = 0.25$, $A_3 = 0.66A_{12}$ and the propagation length is $L = 50$ mm. After Ref. [71].

### 5.6. Soliton steering and dragging in reconfigurable lattices

So far our theoretical and experimental studies of light propagation in modulated photonic lattices assumed fixed lattice geometries, which cannot be changed or modified after the structure has been designed and fabricated. On the other hand, it is clear that the ability to *dynamically modify the lattice geometry* in real time would provide many additional functionalities. Indeed, recent theoretical and experimental studies have demonstrated nonlinear localization of light in the optically-induced photonic lattices where the refractive index is modulated periodically in the transverse direction by an interference pattern of plane waves that illuminate a photorefractive crystal with a strong electro-optic anisotropy [53, 60, 61, 217]. When the lattice-forming waves are polarized orthogonally to the $c$-axis of the photorefractive crystal, the periodic interference pattern propagates in the diffraction-free linear regime, thus creating a refractive-index modulation similar to that in weakly coupled waveguide array structures [28].

Such *optically-induced* photonic lattices have been employed to demonstrate many fundamental concepts of the linear and nonlinear light propagation in periodic photonic systems, including the generation of lattice [60, 217, 200] and spatial gap solitons in defocusing [60, 61] and self-focusing [218] regimes, Bragg scattering and Bloch-wave steering [277], tunable negative refraction [250], etc.

In this Chapter, we study the soliton propagation in dynamic optical lattices and identify novel effects associated with the optically-induced refractive index modulation in the longitudinal direction. Such lattices can be created by several interfering beams, which are inclined at different angles with respect to the crystal.

Photorefractive crystals exhibit a very strong electro-optic anisotropy, e.g. in SBN:75 the electro-optic coefficient for extraordinary polarized waves is more than 20 times higher than the electro-optic coefficient for ordinary polarized waves [61]. Thus, the lattice-writing beams polarized orthogonal to the $c$-axis of the crystal satisfy the linear wave equation, while extraordinary polarized beam will experience a highly nonlinear evolution. Then, each of the broad lattice beams propagates independently, and it can be presented as a linear plane-wave solution in the form

$$E_{\text{lattice}} = A \exp(i\beta z + ik_x x), \qquad (109)$$

where $k_x$ is the transverse wavenumber proportional to the inclination angle, and the propagation constant $\beta = -Dk_x^2$ defines the longitudinal wavevector component $k_z$. The value of diffraction coefficient $D$ can be controlled by varying the wavelength of lattice beams, and also depends on the crystal anisotropy. Following Refs. [109, 71], we consider a lattice induced by three interfering waves [109]: (i) two waves with equal amplitudes $A_{12}$ and opposite inclination angles, with the corresponding wavenumbers $k_{12x}$ and $-k_{12x}$, and (ii) an additional third wave with amplitude $A_3$ and wavenumber $k_{3x}$. Then, the optical lattice is defined through the wave interference pattern $I_p(x, z) = |A_L|^2$, where

$$A_L = A_3 e^{i\beta_3 z + ik_{3x}x - i\varphi} + 2A_{12} e^{i\beta_{12}z} \cos(k_{12x}x), \qquad (110)$$

and $\varphi$ determines the relative phase between the third wave and the other two waves. It follows that additional beam (with $k_{3x} \neq k_{12x}$) always leads to the lattice modulation both in the transverse and longitudinal directions. We show examples of modulated lattices in Figs. 50(a-c) corresponding to the same wave amplitudes but different inclinations of the third beam (defined by $k_{3x}$) as indicated in the insets. We see that for $k_{3x} = 0$ [Fig. 50(a)] the lattice profile in the transverse cross-section becomes double-periodic corresponding to an alternating sequence of deeper and shallower potential wells resembling a binary superlattice [274], however its configuration is periodically inverted due to modulations in the longitudinal direction along $z$. On the other hand, when $k_{3x} \simeq k_{12x}$, the lattice is slowly modulated in both spatial directions and the left-right reflection symmetry is removed [109], see Figs. 50(b) and (c).

One of the original theoretical ideas to employ such modulated lattice is related to the possibility of binary steering of strongly localized solitons, where the soliton propagates in one of two allowed directions when the amplitude $A_3$ is in one of the two stable regions [71].

In numerical modeling reported in Ref. [71] a strongly localized lattice soliton was generated by an input Gaussian beam,

$$E_{\text{in}} = A_{\text{in}} \exp\left\{ -\left[\frac{x - x_0}{w}\right]^2 + ik_{0x}(x - x_0) \right\}, \qquad (111)$$

which is incident on the crystal at normal angle (i.e. $k_{0x} = 0$) and has extra-ordinary linear polarization. When the amplitude of the third wave $A_3$ is relatively small, the generated soliton starts moving between the neighboring lattice sites. As the amplitude $A_3$ of the modulating beam increases, at certain point strongly localized soliton becomes locked at a particular lattice site, and it propagates straight along the lattice, similar to the case of homogeneous structures without longitudinal modulation [215].

When the amplitude $A_3$ grows further, the so-called binary soliton steering occurs due to the substantial change in the geometry of the optical lattice, where the connectivity between high-index lattice sites changes from vertical to diagonal through a disconnected state when we increase the amplitude of the third modulating wave. Such a behavior differs from the dynamics of broad solitons in weakly modulated lattices [109], which feel only spatially averaged, smoothed lattice potential. In contrast, behavior of strongly localized solitons is dominated by the fine geometrical structure of the lattice.

The origin of the soliton switching effect described above is fundamentally different from dragging of broad solitons reported in Ref. [109] which is almost directly proportional to the third beam amplitude $A_3$.

In the experiments [248], a modulated optical lattice was crated by interfering three ordinarily-polarized broad laser beams [see Eq. (110)] from a frequency-doubled Nd:YVO4 cw laser at a wavelength 532 nm inside a 15×5×5 mm SBN:60 crystal. Applying an external bias voltage of 2.2 kV, we produce a refractive index modulation and control the saturation by homogeneously illuminating the crystal with white light. An extraordinarily-polarized Gaussian probe beam with a full width at half maximum (FWHM) of 25$\mu$m (along the $x$ direction and extended in $y$) is launched into the crystal, parallel to the $z$ axis ($k_x = 0$).

First, we characterize the effect of the modulated lattice geometry on the propagation of a low power ($\sim$ 25 nW) probe beam in the linear regime for three different angles of the modulating beam, in the case of strong lattice modulation, $I_3 = 4I_{12}$. In Fig. 51(a) we plot the shift of the beam centre of mass vs. the modulating beam phase $\varphi$. Solid lines represent smoothing spline fitting to the data points (experimental uncertainty in the vertical direction is approximately 10$\mu$m in Fig. 51(a), and up to 20 $\mu$m in Fig. 51(b)). The phase $\varphi$ is adjusted by passing the third beam through a thin glass plate with a variable tilt. For $k_{3x} = 1.01k_{12x}$ the beam shift is virtually zero throughout the entire phase scan [stars in Fig. 51(a)]. The insensitivity to $\varphi$ results from the lattice being fully symmetric when beams 2 and 3 in Fig. 50 are parallel ($k_{3x} = k_{12x}$). On the other hand, as the angle of the third lattice-forming wave is increased, the lattice becomes asymmetrically modulated, and the beam shifts to one or the other side [squares and circles in Fig. 51(a)], depending strongly on the value of $\varphi$ (similar behavior was observed for $k_{3x} < k_{12}$). The observed features were confirmed by numerical simulations (not shown), and they prove that not only the local asymmetric distortion of the lattice, which in this case tends to shift the beam towards positive $x$ [see Fig. 50(c)], but also the broader effective modulation geometry plays an important role for the beam propagation dynamics. We further note that in Fig. 51(a) the beam shift is not symmetric with respect to $x = 0$, and it is strongest in the region $0 < \varphi < \pi$ where the effects of local and broad lattice modulation pull the beam in the same direction.

Figure 51(b) maps the corresponding output beam width (FWHM of Gaussian fit) as a function of $\varphi$. Again, the case $k_{3x} = 1.01k_{12x}$ proves to be relatively insensitive to the phase shift, whereas for larger angles, substantial beam broadening is observed close to $\varphi = \pi$. For $k_{3x} = 1.18k_{12x}$, the maximum output beam width greatly exceeds that of a diffracting beam in the absence of a lattice (57 $\mu$m), and the observed broadening must be attributed to the geometry of the modulated lattice, and not solely to the decreased contrast of the lattice at $\varphi = \pi$. We note that when the phase $\varphi$ is scanned from zero to $2\pi$, the local input beam excitation symmetry changes from on-site to off-site and back, and this may in principle lead to additional beam steering [215]. However, we verified that under our experimental conditions the contributions to centre of mass shift as well as beam broadening due to this effect are negligible ($< 3\mu m$ in both cases). Examples of output beam profiles corresponding to points c, d and e in Figs. 51(a,b) are shown in Figs. 51(c-e).

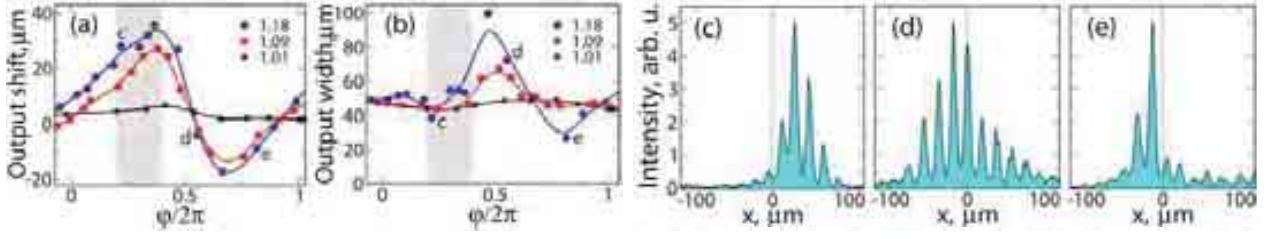

Figure 51: (a,b) Measured shift and width of a linear probe beam output vs. the phase of the modulating lattice beam for three different values of $k_{3x}/k_{12x}$ for $I_3 = 4I_{12}$. Shading marks the region in which the figure of merit is maximized. (c-e) Examples of measured output profiles corresponding to the points (c,d,e) in the plots (a,b). After Ref. [248].

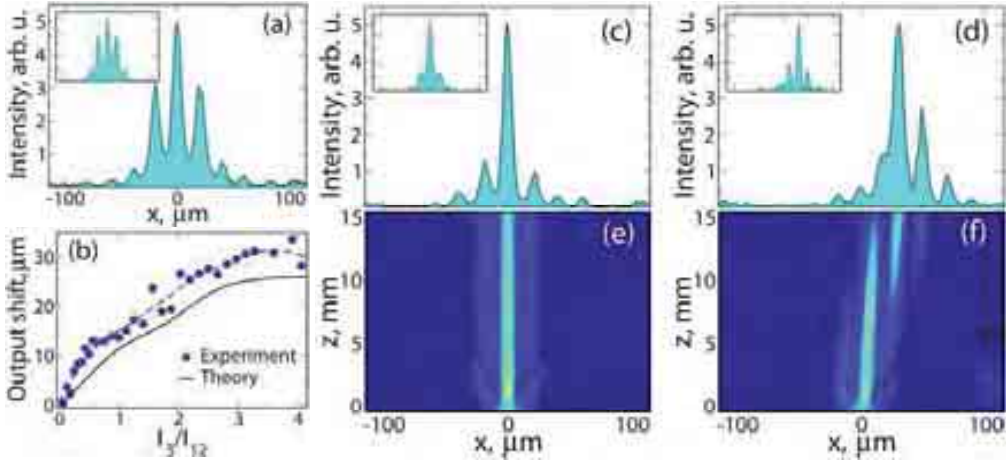

Figure 52: (a) Experimental and theoretical (inset) linear output in a straight lattice ($I_3 = 0$). (b) Shift of the nonlinear probe beam output vs. the modulating beam power for $k_{3x} = 1.18k_{12x}$ and $\varphi/2\pi = 0.22$. (c,d) Experimental and theoretical (inset) nonlinear output in straight and modulated lattices for $I_3 = 0$ and $I_3 = 4I_{12}$, respectively, and $k_{3x} = 1.18k_{12x}$. (e,f) Numerical simulations of the longitudinal propagation. After Ref. [248].

Concentrating now on the nonlinear case, we show in Fig. 52 that increasing the power of the probe beam to $1.5\,\mu W$ leads to strong self-focusing and enhanced beam localization [60] [Fig. 52(c)] while preserving a large beam shift [Fig. 52(d)]. As a result, the figure of merit is increased by approximately a factor of two compared to the linear case, thus exceeding unity. Figure 52(a) shows, for comparison, the broader low power output profile in a regular straight lattice ($I_3 = 0$). In all cases the experimental observations match our numerical simulations, shown in Fig. 52 as the beam profile insets in panels (a,c,d), and the top view of the propagation dynamics in panels (e,f). In Fig. 52(b) we trace the experimental (circles) and theoretical (solid line) nonlinear beam shift as a function of the lattice modulation power. We find that the beam shift gradually increases and, in experiment, saturates at approximately $\Delta x = 30\,\mu m$ for $I_3/I_{12} > 3$. The small (15%) difference between theory and experiment in Fig. 52(b) is attributed to a small self-induced drift of the strongly localized beam [261], that was not taken into account in simulations.

Finally, we mention that the soliton control and dragging can be achieved in modulated optical lattices featuring topological dislocations where solitons experience attractive and repulsive forces around the dislocations [115]. Suitable arrangements of dislocations were shown to form soliton traps, and the properties of such solitons depend crucially on the trap topology.

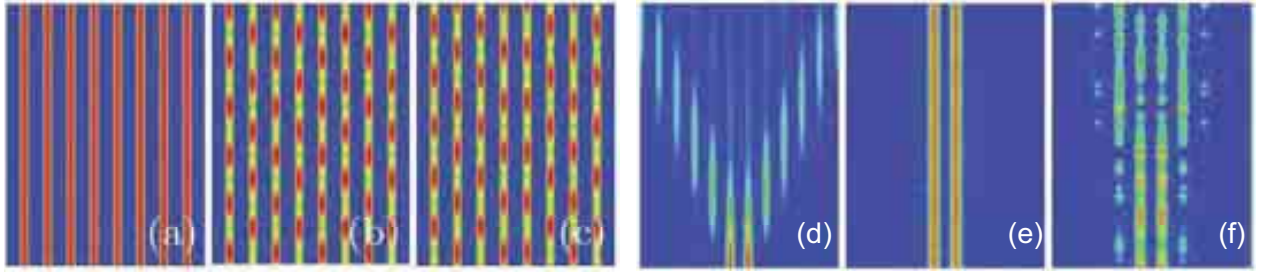

Figure 53: Examples of modulated lattices: (a) Unmodulated, (b) out-of-phase modulated, (c) and in-phase modulated waveguide arrays. (d) Diffraction of the antisymmetric mode in an unmodulated array. Dynamics of propagation of the antisymmetric mode in in-phase modulated waveguide arrays at $\mu = 0.15$ for (e) $A = 0.01$ and (f) 1.59. After Ref. [145].

### 5.7. Modulated lattices and waveguiding structures

In most of the examples discussed above the modulation of the waveguide arrays is achieved by a periodic longitudinal bending of the waveguide axis. However, the control of diffraction is also possible through other types of modulation of waveguide arrays. In general, two major types of the lattice modulations can be considered. In the first case, only linear properties of the lattice are assumed to vary periodically, see e.g. Refs. [145, 117, 294]. Such modulated waveguide structures allow controlling resonant Rabi oscillations and switching dynamics of optical couplers, as well as suppress diffraction in waveguide arrays. In the second case, the linear properties of the lattices do not change and only nonlinear response of the waveguides is modulated, either longitudinally [105] or transversally [114, 110, 106, 107], the former case corresponds more closely to the case of the periodic modulation of the waveguide axes [see Figs. 53(a-c)].

In such structures, modulations along both transversal and longitudinal directions open new routes for diffraction management and make possible a variety of new phenomena such as the possibility of discrete diffraction suppression even in the linear regime.

To discuss one of such effects, we follow Ref. [145] and consider the NLS equation for the dimensionless field amplitude $E$, governing the propagation of the light beam along the $z$ axis of the waveguide array with a longitudinally modulated refractive index

$$i\frac{\partial E}{\partial zi} = -\frac{1}{2}\frac{\partial^2 E}{\partial x^2} - \Delta nV(x,z)E - |E|^2 E, \qquad (112)$$

Here, $x$ and $z$ are the normalized transverse and longitudinal coordinates, while $\Delta n$ stands for the refractive index contrast of the individual waveguide. The refractive index profile of the lattice is given by $V(x,z)$ that describes the longitudinal modulations of the amplitude $\mu$ with the frequency $\Omega$.

Usually, two types of the longitudinal refractive index modulation are considered [see Figs. 53(b,c)]. In the first case (termed out-of-phase modulation) the refractive index oscillates out-of-phase in all waveguides of the array [270]. In the second case (further called in-phase modulation) the refractive index oscillates in-phase in the selected group of several excited waveguides, while in all other waveguides surrounding the selected group it oscillates out-of-phase.

Figures 53(d-f) show an example of the effect of nodulation amplitude of the propagation of the antisymmetric mode in in-phase modulated waveguide arrays. The output intensity pattern remains symmetric even in the delocalization regime [Fig. 53(e)], while for symmetric excitations one observes remarkable asymmetries in output intensity distributions [Fig. 53(f)].

In general, the longitudinal refractive index modulation allows inhibition of light tunneling not only for complex multichannel structures. For example, the antisymmetric fourth-order mode perfectly preserves its structure in the in-phase modulated lattice, while in the case of out-of-phase modulation the structure of this mode is strongly distorted due to the energy exchange between adjacent waveguides, although in both cases the coupling between the central group of excited guides and the surrounding array is inhibited.

### 5.8. Light bullets in modulated lattices

When nonlinear waves are localized both in space and in time as a result of balance between the self-focusing nonlinearity, diffraction, and anomalous dispersion, they form the so-called *spatiotemporal optical solitons* often called *light bullets* [260]. Light bullets are inherently multi-dimensional, and consequently they are unstable in Kerr media [260, 195]. Discreteness in photonic lattices can provide an additional stabilization mechanism, and indeed stable continuous-discrete solitons were found in nonlinear waveguide arrays [129, 6] and optical fiber bundles [134, 135]. Continuous-discrete spatiotemporal surface solitons and collisions between them were also extensively analyzed [205, 204, 206, 207].

Arrays or lattices of periodically curved (or periodically modulated) optical waveguides offer unique opportunities for the control of both the strength and frequency dispersion of the waveguide coupling [56, 183, 289]. In particular, light beams can become localized in the periodically curved nonlinear waveguide arrays in the form of diffraction-managed solitons [1, 73, 288]. These diffraction-managed spatial solitons are reminiscent of dispersion-managed temporal solitons [67, 273, 193].

As discussed below, stable spatiotemporal solitons can exist in lattices of periodically curved [201] and periodically modulated [144] optical waveguides, and such light bullets can be mobile moving across the lattice [201].

*5.8.1. Mobile discrete-continuous light bullets*

First, we consider propagation of light pulses in a one-dimensional array of coupled optical waveguides, where waveguide axes are periodically curved in the longitudinal propagation direction $z$. We follow the paper [201] and model the light bullet dynamics with the modified discrete-continuous nonlinear Schrodinger propagation equation which takes into the periodic waveguide bending [152, 75] and temporal dispersion [232, 47, 131, 87]:

$$i\frac{\partial \Psi_n}{\partial z} + C\left(\exp^{-ip(z)} \Psi_{n+1} + \exp^{ip(z)} \Psi_{n-1}\right) + \frac{D}{2}\frac{\partial^2 \Psi_n}{\partial t^2} + \gamma|\Psi_n|^2\Psi_n = 0, \tag{113}$$

where $\Psi_n(t)$ are the complex amplitudes of individual waveguide modes, $t$ is time, $C$ is the coupling between the modes of straight neighboring waveguides, $D$ is the second-order dispersion coefficient, and $p(z)$ is the phase coefficient related to the waveguide bending [75]. We apply the following normalization $\Psi_n = \Phi_n/\sqrt{I_s}$, $z = z_d/z_s$, $t = t_d/t_s$, $C = C_d z_s$, $D = D_d z_s t_s^{-2}$, and $\gamma = \gamma_d I_s z_s$, where $\Phi_n$ and variables with the subscript $d$ denote the values in physical dimensions, and subscript $s$ denotes the scaling coefficients. With these notations, the coupling phase coefficient is defined as

$$p(z) = \frac{2\pi n_0 d x_s^2}{\lambda_d z_s} x_0' \tag{114}$$

where $x_0(z)$ is the waveguide bending profile normalized to $x_s$, prime stands for the derivative, $d_x = d_d/x_s$ is the normalized transverse separation between the neighboring waveguides, $\lambda_d$ is the laser wavelength in vacuum, and $n_0$ is the average medium refractive index.

In the continuous approximation, we describe the wave profiles with a continuous function, $\Psi_n(t,z) = \tilde{u}(x = nd, t, z)$, and expand $\tilde{u}$ in Taylor series deriving the equation (see details in Ref. [201]),

$$i\frac{\partial u}{\partial z} + \frac{\lambda(z)}{2}\frac{\partial^2 u}{\partial x^2} + \frac{\sigma_D}{2}\frac{\partial^2 u}{\partial t^2} + \sigma_\gamma |u|^2 u = 0, \tag{115}$$

where

$$\lambda(z) = \cos(B \sin z), \quad B = -x_{0m}\frac{4\pi^2 n_0 d_x x_s^2}{\lambda_d z_s}. \tag{116}$$

and $\sigma_D = \pm 1$, $\sigma_\gamma = \pm 1$.

In self-focusing media we have $\sigma_\gamma = 1$, whereas $\sigma_\gamma = -1$ in case of self-defocusing nonlinearity. In order to achieve soliton localization in temporal dimension we require that the temporal dispersion is either anomalous in self-focusing media [260], or dispersion is normal if the nonlinearity is self-defocusing, i.e. $\sigma_D \sigma_\gamma > 0$.

To analyze further Eq. ((115)), Matuszewski *et al.* [201] employed the variational method [192]. with the Gaussian ansatz,

$$u = A(z) \exp\left\{i\phi(z) - \frac{1}{2}\left[\frac{x^2}{W^2(z)} + \frac{t^2}{T^2(z)}\right] + \frac{i}{2}\left[b(z) x^2 + \beta(z) t^2\right]\right\}, \tag{117}$$

where $A$ and $\phi$ are the amplitude and phase of the soliton, $T$ and $W$ are its temporal and transverse spatial widths, and $\beta$ and $b$ are the temporal and spatial chirps.

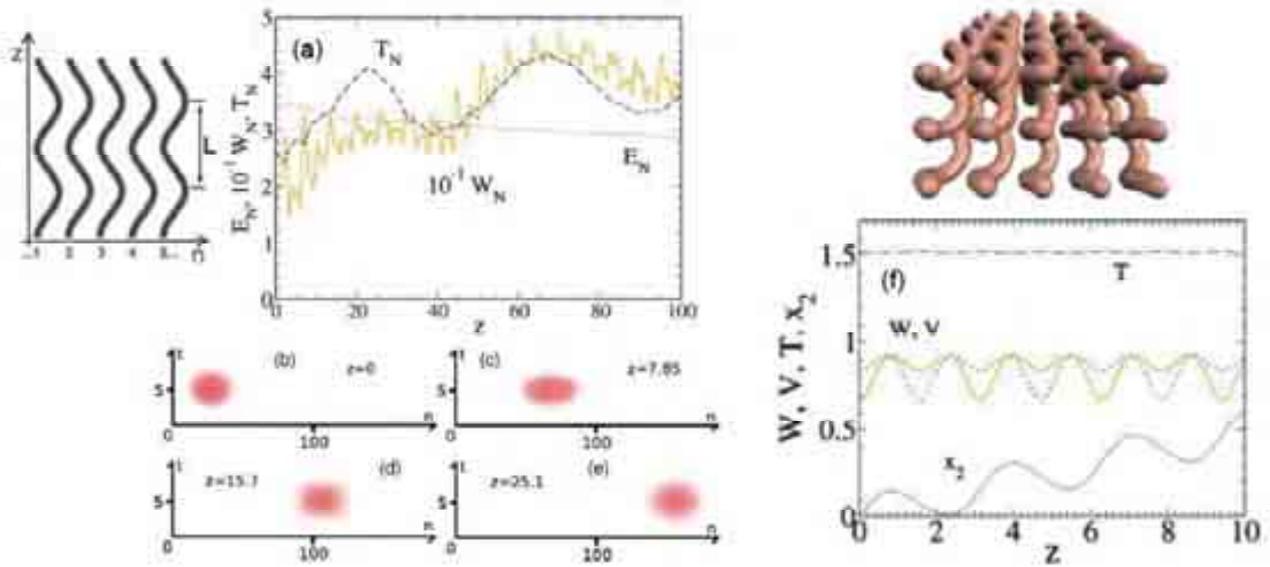

Figure 54: (a-e) Stable two-dimensional light bullet moving across a one-dimensioual array of curved waveguides (shown in the insert). Parameters are $B = 2.1$, $C = 100$, $D = \gamma = 1$, and the initial Gaussian shape corresponds to the variational solutions with $E = 1.1$, $W(0) = 0.68$, $T(0) = 1.55$, $b(0) = 0$, $T'(0) = 0$, multiplied by a spatial chirp function $\exp(ikn)$ with $k = \pi/20$. (a) The spatial and temporal widths $W_N$ and $T_N$, and the total energy $E_N$ of the propagating pulse found by the numerical integration of the discrete propagation equation. (b)-(e) Snapshots of the spatiotemporal pulse shape at several propagation distances. (f) Stable three-dimensional light bullet moving across the helical array. The insert shows sketch of helical waveguide array in three dimensions. The transverse spatial widths $W$ (solid), $V$ (dotted) and the temporal width $T$ (dashed) of the light bullet are shown as functions of the propagation distance $z$. Solid line below shows the position of the moving pulse according to Eq. (119) for $k_x = 0.3$. After Ref. [201].

The systematic results obtained using the variational model suggest that for a fixed value of the structural parameter $B$, there exists a family of soliton solutions. These solutions can be characterized by the pules energy $E$, the phase accumulated over one period $\Delta\phi$ (this parameter has a similar meaning to the propagation constant for solitons in homogeneous structures), and the diffraction map strength $S$ defined in analogy with the dispersion map strength of temporal dispersion-managed solitons [11, 194]:

$$S = \int_0^L |\lambda(z)| \frac{dz}{W_{\text{FWHM}}^2}, \qquad (118)$$

where $W_{\text{FWHM}}$ is the full width at half maximum of the wave packet at the point it is the narrowest along the $x$ direction. The value of $E$ at $\Delta\phi \to 0$ is equal to the critical energy of the Townes soliton $E_{cr}$. The energy is a monotonously growing

We note that the equation (115) admits solutions moving in both spatial and temporal dimensions. For example, any solution $u(x,t,z)$ can be multiplied by a linear phase which gives a new solution $u(x - x_2)\exp[ikx + i\phi_2(z)]$, where

$$\frac{dx_2}{dz} = k\lambda(z). \qquad (119)$$

Hence, pulses can move along $x$-direction during the propagation, and this movement gives rise to an averaged soliton drift if the mean diffraction $\langle\lambda\rangle$ is nonzero.

In Fig. 54 we present an example of a stable light bullet which moves across the array according to numerical simulations of the full discrete-continuous propagation equation (113). Here, the initial condition was taken from the periodic solution of the variational equations multiplied by a spatial chirp function $\exp(ikn)$ with $k = \pi/20$, which corresponds to the initial beam tilt. As depicted in Fig. 54(a), the spatial and temporal widths of the soliton oscillate during the propagation in an irregular pattern, but dynamical stability is achieved over long propagation distances. The widths in the numerical simulations are calculated according to the formula

$$W_N = 3\langle |n - \langle n\rangle| \rangle, \qquad (120)$$

where $\langle n \rangle = \sum_n \int n|a_n(t,z)|^2 dt / \sum_n \int |a_n(t,z)|^2 dt$. It gives approximately the FWHM value for a Gaussian pulse. The numerically calculated energy is defined as

$$E_N = d_x \sum_n \int |\Psi_n(t,z)|^2 dt. \tag{121}$$

Moving solitons in lattices are known to exhibit radiative losses [325]. In our simulations, we observe that the energy of the light bullet slowly decreases mainly due to radiation in the waveguide sections with the highest curvature. This radiation is more pronounced for tighter bending. In Fig. 54(b)-(e) the spatiotemporal profiles of the light bullet are shown for several values of the propagation distance $z$. Here, the solution is shown in the reference frame moving along the periodic translation with zero mean shift resulting from the waveguide bending. The stable evolution presented in Fig. 54 corresponds to typical physical parameters for optical pulses $W_d \simeq 100\,\mu$m and $T_d \simeq 50$ fs in photonic lattice with the physical characteristics of $d_d = 5\,\mu$m, $L_d = 10$ cm, $C_d \simeq 6.3$ mm$^{-1}$, and $D_d = 175$ fs$^2$ cm$^{-1}$.

This concept can be extended to higher dimensions [201] with the possibility to stabilize three-dimensional light bullets, while preserving their mobility. This is a more challenging problem, since two-dimensional light bullets in media with cubic nonlinearity exhibit critical collapse, and even weak perturbations can be sufficient to prevent the collapse effect [9]. On the other hand, the three-dimensional light bullets can exhibit super-critical collapse, which can only be prevented through strong changes in the physical system [9].

Matuszewski *et al.* [201] performed numerical analysis to identify periodic solutions of the variational equations. Such solutions correspond to three-dimensional light bullets, where the light-bullet extension along all dimensions $(x, y, t)$ is restored after each bending period. An example of such solution is presented in Fig. 54(f), which illustrates a stable and mobile three-dimensional light bullet. Indeed, the light bullet spatial and temporal localization widths are restored to the same values after each bending period. The solid line at the bottom shows the transverse position $x_2$ of the pulse center as a function of the propagation distance $z$ for a finite initial beam tilt. The light bullet wobbles in the transverse direction during the propagation, and it is clear that on average the pulse moves across the array.

*5.8.2. Diffraction-dispersion matching*

A somewhat similar approach to generate stable, fully three-dimensional light bullets was suggested by Lobanov *et al.* [144] who employed the matching of the intrinsic material dispersion with a suitable effective diffraction. This matching was achieved in a two-dimensional lattice periodically modulated along the direction of light propagation. It was shown that by using nonconventional, out-of-phase longitudinal modulation of the refractive index of neighboring waveguides, it is possible to tune the effective diffraction to match the intrinsic material group velocity dispersion. As a result, three-dimensional light bullets were shown to form at reduced energy levels, in settings where the dispersion would be far too weak to generate bullets in the absence of array.

In comparison with the discrete mobile bullets in curved waveguide arrays discussed in Ref. [201] and above, where the bullets extend over multiple waveguides and can move across the lattice, the approach developed by Lobanov *et al.* [144] leads to strongly localized states pinned by the lattice discreteness.

The model considered in Ref. [144] is based on the continuous NLS equation for the dimensionless field amplitude q, governing the propagation of spatiotemporal wave packets along the $z$ axis of the waveguide array with a longitudinally modulated refractive index,

$$i\frac{\partial E}{\partial \xi} = -\frac{1}{2}\left(\frac{\partial^2 E}{\partial x^2} + \frac{\partial^2 E}{\partial y^2}\right) - \frac{\beta}{2}\frac{\partial^2 E}{\partial \tau^2} - \Delta n V E - |E|^2 E. \tag{122}$$

The function $V \equiv V(x, y, z)$ describes the refractive index distribution describing the array of Gaussian waveguides with centers $(x_k, y_k)$ placed in the nodes of a honeycomb lattice. The refractive index in neighboring waveguides is modulated out-of-phase along the $z$ axis; i.e., if in central waveguide the refractive index oscillates as $(1 + \mu \sin(\Omega z))$, in all adjacent waveguides it changes as $(1 - \mu \sin(\Omega z))$; where $\Omega$ is the modulation frequency. For $\Omega = \Omega_r$

In the absence of dispersion ($\beta = 0$), diffraction is almost completely inhibited even for linear single-channel excitations for certain resonant modulation frequencies, the primary frequency is denoted as $\Omega_r$.

Figure 55 demonstrates the dynamics of bullet excitation in a nonresonant case. If the input amplitude is not high enough then one observes simultaneous spreading in time and in space [Figs. 55(a,b)]. If the amplitude is close to the optimal one so that the value of distance-averaged energy is sufficiently high, the light bullet forms [Fig. 55(c)]. The corresponding amplitude is notably larger than the amplitude required for bullet formation in a resonant system—clear evidence of the strong effect of diffraction engineering on bullet formation in such structures. Similar results were also obtained for different pulse durations and refractive index modulation depths [144].

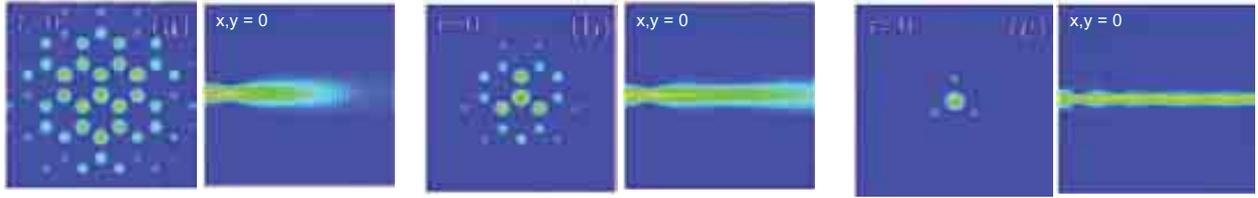

Figure 55: Field modulus distributions at $\tau = 0$ (top panels) and temporal dynamics at $x, y = 0$ (bottom panels) for single-site excitations in modulated honeycomb lattice at $(\Omega - \Omega_r)/\Omega_r = -0.2$, and (a) the input amplitude $E_0 = 0.45$, (b) 0.47, and 0.45, (b) 0.47, and (c) 0.53. [144].

Finally, we mention the stable light bullets can be supported by other types of spatially modulated lattices such as Bessel optical lattices with out-of-phase modulation of the linear and nonlinear refractive indices [321]. As was shown by Fangwei Ye *et al.* [321], spatial modulation of the nonlinearity significantly modifies the shapes and stability domains of the light bullets. The addressed bullets can be stable, provided that the peak intensity does not exceed a critical value, and the width of the stability domain in terms of the propagation constant may be controlled by varying the nonlinearity modulation depth. The maximum energy of the stable bullets was shown to grow with increasing nonlinearity modulation depth.

## 6. Summary

As we discussed above, the recent progress in the field of modulated photonic lattices resulted in a series of pioneering demonstrations of tailored diffraction and dispersion management (Sections 2 and 3), engineered surface and defect-induced localization (Sec. 4), as well as tunable nonlinear interaction (Sec. 5). The unique opportunities which the modulated photonic lattices offer for the control and manipulation of light are expected to find many practical applications, in particular in optical data processing and storage photonic devices that can be used in optical communication networks and integrated photonic signal processors.

However, up to now most of the studies of light propagation in modulated photonic lattices were done for the regime of light-beam propagation and a constant cw excitation. While several recent works considered the propagation of optical pulses in straight waveguide arrays, predicting novel intriguing phenomena such as the generation of optical X-waves [47, 46, 93] and and spatiotemporal self-trapped states known as light bullets [209], an interplay between the spatial and temporal degrees of freedom in modulated photonic lattices is largely unexplored. Additional diffraction and dispersion management which is possible in modulated photonic lattices can strongly affect the propagation of spatiotemporal wavepackets, and this direction deserves further attention. Temporal effects may also be important, e.g., for a nonlinear switching of light pulses in broadband couplers which were considered briefly in Sec. 5.5 above.

Modulated photonic lattices with engineered diffraction which were discussed in Sec. 2 are very much similar to other types of periodic photonic structures with engineered dispersion, including photonic crystals. In particular, an all-angle self-collimation was recently demonstrated in specially designed photonic crystals [318, 85], which dispersion is similar to the dispersion of the modulated lattices with defect-free surface states considered in Sec. 4. Therefore, a similar type of surface-mediated localization could also be expected to take place in the photonic crystals with broadband self-collimation.

The experimental studies of modulated photonic lattices were done using conventional fabrication platforms, such as femtosecond laser direct-writing in silica glass and titanium indiffusion in lithium niobate crystals, which allow fabrication of micron-size waveguides. On the other hand, the tremendous progress in the field of nanotechnology made possible fabrication of photonic nanowaveguides with subwavelength dimensions less than 100 nm [241, 62, 141]. Strong power enhancement in a tiny volume leads to very high optical intensities in nanowaveguides. Additionally, dispersion in nanowaveguides is dominated by the geometry of the system, and it can be tailored to achieve extremely large positive and negative values not available in conventional micron-size waveguides. It was shown recently that the well-known modulation effects can be dramatically modified at the nanoscale, such as recently discovered diverging Rabi oscillations in nanowaveguide arrays [4]. Nonlinear waves in subwavelength waveguide arrays were shown to exhibit unique features, such as recently discovered self-reviving phoenix solitons originating solely form evanescent bands [225], and novel subwavelength discrete solitons in nonlinear metamaterials [143]. Self-imaging and diffraction management was recently achieved in arrays of coupled metal-dielectric nanowaveguides with a periodically curved axis by a suitable engineering of the bending geometry [307]. Thus, it is most interesting to combine modulation-driven tailored diffraction and dispersion characteristics with the unprecedented enhancement of nonlinear effects in nanowaveguide lattices to realize ultra-fast all-optical control of optical pulses, including spatiotemporal pulse shaping and switching between multiple array ports.

It should be mentioned that modulation effects play an important role in other photonic structures, such as recently demonstrated Bloch oscillations and Zener tunneling in photonic topological crystals [285], and resonant tunneling and frustrated total internal reflection at periodically curved interfaces [151]

Finally, we would like to mention that many effects revealed for modulated photonic lattices are of a general nature, and they can also be applied to other physical systems supporting propagation of waves in periodic potentials. For example, dynamic localization of a Bose-Einstein condensate in a shaken optical lattice was recently experimental observed for sinusoidal and square-wave forcing [50]. There also exists a deep analogy between photon propagation in photonic lattices and electron motion in metals and semiconductor crystals. These analogies may expand a range of systems where the effects suggested for modulated photonic lattices may also be observed, facilitating the knowledge exchange between seemingly different fields.

## 7. Appendices

### 7.1. Generalized tight-binding approximation
#### 7.1.1. Two-dimensional curved waveguide arrays

We consider propagation of optical beams in a two-dimensional array of coupled optical waveguides, where the waveguide axes are periodically curved in the longitudinal propagation direction. In the linear regime, the beam dynamics is defined by the independent evolution of the complex envelopes $E(x, y, z)$ of the electric field at the different optical wavelengths $\lambda$. In the case of weak refractive index contrast, which is satisfied, e.g., for the laser-written structures in glass [282], the field evolution is governed by the normalized paraxial equation,

$$i\frac{\partial E}{\partial z} + \frac{\lambda}{4\pi n_0}\left(\frac{\partial^2 E}{\partial x^2} + \frac{\partial^2 E}{\partial y^2}\right) + \frac{2\pi}{\lambda} V\left[x - x_0(z), y - y_0(z)\right] E = 0. \tag{123}$$

Here $x$ and $y$ are the transverse coordinates, $z$ is the propagation coordinate, $\lambda$ is the vacuum wavelength, $c$ is the speed of light, and $n_0$ is the average refractive index of the medium. The functions $x_0(z)$ and $y_0(z)$ determine the transverse shift of the whole lattice depending on the propagation distance $z$. The function $V(x, y)$ describes the refractive index modulation in the transverse cross-section. For the waveguide array,

$$V(x, y) = \sum_{n,m} V_0(x - x_{n,m}, y - y_{n,m}), \tag{124}$$

where $V_0(x, y)$ is the refractive index profile of a single waveguide, $(x_{n,m}, y_{n,m})$ are the waveguide positions at the input facet, and $n$ and $m$ are the discrete waveguide numbers.

When the tilt of beams and waveguides at the input facet is less than the Bragg angle, the beam propagation is primarily characterized by coupling between the fundamental modes of the individual waveguides, and it can be described by the tight-binding coupled equations [152, 183]. Specifically, we represent the field as a sum of individual waveguide modes,

$$\begin{aligned}
E = \sum_{n,m} \Psi_{n,m}(z) E_0 \left[x - x_{n,m} - x_0(z), y - y_{n,m} - y_0(z), z\right] \\
\exp\left\{2ip\dot{x}_0(z)[x - x_{n,m} - x_0(z)] + 2ip\dot{y}_0(z)[y - y_{n,m} - y_0(z)]\right\} \\
\exp\left\{ip \int_0^z [\dot{x}_0(\xi)]^2 d\xi + ip \int_0^z [\dot{y}_0(\xi)]^2 d\xi\right\},
\end{aligned} \tag{125}$$

where $\Psi_{n,m}$ are the mode amplitudes, the dots stand for the derivatives, $E_0$ is the mode of individual straight waveguide, and $p = \pi n_0/\lambda$.

Following the standard procedure, we substitute Eq. (125) into Eq. (123), multiply the resulting expressions by $E_0^*[x - x_{n',m'} - x_0(z), y - y_{n',m'} - y_0(z), z]$, and integrate over the transverse dimensions $(x, y)$. Then, in the leading order approximation we derive a set of coupled equations for the mode amplitudes,

$$i\frac{d\Psi_{n,m}}{dz} + \sum_{n',m' \neq n,m} \widetilde{C}_{n,n',m,m'} \Psi_{n',m'} = 0, \tag{126}$$

where

$$\begin{aligned}
\widetilde{C}_{n,n',m,m'} &= \exp[-2ip\dot{x}_0(z)(x_{n',m'} - x_{n,m}) - 2ip\dot{y}_0(z)(y_{n',m'} - y_{n,m})] \\
&\quad \int\int E_0[x - x_{n,m} - x_0(0), y - y_{n,m} - y_0(0), 0] \\
&\quad E_0^*[x - x_{n',m'} - x_0(0), y - y_{n',m'} - y_0(0), 0] \\
&\quad [V(x, y) - V_0(x - x_{n,m}, y - y_{n,m})]\, dx\, dy \\
&\quad \left[\int\int |E_0[x - x_{n,m} - x_0(0), y - y_{n,m} - y_0(0), 0]|^2 dx\, dy\right]^{-1}.
\end{aligned} \tag{127}$$

These expressions show that the effect of periodic bending appears through the modifications of phases of the coupling coefficients along the propagation direction $z$.

*7.1.2. One-dimensional curved waveguide arrays*

In the case of one-dimensional periodically curved waveguide arrays, Eq. (126) simplifies,

$$i\frac{d\Psi_n}{dz} + \sum_{n' \neq n} \widetilde{C}_{n,n'} \Psi_{n'} = 0, \qquad (128)$$

where

$$\widetilde{C}_{n,n'} = \exp[-2ip\dot{x}_0(z)(x_{n'} - x_n)]$$
$$\int E_0[x - x_n - x_0(0), 0] E_0^*[x - x_{n'} - x_0(0), 0][V(x) - V_0(x - x_n)] \, dx \qquad (129)$$
$$\left[\int |E_0[x - x_n - x_0(0), 0]|^2 \, dx\right]^{-1}.$$

We note, that by the substitution $z \to t$ in Eq. (128) the spatial light evolution in a curved optical waveguide array becomes fully analogous to the temporal evolution of a quantum particle in an externally driven periodic potential, where $\Psi_n(t)$ play the role of the probability amplitudes, $C_{n,n'}$ is replaced by the nearest-neighbor transfer-matrix element, and the driving force $x_0(t)$ is determined by the external electric field [49, 168].

If the waveguide array is composed of identical waveguides with the same center-to-center spacing $d$, such that $x_n = nd$, then

$$\widetilde{C}_{n,n'} \equiv \widetilde{C}_{n-n'} \qquad (130)$$

as a result of the translation invariance in the transverse direction.

If we further assume that the coupling between the waveguides is weak so that the overlap integrals vanish for $|n - n'| > 1$, we finally arrive at the following expression for the nearest-neighbor coupling in the curved waveguide array,

$$\widetilde{C}_{n,n'} = \begin{cases} C \exp\left[-2ipd\dot{x}_0(z)(n' - n)\right], & \text{if } |n - n'| = 1 \\ 0, & \text{if } |n - n'| > 1, \end{cases} \qquad (131)$$

where

$$C = \frac{\int E_0[x, 0] E_0^*[x + d, 0][V(x) - V_0(x)] \, dx}{\int |E_0[x, 0]|^2 \, dx} \qquad (132)$$

is the coupling coefficient which in our approximation is equal to the coupling coefficient in the straight waveguide array.